%% file: SUS-13-006_temp.tex
\pdfoutput=1

\documentclass[11pt,twoside,a4paper,cmspaper,final,collab]{cms-tdr}
\def\svnVersion{266929}\def\cmsCernNo{2013-037}\def\cmsCernDate{\today}\def\cmsMessage{Published in the European Physical Journal C as \href{http://dx.doi.org/10.1140/epjc/s10052-014-3036-7}{\doi{10.1140/epjc/s10052-014-3036-7.}}}

\begin{document}\cmsNoteHeader{SUS-13-006}

\hyphenation{had-ron-i-za-tion}
\hyphenation{cal-or-i-me-ter}
\hyphenation{de-vices}
\RCS$Revision: 255568 $
\RCS$HeadURL: svn+ssh://svn.cern.ch/reps/tdr2/papers/SUS-13-006/trunk/SUS-13-006.tex $
\RCS$Id: SUS-13-006.tex 255568 2014-08-11 15:15:01Z alverson $
\ifthenelse{\boolean{cms@external}}{\providecommand{\cmsLeft}{top}}{\providecommand{\cmsLeft}{left}}
\ifthenelse{\boolean{cms@external}}{\providecommand{\cmsRight}{bottom}}{\providecommand{\cmsRight}{right}}
\newlength\cmsFigWidth
\ifthenelse{\boolean{cms@external}}{\setlength\cmsFigWidth{0.95\columnwidth}}{\setlength\cmsFigWidth{0.6\textwidth}}
\ifthenelse{\boolean{cms@external}}{\providecommand{\cmsRow}[1]{\relax{#1}}}{\providecommand{\cmsRow}[1]{\resizebox{\columnwidth}{!}{#1}}}
\ifthenelse{\boolean{cms@external}}{\providecommand{\breakhere}{\linebreak[4]}}{\providecommand{\breakhere}{\linebreak[4]}}
\ifthenelse{\boolean{cms@external}}{\providecommand{\cmsTLeft}{top}}{\providecommand{\cmsTLeft}{upper left}}
\ifthenelse{\boolean{cms@external}}{\providecommand{\cmsTCenter}{center}}{\providecommand{\cmsTCenter}{upper right}}
\ifthenelse{\boolean{cms@external}}{\providecommand{\cmsTRight}{bottom}}{\providecommand{\cmsTRight}{bottom}}

\newcommand{\slep}{\ensuremath{\widetilde{\ell}}\xspace}
\newcommand{\tauh}{\ensuremath{\tau_\mathrm {h}}\xspace}
\newcommand{\xslep}{\ensuremath{x_{\slep}}\xspace}
\newcommand{\snu}{\ensuremath{\widetilde{\nu}}\xspace}
\newcommand{\Irel}{\ensuremath{I_\text{rel}}\xspace}
\newcommand{\W}{\ensuremath{\cmsSymbolFace{W}}\xspace} % \Z is defined by default but not \W; there is \PW in PENNAMES
\newcommand{\ZZ}{\ensuremath{\cmsSymbolFace{ZZ}}\xspace}
\newcommand{\bjet}{\ensuremath{\cPqb\text{-jet}}\xspace}
\newcommand{\bjetnohyphen}{\ensuremath{\cPqb\ \text{jet}}\xspace}
\newcommand{\bjets}{\ensuremath{\cPqb\text{-jets}}\xspace}
\newcommand{\bjetsnohyphen}{\ensuremath{\cPqb\ \text{jets}}\xspace}

\newcommand{\bquark}{\ensuremath{\cPqb\text{ quark}}\xspace}
\newcommand{\cquark}{\ensuremath{\cPqc\text{ quark}}\xspace}
\newcommand{\udsgparton}{\ensuremath{\cPqu\cPqd\cPqs\cPg\text{ parton}}\xspace}
\newcommand{\zdijet}{\ensuremath{\Z+\mathrm{dijet}}}
\newcommand{\CTEQ} {{\textsc{cteq}}}
\newcommand{\mctp}{\ensuremath{M_{\mathrm{CT}\perp}}}
\newcommand{\mcpt}{\ensuremath{M_{\mathrm{CT}\perp}}}
\newcommand{\mct}{\ensuremath{M_{\mathrm{CT}}}}

\providecommand{\MT}{\ensuremath{M_\mathrm{T}}\xspace}
\newcommand{\MZ}{\ensuremath{M_{\Z}}\xspace}
\newcommand{\mdil}{\ensuremath{M_{\ell\ell}}\xspace}
\newcommand{\mjj}{\ensuremath{M_\text{jj}}\xspace}

\newcommand{\ST}{\ensuremath{S_{\mathrm{T}}}\xspace}

\newcommand{\zjets}{\ensuremath{\Z+\text{jets}}\xspace}
\newcommand{\wwjets}{\ensuremath{\W\W+\text{jets}}\xspace}
\newcommand{\wzjets}{\ensuremath{\W\Z+\text{jets}}\xspace}
\newcommand{\gjets}{\ensuremath{\gamma+\text{jets}}\xspace}

\newcommand{\zzmet}{$\Z\Z+\MET$\xspace}
\newcommand{\wzmet}{$\W\Z+\MET$\xspace}
\newcommand{\whmet}{$\W\PH+\MET$\xspace}
\newcommand{\zhmet}{$\Z\PH+\MET$\xspace}
\newcommand{\hhmet}{$\PH\PH+\MET$\xspace}

\newcommand{\wzzmet}{$\W\Z/\Z\Z+\MET$\xspace}
\newcommand{\cls}{CL$_\text{s}$\xspace}

\newcommand{\nbjets}{\ensuremath{N_{\text{b jets}}}\xspace}
\newcommand{\njets}{\ensuremath{N_{\text{jets}}}\xspace}

\newcommand{\chipmo}{\ensuremath{\chipm_{1}}}
\newcommand{\chitn}{\ensuremath{\chiz_{2}}}
\newcommand{\lsp}{\ensuremath{\chiz_{1}}}
\newcommand{\mchipmo}{\ensuremath{m_{\chipmo}}}
\newcommand{\mchitn}{\ensuremath{m_{\chitn}}}
\newcommand{\mlsp}{\ensuremath{m_{\lsp}}}
\newcommand{\mchi}{\ensuremath{m_{\widetilde{\chi}}}}
\newcommand{\hbb}{$\PH\to\bbbar$}
\newcommand{\hww}{$\PH\to\PWp\PWm$}
\newcommand{\hzz}{$\PH\to\Z\Z$}
\newcommand{\htt}{$\PH\to\tau^+\tau^-$}
\newcommand{\wjets}{\ensuremath{\PW+\text{jets}}}
\newcommand{\wl}{\ensuremath{\PW+\text{light-flavor jets}}}
\newcommand{\wbb}{\ensuremath{\PW+\bbbar}}
\newcommand{\wzbb}{$\PW\Z\to\ell\nu\bbbar$}
\newcommand{\whbb}{$\PW\PH\to\ell\nu\bbbar$}
\newcommand{\mtbl}{\ensuremath{M_{\mathrm{T2}}^{\mathrm{bl}}}}
\newcommand{\mtj}{\ensuremath{M_{\mathrm{T2}}^{\mathrm{J}}}}
\newcommand{\mt}{\ensuremath{M_{\mathrm{T}}} }
\newcommand{\Ht}{\ensuremath{H_{\mathrm{T}}} }
\newcommand{\mbb}{\ensuremath{M_{\bbbar}}}
\newcommand{\ttsl}{\ensuremath{\ttbar\to\ell+\mathrm{jets}}}
\newcommand{\mljj}{\ensuremath{M_{\ell \text{j j} }}}
\newcommand{\signal}{\ensuremath{ \chipmo\chitn \to (\PW \lsp)(\PH \lsp)}}

\cmsNoteHeader{SUS-13-006} % This is over-written in the CMS environment: useful as preprint no. for export versions
\title{Searches for electroweak production of charginos, neutralinos, and sleptons decaying to leptons and W, Z, and Higgs bosons in pp collisions at 8\TeV}

\titlerunning{Searches for electroweak production of supersymmetric particles decaying to leptons and W, Z, H}
\date{\today}

\abstract{
Searches for the direct electroweak production of supersymmetric charginos, neutralinos, and sleptons in a variety of signatures
with leptons and \PW, \Z, and Higgs bosons are presented.
Results are based on a sample of proton-proton collision data collected at center-of-mass
energy $\sqrt{s}=8\TeV$ with the CMS detector in 2012, corresponding to an integrated luminosity of 19.5\fbinv.
The observed event rates are in agreement with expectations from the standard model. These results
probe charginos and neutralinos with masses up to 720\GeV, and sleptons up to 260\GeV, depending on the model details.
}

\hypersetup{%
pdfauthor={CMS Collaboration},%
pdftitle={Searches for electroweak production of charginos, neutralinos, and sleptons decaying to leptons and W, Z, and Higgs bosons in pp collisions at 8 TeV},
pdfsubject={CMS},%
pdfkeywords={CMS, physics, supersymmetry}}

\maketitle

\section{Introduction}
\label{introduction}
Many searches for supersymmetry
(SUSY)~\cite{Golfand:1971iw,Wess:1973kz,Wess:1974tw,Fayet1,Fayet2}
carried out at the CERN Large
Hadron Collider (LHC) have focused on models with cross sections
dominated by the production of strongly interacting new particles
in final states with high levels of hadronic activity~\cite{CMSSUSY1,CMSSUSY2,CMSSUSY3,SUS-11-021-paper,CMSSUSY5,CMSSUSY6,CMSSUSY7,SUS-11-013-paper,SUS13011,CMSSUSY9,CMSSUSY10,CMS-PAS-SUS-13-013}.
Null results from these searches constrain the squarks and gluinos to be heavier than several hundred\GeV.
In contrast, in this paper, we describe searches motivated by the
direct electroweak production of charginos $\chipm$ and neutralinos $\chiz$,
mixtures of the SUSY partners of the gauge and Higgs bosons, and of sleptons $\slep$,
the SUSY partners of leptons.
These production modes may dominate at the LHC if the strongly interacting SUSY particles are heavy.
The corresponding final states do not necessarily contain much hadronic activity and thus may have eluded detection.

The smaller cross sections typical of direct electroweak SUSY production require dedicated searches
targeting the wide variety of possible signal topologies.
Depending on the mass spectrum, the charginos and neutralinos can have significant decay branching
fractions to leptons or $\W$, $\Z$, and Higgs bosons (H), yielding
final states with at least one isolated lepton.  Similarly, slepton pair production gives rise to
final states with two leptons.
In all these cases, and under the assumption of R-parity conservation~\cite{Fayet2}, two stable, lightest SUSY particles (LSP) are produced,
which are presumed to escape without detection,
leading to significant missing transverse energy $\MET$.
We thus search for SUSY in a variety of final states with one or more leptons and $\MET$.

\begin{figure}[htb]
\centering
\includegraphics[width=0.32\textwidth]{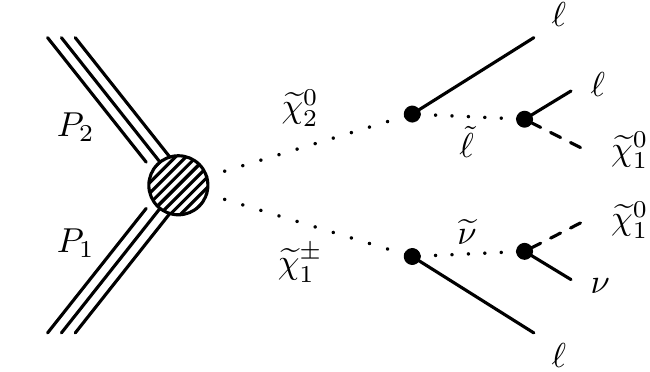}
\includegraphics[width=0.32\textwidth]{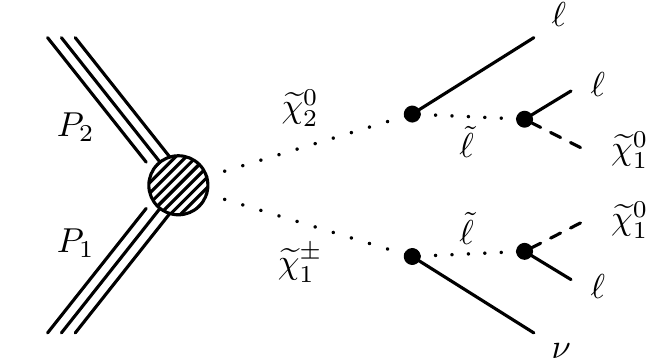}
\caption{Chargino-neutralino pair production with decays mediated by sleptons and sneutrinos, leading to a
three-lepton final state with missing transverse energy \MET.
\label{fig:charginos-slep}
}
\end{figure}

\begin{figure}[htb]
\centering
\includegraphics[width=0.32\textwidth]{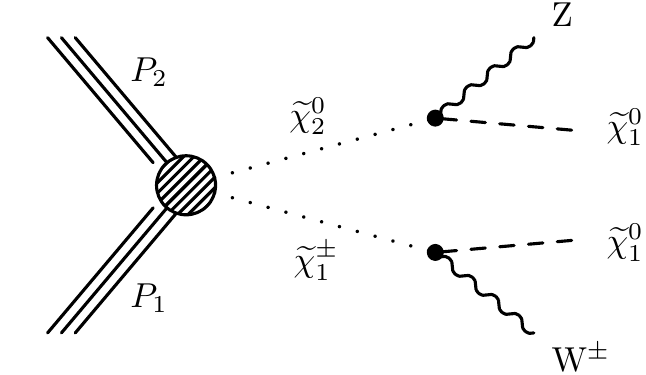}
\includegraphics[width=0.32\textwidth]{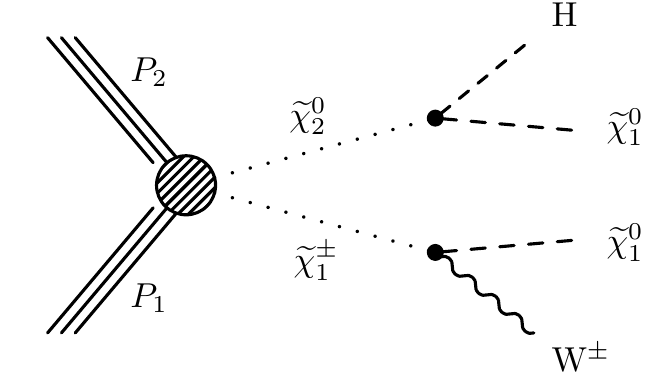}
\includegraphics[width=0.32\textwidth]{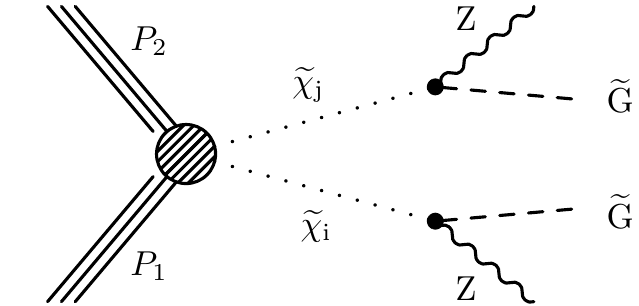}
\caption{Chargino-neutralino production, with the chargino decaying to a \PW\ boson
and the LSP, and with the neutralino decaying to (\cmsLeft) a \Z boson and the LSP or
(center) a Higgs boson and the LSP; (\cmsRight) a GMSB model with higgsino pair production, with $\tilde{\chi}_i$ and $\tilde{\chi}_j$ indicating nearly mass-degenerate charginos and neutralinos, leading to the \zzmet final state.
\label{fig:charginos-wz}
}
\end{figure}

\begin{figure}[htb]
\centering
\includegraphics[width=0.32\textwidth]{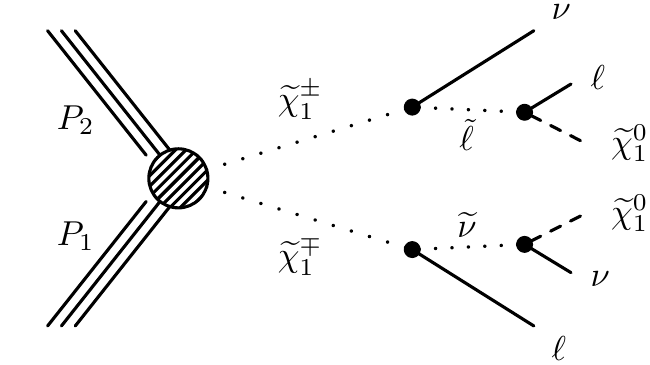}
\includegraphics[width=0.32\textwidth]{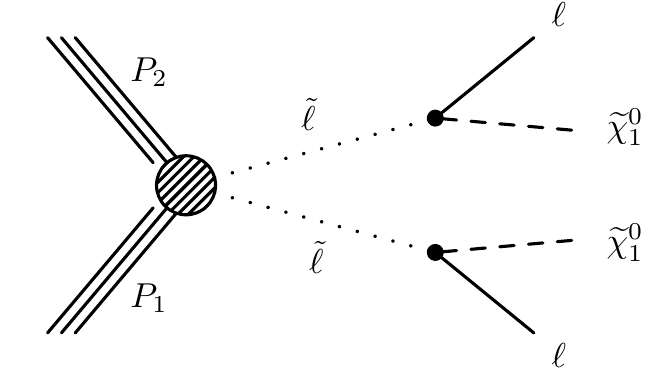}
\caption{(\cmsLeft) Chargino, and (\cmsRight) slepton pair production
leading to opposite-sign lepton pairs with $\MET$.
\label{fig:charginos-ll}}
\end{figure}

The searches are based on a sample of proton-proton (pp) collision data collected
at $\sqrt{s}=8\TeV$ with the Compact Muon Solenoid (CMS) detector at the LHC in 2012,
corresponding to an integrated luminosity of 19.5\fbinv.
The study is an update of Ref.~\cite{SUS-12-006-paper},
with improvements to the analysis techniques and the addition of new signal scenarios and search channels.
Similar studies in the two-lepton, three-lepton, and four-lepton final states have been performed by the ATLAS Collaboration~\cite{ATLAS1,ATLAS2,ATLAS3}.
The new-physics scenarios we consider are shown in
Figs.~\ref{fig:charginos-slep}--\ref{fig:charginos-ll}.
These figures are labeled using SUSY nomenclature,
but the interpretation of our results can be extended to
other new-physics models.
In SUSY nomenclature,
$\chiz_1$ is the lightest neutralino,
presumed to be the LSP,
$\chiz_2$ is a heavier neutralino,
$\chipm_1$ is the lightest chargino,
and $\slep$ is a slepton. We also consider a model in which the gravitino ($\sGra$) is the LSP.

The results are interpreted considering each diagram in
Figs.~\ref{fig:charginos-slep}--\ref{fig:charginos-ll}
individually. The masses of the new-physics particles are treated
as independent parameters.
SUSY models with a bino-like $\chiz_1$ and wino-like $\chiz_2$
and $\chipm_1$
motivate the simplifying assumption $\mchi \equiv \mchipmo = \mchitn$
since these two gauginos belong to the same gauge group multiplet.
We thus present results as a function of the common
mass $\mchi$ and the LSP mass $m_{\chiz_1}$.

In the models shown in Figs.~\ref{fig:charginos-slep} and~\ref{fig:charginos-ll}(\cmsLeft),
the slepton mass $m_{\slep}$ is less than the common mass $\mchi$, and the sleptons
are produced in the decay chains of the charginos and neutralinos.
The results in these scenarios also depend on the mass $m_{\slep}$
of the intermediate slepton (if left-handed, taken to be the
same for its sneutrino $\snu$), parametrized in terms
of a variable $\xslep$ as:
\begin{equation}
  m_{\slep} = m_{\snu} = m_{\chiz_1} +  \xslep\, ( \mchi - m_{\chiz_1}\,),
\end{equation}
where $0<\xslep<1$. We present results for $\xslep=0.50$, \ie, the slepton mass equal to the mean of the LSP and the $\widetilde{\chi}$
masses, and in some cases for more compressed spectra with $\xslep=0.05$ or~0.95, \ie, the
slepton mass close to either the LSP or the $\widetilde{\chi}$ mass, respectively.

For the models in Fig.~\ref{fig:charginos-wz}, we
assume that sleptons are so massive that diagrams containing virtual or real sleptons
in the chargino or neutralino decay process can be ignored.
In Figs.~\ref{fig:charginos-wz}(\cmsLeft) and \ref{fig:charginos-wz}(center),
the chargino decays to a \PW\ boson and the LSP, while the neutralino
may decay either to a \Z or \PH boson and the LSP, with branching fractions that
depend on model details. The \PH boson is identified with the lightest neutral CP-even state of extended Higgs sectors.
The \PH boson is expected to have SM Higgs boson properties if all other Higgs bosons are much heavier~\cite{SUSYprimer}.
We thus search in both the \wzmet\ and \whmet\ signatures.
There is little sensitivity to the \ZZ channel of
Fig.~\ref{fig:charginos-wz}(\cmsRight) if the $\chiz_2$ and $\chipm_1$
are wino-like,
in which case neutralino pair production is suppressed relative to
neutralino-chargino production.
Therefore, for the \ZZ signature, we
consider a specific gauge-mediated supersymmetry breaking (GMSB)
model with higgsino next-to-lightest SUSY particles (NLSP) and a gravitino LSP~\cite{Matchev:1999ft,Meade:2009qv,ref:ewkino},
which enhances the $\ZZ + \MET$ production rate. In this model, the $\chiz_2$ and $\chipm_1$ particles are nearly mass degenerate with the $\chiz_1$ NLSP, and each decay to the $\chiz_1$ through the emission of low-\pt, undetected SM particles.  The $\chiz_1$ then decays to a \Z boson and the gravitino LSP. The production of the \hhmet\ and \zhmet\ final states is also possible in the GMSB model, depending on the character of the NLSP.  These latter two final states are not considered in the current study.

Figure~\ref{fig:charginos-ll}(\cmsLeft) depicts chargino pair production.
For this process, each chargino can decay via either of the two modes shown.
Thus, there are four different decay pairs, but all yield a similar
final state, with two opposite-sign leptons and $\MET$.
For this model, we consider $\xslep=0.5$ only.
Figure~\ref{fig:charginos-ll}(\cmsRight) illustrates slepton
pair production, where each slepton decays to a lepton of
the same flavor and to the LSP. We
consider left- and right-handed slepton production separately, and assume
a universal mass for both the selectron and smuon.
The results of this analysis are not sensitive to the direct production of $\tau$-slepton pairs.

This paper is organized as follows. In Section~\ref{detector}, we describe the detector, data and simulated samples,
and event reconstruction procedures. Section~\ref{trilepton} presents
a search based on
the three-lepton final states
of Figs.~\ref{fig:charginos-slep} and~\ref{fig:charginos-wz}(\cmsLeft). A search based on
the four-lepton final state, which is sensitive to the diagram of Fig.~\ref{fig:charginos-wz}(\cmsRight),
is presented in Section~\ref{quadlepton}. Section~\ref{dilepton}
describes a search in a channel with exactly
two same-sign dileptons, which enhances sensitivity to the diagrams of
Fig.~\ref{fig:charginos-slep}
in cases where one of the three leptons is not identified. In
Section~\ref{diboson} we present a search based on the \wzzmet signature,
which is sensitive to the diagrams shown in Figs.~\ref{fig:charginos-wz}(\cmsLeft) and \ref{fig:charginos-wz}(\cmsRight).
Section~\ref{sec:wh} presents a set of searches targeting \whmet\ production
in the single-lepton, same-sign dilepton, and three-or-more-lepton channels,
probing the diagram
of Fig.~\ref{fig:charginos-wz}(center).
In Section~\ref{osdilepton},
we present a search based on an opposite-sign, non-resonant
   dilepton pair (electrons and muons), which is sensitive to the processes of
Fig.~\ref{fig:charginos-ll}.
Section~\ref{sec:interpretation} presents interpretations of these searches and Section~\ref{sec:summary} a summary.

\section{Detector, trigger, and physics object selection}
\label{detector}

The central feature of the CMS apparatus is a superconducting
solenoid, of 6\unit{m} internal diameter, providing a magnetic field of
3.8\unit{T}. Within the field volume are a silicon pixel and strip
tracker, a crystal electromagnetic calorimeter, and a
brass-scintillator hadron calorimeter. Muons are measured with
gas-ionization detectors embedded in the steel flux-return yoke of the solenoid.
A detailed description can be found in Ref.~\cite{:2008zzk}.

The origin of the coordinate system is the nominal interaction point.
The $x$ axis points to the center of the LHC ring and the $y$ axis
vertically upwards.
The $z$ axis lies in the direction of the counterclockwise proton beam.
The polar angle $\theta$ is measured from the positive $z$
axis, and the azimuthal angle $\phi$ (in radians) is measured in the
$x$-$y$ plane. The pseudorapidity $\eta$ is
defined by $\eta = -\ln [\tan (\theta/2)]$.

Events from pp interactions must satisfy the requirements of a
two-level trigger system. The first level performs a fast selection
of physics objects (jets, muons, electrons, and photons) above
certain thresholds.  The second level performs a full event
reconstruction.
The principal trigger used for the searches with two or more leptons
is a dilepton trigger.
It requires at least one electron or muon with transverse momentum
$\pt>17\GeV$ and another with $\pt>8\GeV$.  The trigger used for the single-lepton final
state requires a single electron (muon) with $\pt>27\,(24)\GeV$.
All leptons must satisfy $\abs{\eta}<2.4$.

Simulated event samples are used to study the characteristics of
signal and standard model (SM) background processes,
using the CTEQ6L1~\cite{PhysRevD.78.013004} parton distribution functions.
The main backgrounds are from top-quark pair (\ttbar), diboson, \zjets, and \wjets\ processes,
depending on the channel considered.
Most of the simulated SM background
samples are produced with the \MADGRAPH 5.1.5.4~\cite{Alwall:2011uj}
event generator, with parton
showering and hadronization performed with the \PYTHIA 6.4.26~\cite{Sjostrand:2006za} program.
We use the most accurate calculations of the cross sections available,
generally with next-to-leading-order (NLO)
accuracy~\cite{Campbell:2012top,Campbell:2011bn,xsec_WZ}.
The detector response is
modeled with the \GEANTfour~\cite{Geant}
library, followed by the same event reconstruction
as used for data.

Signal samples are generated with
the \MADGRAPH 5.1.5.4 generator including up to two additional partons at the matrix element level.
Parton showering, hadronization, and the decay of particles, including SUSY particles, are described with the \PYTHIA 6.4.26~\cite{Sjostrand:2006za} program.
Signal cross sections are calculated at NLO+NLL using the \textsc{Resummino}~\cite{Fuks:2012qx,Fuks:2013vua,Fuks:2013lya} calculation,
where NLL refers to the next-to-leading-logarithmic precision.
For the SUSY samples with a Higgs boson (\PH) in the final state, a mass of $m_\PH = 126$\GeV~\cite{ZZ4l}
is assumed, along with SM branching fractions.
Here the \PH particle indicates the lightest neutral CP-even SUSY Higgs boson, which is expected to have SM-like properties
if the other SUSY Higgs bosons are much heavier~\cite{SUSYprimer}.
To reduce computational requirements, the simulation of detector response for signal samples
is based on the CMS fast simulation program~\cite{Abdullin:1328345} in place
of \GEANTfour.

Events are reconstructed using the particle-flow (PF)
algorithm~\cite{PFT-09-001,PFT-10-001}, which provides a
self-consistent global assignment of momenta and energies to the physics objects.  Details of
the reconstruction and identification procedures  for electrons, muons, and photons are given in
Refs.~\cite{EGM-10-004, MUO-10-004,EGM-10-005}.
Lepton ($\Pe, \mu$) candidates are required to be consistent with
the primary event vertex, defined as the vertex with the largest
value of $\Sigma (\pt^{\text{track}})^2$, where the summation includes all
tracks associated to a given vertex.
In the searches with two or more leptons,
events with an opposite-sign $\Pe\Pe$, $\mu\mu$, or $\Pe\mu$
pair with an invariant mass below
12\GeV
are rejected in order to exclude quarkonia
  resonances, photon conversions, and low-mass continuum events.
  To reduce contamination due to leptons from heavy-flavor decay
  or misidentified hadrons in jets, leptons are required to be
  isolated and to have a transverse impact parameter with respect to the
primary vertex satisfying $d_0 < 0.2$\unit{mm}.
Electron and muon candidates are considered
  isolated if the ratio $\Irel$ of the scalar sum of the transverse momenta of charged
  hadrons, photons, and neutral hadrons in a cone of
$\Delta R = \sqrt{\smash[b]{(\Delta\eta)^2+(\Delta\phi)^2}}=0.3$
around the candidate, divided by the lepton
$\pt$
value, is less than 0.15. The requirements on the $d_0$ and $\Irel$ variables are more stringent
in the searches utilizing same-sign dileptons and are
described in Section~\ref{dilepton}.

The ``hadrons-plus-strips'' algorithm~\cite{CMS-PAS-TAU-11-001}, which combines PF photon and electron candidates to form neutral pions, and then the neutral  pions with charged hadrons, is used to identify hadronically decaying $\tau$-lepton candidates (\tauh).

Jets are reconstructed with the anti-\kt clustering
algorithm~\cite{Cacciari:2008gp} with a distance parameter of 0.5. We apply
\pt- and $\eta$-dependent corrections to account for residual
effects of non-uniform detector response~\cite{Chatrchyan:2011ds}.
A correction to account for multiple pp collisions within the same or a nearby
bunch crossing (pileup interactions) is estimated on an event-by-event basis using the
jet-area method described in Ref.~\cite{cacciari-2008-659}, and is
subtracted from the reconstructed jet \pt. We reject jets that are
consistent with anomalous noise in the calorimeters~\cite{HCAL}.
Jets must satisfy $\abs{\eta} < 2.5$ and $\pt > 30\GeV$ and
be separated by $\Delta R > 0.4$ from lepton candidates.
The searches presented below make use of the missing
transverse energy \MET, where \MET is defined as the modulus of
the vector sum of the transverse momenta of all PF objects.
The \MET vector is the negative of that same vector sum.
Similarly, some of the searches use the quantity \HT, defined as
the scalar sum of jet \pt values.

Most signal topologies considered do not
have jets from bottom quarks (``\bjetsnohyphen''); for these topologies, events containing \bjetsnohyphen are rejected to reduce
the background from \ttbar production.
Jets originating from b quarks are identified using
the combined secondary vertex algorithm (CSV)~\cite{btag}. Unless otherwise stated, we use the "medium" working point,
denoted CSVM, which has an average \bjet tagging efficiency of 70\%, light-quark jet
misidentification rate of 1.5\%, and $\cPqc$-quark jet misidentification rate of 20\% for jets with a \pt\ value greater than 60\GeV.
Corrections are applied to simulated samples to match the expected efficiencies and misidentification rates measured in data.
With the exception of the searches described in Sections~\ref{dilepton} and~\ref{sec:wh},
the searches reject events containing CSVM-identified \bjetsnohyphen with $\pt>30\GeV$.

\section{Search in the three-lepton final state}
\label{trilepton}
Three-lepton channels have sensitivity to models with signatures like
those shown in Figs.~\ref{fig:charginos-slep}
and \ref{fig:charginos-wz}. For the three-lepton search,
we use reconstructed electrons, muons, and \tauh leptons, all within $\abs{\eta}<2.4$,
requiring that there be exactly three leptons in an event.  There must be at
least one electron or muon with $\pt > 20\GeV$. Other electrons or muons must have $\pt >
10\GeV$. At most one \tauh candidate is allowed and it must have $\pt
> 20\GeV$. Events with multiple \tauh leptons have large backgrounds and are not considered in the present analysis.  %Events with an identified \bjetnohyphen are rejected.
The principal backgrounds are from $\W\Z$ diboson production
with three genuine isolated leptons that are ``prompt'' (created at
the primary vertex), and from $\ttbar$ production with two genuine
prompt leptons and a third non-prompt lepton that is misclassified as prompt.

Events are required to have $\MET > 50\GeV$. We consider events both with
and without an opposite-sign-same-flavor (OSSF) electron or muon pair.  Events
with an OSSF pair are characterized by the invariant mass $\mdil$
of the pair
and by the transverse mass \breakhere
$\MT \equiv \sqrt{\smash[b]{2\MET \pt^{\ell}[1-\cos(\Delta \phi)]}}$ formed from the \MET\ vector, the
transverse momentum $\PT^{\ell}$ of the remaining lepton, and the corresponding difference $\Delta\phi$ in azimuthal angle.
For the three-muon and three-electron events, the OSSF pair with $\mdil$ closer to the $\Z$ mass is used.
For events without an OSSF pair, which might arise
from events with a $\Z \to \tau \tau$
decay, $\mdil$ is calculated by combining opposite-sign leptons and choosing
the pair closest to the corresponding mean dilepton mass determined from
$\Z \to \tau \tau$ simulation (50\GeV for an $\Pe\mu$ pair, and 60\GeV for a $\tauh \mu$ or $\tauh\Pe$
pair).

Events are examined in exclusive search regions (``bins'') based
on their values of $\mdil$, \MET, and \MT, as presented below.
The $\mdil$ regions for OSSF dilepton pairs are
$\mdil<75\GeV$ (``below-Z''), $75<\mdil<105\GeV$ (``on-Z''), and $\mdil>105\GeV$ (``above-Z'').
Further event classification is in \MET\ bins of
50--100, 100--150, 150--200, and $>$200\GeV.
Finally, the \MT\ regions
are $<$120, 120--160, and $>$160\GeV.

\subsection{Background estimation}
\label{sec:SMtri}

The main backgrounds in this search are due to $\W\Z$ and \ttbar production,
while the background from events with \zjets and Drell-Yan production is strongly suppressed by the requirement on \MET.
The evaluation of these backgrounds is described in Sections~\ref{sec:WZ} and \ref{sec-ttbar}.
Less important backgrounds from $\Z\Z$ production and from rare SM processes such as $\ttbar\Z$, $\ttbar\W$, $\ttbar\PH$,
and triboson production are estimated from simulation using leading-order (LO) generators and are normalized to the NLO production
  cross sections~\cite{Campbell:2012dh,ttzNLO,Garzelli:2011is}.
A 50\% systematic uncertainty is assigned to these backgrounds to account
 both for the theoretical uncertainty of the cross section calculation
and for the differences of the ratio between the LO and NLO cross sections
as a function of various physical observables~\cite{Campbell:2012dh}.

The systematic uncertainty for backgrounds determined using data control samples is estimated from
the difference between the predicted and genuine yields when the methods are applied
to simulation.

\subsubsection{Background due to $\W\Z$ production}
\label{sec:WZ}

The three-lepton analysis relies on the \MET\ and \MT\ variables to
discriminate between signal and background.
The largest background is from $\W\Z$ production.
For our previous study~\cite{SUS-12-006-paper}, based on the CMS data collected in 2011, we calibrated the
hadronic recoil of the WZ system using a generalization of the
Z-recoil method discussed in Ref.~\cite{WZpaper}. This calibration led to corrections to the \MET\ and \MT\ distributions
in simulated $\W\Z$ events.  For the data collected in 2012, the
rate of pileup interactions increased.  We therefore developed
a second method, described below, designed to specifically
account for jet activity and pileup. The two methods yield consistent
results and have similar systematic uncertainties; hence we use the
average prediction as our $\W\Z$ background estimate.

In the new method, we subdivide the \MET\ distribution in a \zjets sample as a function
of \HT\ and of the number of reconstructed vertices in the event.  A large number of
vertices corresponds to large pileup, which causes
extraneous reconstruction of energy, degrading the \MET\
resolution. Larger \HT\ implies greater jet activity, which
degrades the \MET\ resolution as a consequence of the possible jet energy mismeasurement.

In a given two-dimensional bin of the number of reconstructed vertices and \HT, the $x$ and
$y$ components of \MET\ are found to be approximately Gaussian.
Therefore the \MET\ distribution
is expected to follow the Rayleigh distribution, given by:
\begin{equation}
p(\MET)=\sum_{ij} W_{ij}\frac{\MET}{\sigma^2_{ij}}\re^{-(\MET)^2/2\sigma^2_{ij}},
\end{equation}
where $i$ represents the number of vertices in the event, $j$ is the $\HT$ bin number,
$W_{ij}$ is the fraction of events in the bin, and
$\sigma_{ij}$ characterizes the \MET resolution.
We then adjust the $\sigma_{ij}$ terms in simulation
to match those found in data.
The magnitude of the correction varies from a few
percent to as high as 30\%. To evaluate a systematic uncertainty for
this procedure, we vary the level of \MET smearing and determine the
migration between different $\MET$ and $\MT$ bins in the simulated $\W\Z$
sample. We find the uncertainty of the $\W\Z$ background
to be 20--35\%, depending on the search region.
The final WZ estimate is obtained by normalizing the corrected \MET and \MT shape to the theoretical cross section.  The theoretical cross section is used to evaluate the SM background from WZ events because the contributions of signal events to WZ data control samples are expected to be significant.

\subsubsection{Background due to non-prompt leptons}
\label{sec-ttbar}
Non-prompt lepton backgrounds arise from \zjets, Drell-Yan, \ttbar,  and \wwjets events that have two genuine
isolated prompt leptons. The third lepton can be a non-prompt lepton from a heavy-flavor decay that is
classified as being prompt, or a hadron from a jet that is misidentified as a lepton.
This background is estimated using auxiliary data samples. The probability for a non-prompt
lepton to satisfy the isolation requirement ($\Irel < 0.15$) is measured in a data sample enriched
with dijet events, and varies
as a function of lepton \PT. Alternatively, the isolation
probability is studied using $\Z$-boson and $\ttbar$-enriched data samples.
These probabilities, applied to the three-lepton events with the isolation requirement on one of the leptons
inverted, are used to estimate background due to such non-prompt leptons. We average the results of the two
methods taking into account the precision of each method and the correlations between the individual inputs.

\subsubsection{Background due to internal conversions}
Another background, estimated from data, is due to events with a \Z boson and an initial- or final-state photon in which the photon undergoes an asymmetric internal conversion, leading to
a reconstructed three-lepton state~\cite{SUS-11-013-paper}.
To address this background, we
measure the rates of $\Z \to \ell^{+} \ell^{-} \gamma$ and
$\Z \to \ell^{+} \ell^{-} \ell^{\pm}$ events in an off-peak control
region defined by $|\mdil - \MZ| > 15\GeV$ and $\MET < 50\GeV$. The background estimate
is obtained by multiplying the ratio of these rates by the measured rate of
events with two leptons and a photon in the search regions. Note that external conversions
are strongly suppressed by our electron selection requirements.

\subsection{Three-lepton search results}
\label{sec:FinalResults}

\begin{figure}[bhtp]
\centering
\includegraphics[width=\cmsFigWidth]{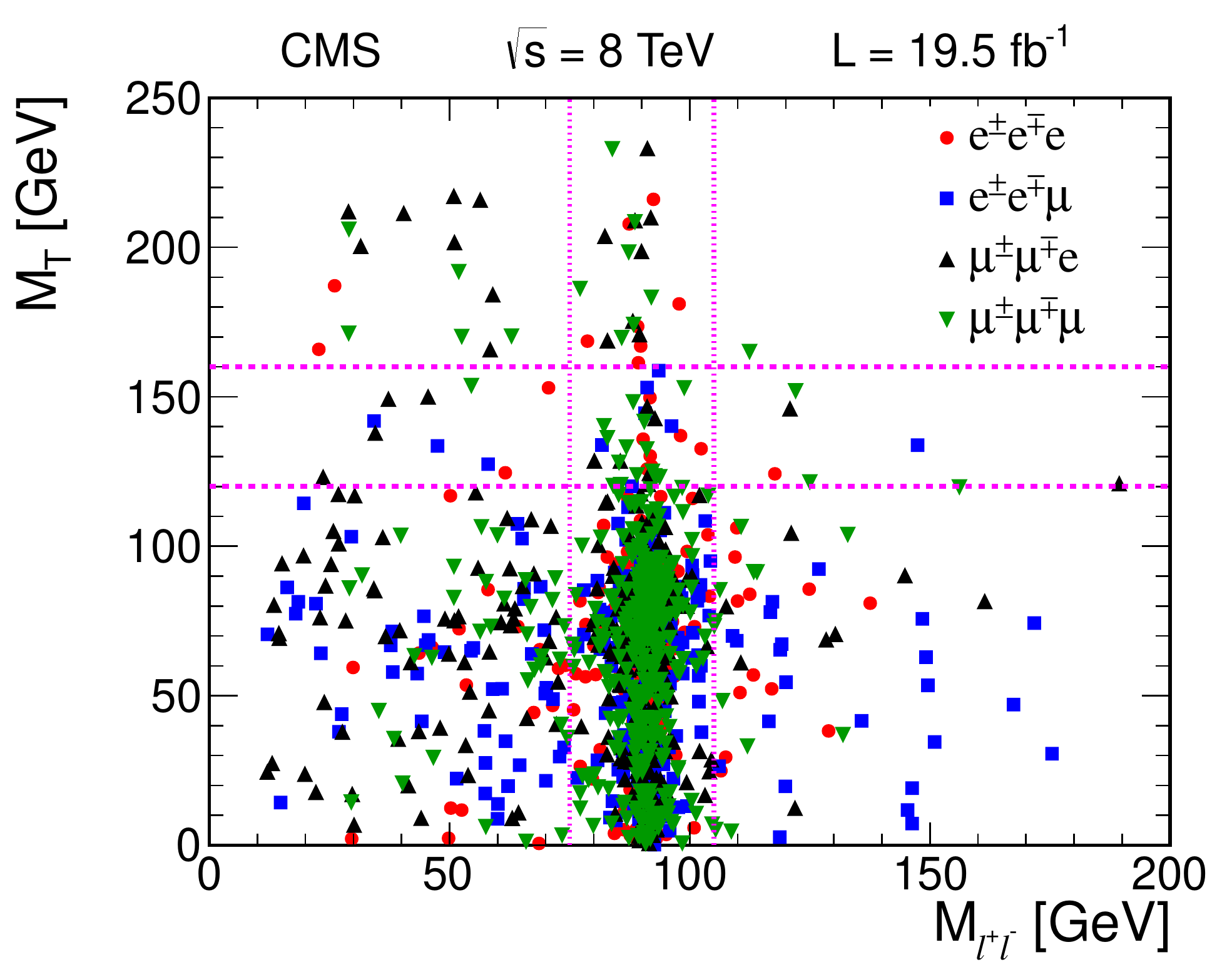} \\
\caption{$\MT$ versus $\mdil$ for three-lepton events in data with an $\Pe\Pe$ or $\mu\mu$ OSSF
dilepton pair, where the third lepton is either an electron or a muon.
Events outside of the plotted range are not indicated.
\label{fig:OSSFscatter}
}
\end{figure}

\begin{figure*}[bhtp]
\centering
\includegraphics[width=\textwidth]{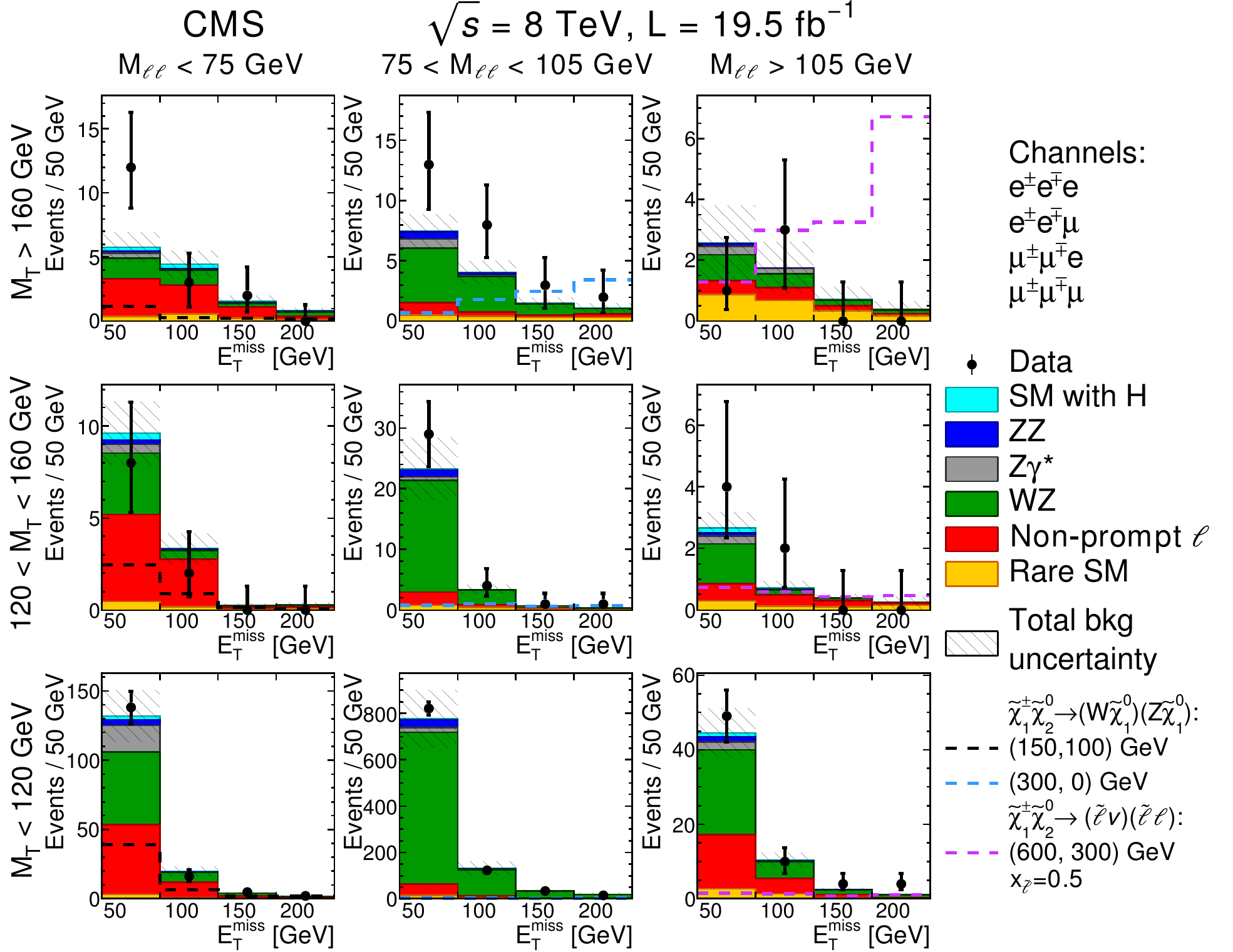} \\
\caption{$\MET$ distributions, in bins of $\MT$ and $\mdil$, for three-lepton events with an $\Pe\Pe$ or $\mu\mu$ OSSF
dilepton pair, where the third lepton is either an electron or a muon.
The SM expectations are also shown. The $\ETmiss$ distributions for example signal scenarios are overlaid.
The first (second) number in parentheses indicates the value of \mchi\ (\mlsp).
\label{fig:OSSFMET}
}
\end{figure*}

Figure~\ref{fig:OSSFscatter} shows the distribution of $\MT$ versus $\mdil$
for data events with an $\Pe\Pe$ or $\mu\mu$ OSSF pair, where the third lepton
is either an electron or muon.  The
dashed lines delineate nine two-dimensional search regions in the
$\MT$--$\mdil$ plane.
The corresponding \MET\ distributions
are shown in comparison to the SM expectations in
Fig.~\ref{fig:OSSFMET}.
Table~\ref{tab:L3OSSF} lists the results as a function of $\MET$, $\MT$, and $\mdil$.
The data are broadly consistent with SM expectations.
In the search regions with $\MT>160\GeV$ and an on-Z OSSF dilepton pair, and in the search region with $\MT>160\GeV$, $50<\MET<100\GeV$, and a below-Z OSSF pair, the data
exceed the expected background with a local significance at the level of approximately two standard deviations.

The corresponding results for $\Pe\Pe\mu$ and $\Pe\mu\mu$ events without
an OSSF pair, for events with a same-sign $\Pe\Pe$ , $\Pe\mu$, or $\mu\mu$ pair
and one \tauh candidate, and for events with an opposite-sign $\Pe\mu$ pair
and one \tauh candidate, are presented in Appendix~\ref{app:3lplots}.
The different leptonic content in these search channels provides
sensitivity to various classes of SUSY models (Section~\ref{sec:interpretation}).

\begin{table*}
\begin{center}
\topcaption{
\label{tab:L3OSSF}
Observed yields and SM expectations for three-lepton events with an $\Pe\Pe$ or $\mu\mu$ OSSF pair, where the third lepton
is either an electron or muon. The uncertainties include both the statistical and
systematic components.
}
\cmsRow{
\begin{tabular}{ c | c | c c | c c | c  c  }\hline\hline
\multirow{2}{*}{$\MT$ (\GeVns)} &\multirow{2}{*}{$\ETm$ (\GeVns)}  & \multicolumn{2}{c|}{$M_{\ell\ell} < 75$\GeV} & \multicolumn{2}{c|}{$75 < M_{\ell\ell} < 105\GeV$} & \multicolumn{2}{c}{$M_{\ell\ell} > 105$\GeV}\\\cline{3-8}
&& Total bkg & Observed & Total bkg & Observed & Total bkg & Observed \\
\hline\hline
\multirow{4}{*}{$>$160}& 50--100&5.8 $\pm$ 1.1&12&7.5 $\pm$ 1.4&13&2.6 $\pm$ 1.2&1\\
& 100--150&4.5 $\pm$ 1.1&3&4.0 $\pm$ 1.0&8&1.8 $\pm$ 0.9&3\\
& 150--200&1.5 $\pm$ 0.4&2&1.5 $\pm$ 0.5&3&0.7 $\pm$ 0.4&0\\
& $>$200 &0.81 $\pm$ 0.21&0&1.1 $\pm$ 0.4&2&0.40 $\pm$ 0.24&0\\
\hline\hline
\multirow{4}{*}{120--160}& 50--100&9.6 $\pm$ 1.7&8&23 $\pm$ 5\phantom{0}&29&2.7 $\pm$ 0.5&4\\
& 100--150&3.3 $\pm$ 0.8&2&3.4 $\pm$ 0.7&4&0.71 $\pm$ 0.22&2\\
& 150--200&0.26 $\pm$ 0.10&0&0.72 $\pm$ 0.19&1&0.38 $\pm$ 0.14&0\\
& $>$200 &0.29 $\pm$ 0.11&0&0.36 $\pm$ 0.12&1&0.24 $\pm$ 0.20&0\\
\hline\hline
\multirow{4}{*}{0--120}& 50--100&132 $\pm$ 19\phantom{0}&138&776 $\pm$ 125&821&45 $\pm$ 7\phantom{0}&49\\
& 100--150&20 $\pm$ 4\phantom{0}&16&131 $\pm$ 30\phantom{0}&123&10.0 $\pm$ 1.9\phantom{0}&10\\
& 150--200&4.0 $\pm$ 0.8&5&34 $\pm$ 8\phantom{0}&34&2.5 $\pm$ 0.5&4\\
& $>$200 &1.9 $\pm$ 0.4&2&21 $\pm$ 7\phantom{0}&14&1.2 $\pm$ 0.3&4\\
\hline\hline
\end{tabular}
}
\end{center}
\end{table*}

\section{Search in the four-lepton final state}
\label{quadlepton}
As mentioned in the introduction, we interpret our four-lepton final
state results in the context of a GMSB model, in combination with
results from a study with two leptons and at least two jets,
which is presented in Section~\ref{diboson}.
This situation motivates the use
  of four-lepton channels with at least one OSSF pair that is consistent
with a $\Z$ boson decay. The data
  are binned in intervals of \MET in order to discriminate between
  signal and background.

We use the same object selection as for the three-lepton final state,
requiring exactly four leptons (electrons, muons, and at most one \tauh candidate).
We require that there be an
$\Pe\Pe$ or $\mu\mu$ OSSF pair with an invariant mass within
$15\GeV$ of the nominal \Z boson mass. The background determination
methods are also the same as described for the three-lepton final state.
The main background, from $\Z\Z$
production, is thus estimated from
simulation, with corrections applied to the predicted \MET spectrum
as described in Section~\ref{sec:WZ}. Backgrounds from hadrons that
are misreconstructed as leptons or from non-prompt leptons
are evaluated using control samples in the data as described in Section~\ref{sec-ttbar}.

Table~\ref{tab:L4results2012} summarizes the
results. We consider events with exactly one
  OSSF pair and no \tauh candidate, with exactly one OSSF pair
  and one \tauh candidate, and with exactly two OSSF pairs
  and no \tauh candidate.
The distribution of \MET versus $\mdil$ for events without a \tauh candidate is presented
in Fig.~\ref{fig:quadscatter} of Appendix~\ref{app:3lplots}.

\begin{table}[htp]
\begin{center}
\topcaption{
Observed yields and SM expectations for exclusive channels of four-lepton final states.
All categories require four leptons including an OSSF ($\Pe\Pe$ or $\mu\mu$)
pair consistent with a $\Z$ boson.
The three sections refer, respectively,
to events with one OSSF pair and no \tauh candidate, one OSSF pair and one \tauh candidate, and two OSSF pairs and no \tauh candidate.
The uncertainties include both the statistical and systematic components.
}
\label{tab:L4results2012}
\begin{tabular}{c|c|c}\hline\hline
$\MET$ (\GeVns) & Observed & Total background  \\
\hline
\multicolumn{3}{c}{1 OSSF pair, 0 \tauh} \\
\hline
0--30   & 1 & 2.3 $\pm$ 0.6  \\
30--50  & 3 & 1.2 $\pm$ 0.3  \\
50--100 & 2 & 1.5 $\pm$ 0.4  \\
$>$100 & 2 & 0.8 $\pm$ 0.3  \\
\hline
\multicolumn{3}{c}{1 OSSF pair, 1 \tauh} \\
\hline
0--30   & 33 & 25  $\pm$ 12   \\
30--50  & 11 & 11  $\pm$ 3.1  \\
50--100 &  9 & 9.3 $\pm$ 1.9  \\
$>$100 &  2 & 2.9 $\pm$ 0.6  \\
\hline
\multicolumn{3}{c}{2 OSSF pairs, 0 \tauh} \\
\hline
0--30   & 142 & 149 $\pm$ 46  \\
30--50  &  25 & 28  $\pm$ 11  \\
50--100 &   4 & 4.5 $\pm$ 2.7 \\
$>$100 &   1 & 0.8 $\pm$ 0.3 \\
\hline
\hline
\end{tabular}
\end{center}
\end{table}

\section{Search in the same-sign two-lepton final state}
\label{dilepton}
Three-lepton final states are not sensitive to the chargino-neutralino pair production processes of
Fig.~\ref{fig:charginos-slep} if one of the leptons is unidentified, not isolated, or outside the
acceptance of the analysis.   For small mass differences between the SUSY particle states in
Fig.~\ref{fig:charginos-slep}, one of the leptons might be too soft to be included in the analysis.
Some of these  otherwise-rejected events can be recovered by requiring only two leptons. These leptons
should have the same sign (SS) to suppress the overwhelming background from opposite-sign lepton pairs.

We therefore perform a search for events with an SS lepton pair,
using the selection and methodology presented in Ref. \cite{CMS-PAS-SUS-13-013}.
We require events to contain exactly one SS $\Pe\Pe$, $\Pe\mu$,  or $\mu\mu$ pair, where the $\Pe$ and $\mu$ candidates must satisfy $\pt>20\GeV$ and $\abs{\eta}<2.4$.  To better reject background from fake leptons, we tighten the $\Pe$ ($\mu$)
isolation requirement to $\Irel<0.09$ $(0.10)$ and the $d_0$
requirement to 0.1 (0.05) mm.

Background from processes such as  WZ and {\ttbar}Z production is reduced by requiring $\ETmiss > 120\GeV$.
This background is further reduced by rejecting events that,  after applying  looser $\Pe$ and $\mu$ selection
criteria, contain an OSSF pair within 15\GeV of  the $\Z$ boson mass.

We evaluate the background from $\W\Z$ events using simulated events and assign a 15\% systematic
uncertainty, which accounts for the difference between the observed and simulated yields in a
$\W\Z$-event-enriched data control sample obtained by inverting the $\Z$-boson veto. A second
background is from events containing a prompt lepton from a W boson decay and
a genuine lepton of the same sign from heavy-flavor decay or a misidentified hadron
(mainly from \ttbar events). We evaluate this background by determining the probability for a loosely
identified electron or muon to satisfy the selection criteria in a background-enriched control region~\cite{CMS-PAS-SUS-13-013}.
We assign a 50\% systematic uncertainty to this background based on the difference in sample composition between
the control regions used to measure this probability and the signal regions.
A third background
is from events with two opposite-sign leptons, in which  one of the leptons is an electron with an incorrect charge assignment caused by severe bremsstrahlung.  To evaluate this background, we select
opposite-sign events that satisfy the selection, weighted by the probability of
electron-charge misassignment, determined using $\Z \to \Pe\Pe$ events. Finally, background
from rare SM processes, such as those described in Section~\ref{sec:SMtri}, is estimated from simulation
and assigned an uncertainty of 50\%.

Two search regions are defined, one by $\ETmiss > 200\GeV$, and the other by $120 < \ETmiss < 200\GeV$ and $\njets=0$, where $\njets$ for this purpose denotes the number of jets with $\pt > 40\GeV$ and $\abs{\eta} < 2.5$.
The jet veto enhances the sensitivity to the signal models targeted here by suppressing backgrounds with
large hadronic activity, such as \ttbar events.

\begin{figure}[thbp]
\centering
\includegraphics[height=0.48\textwidth]{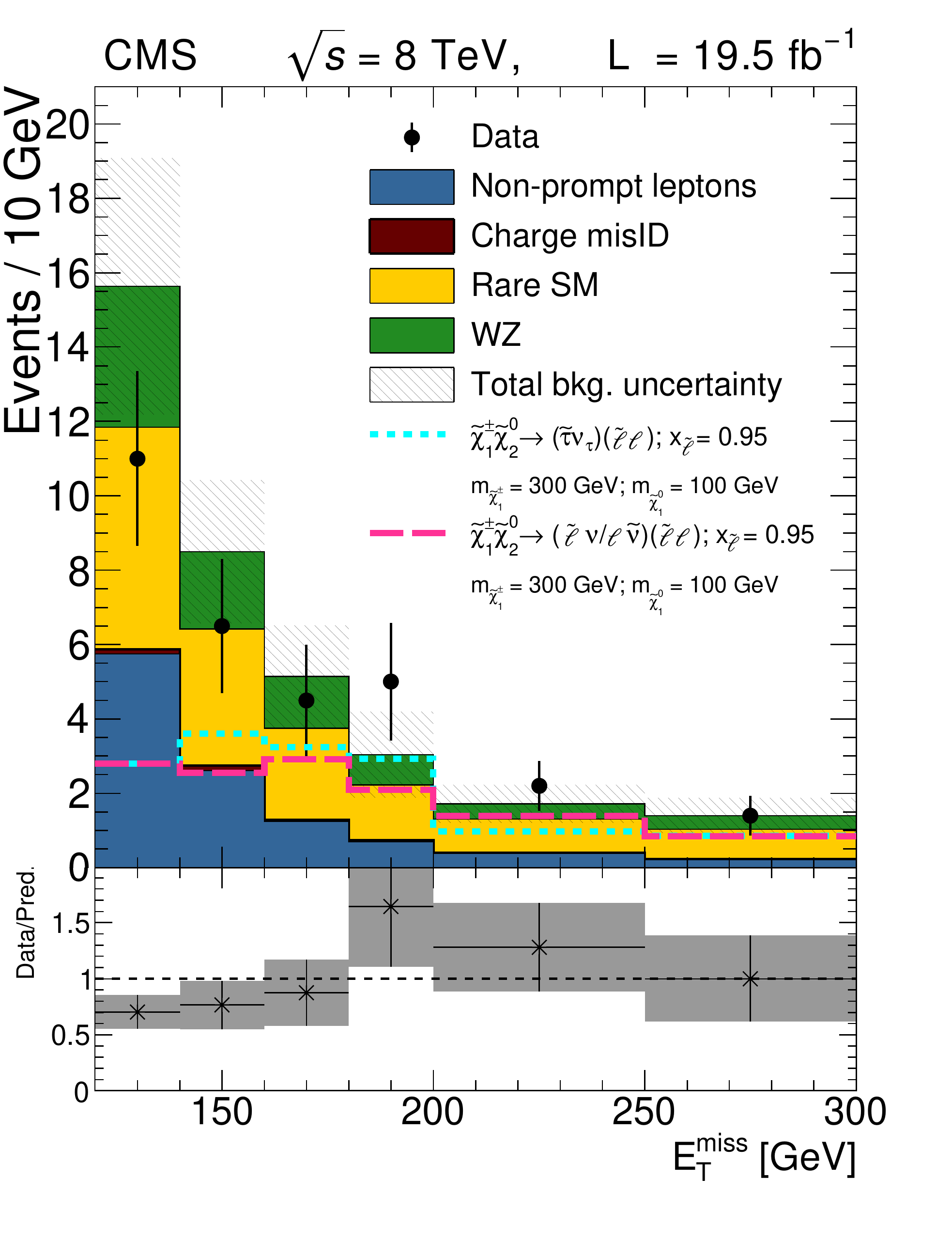}
\includegraphics[height=0.48\textwidth]{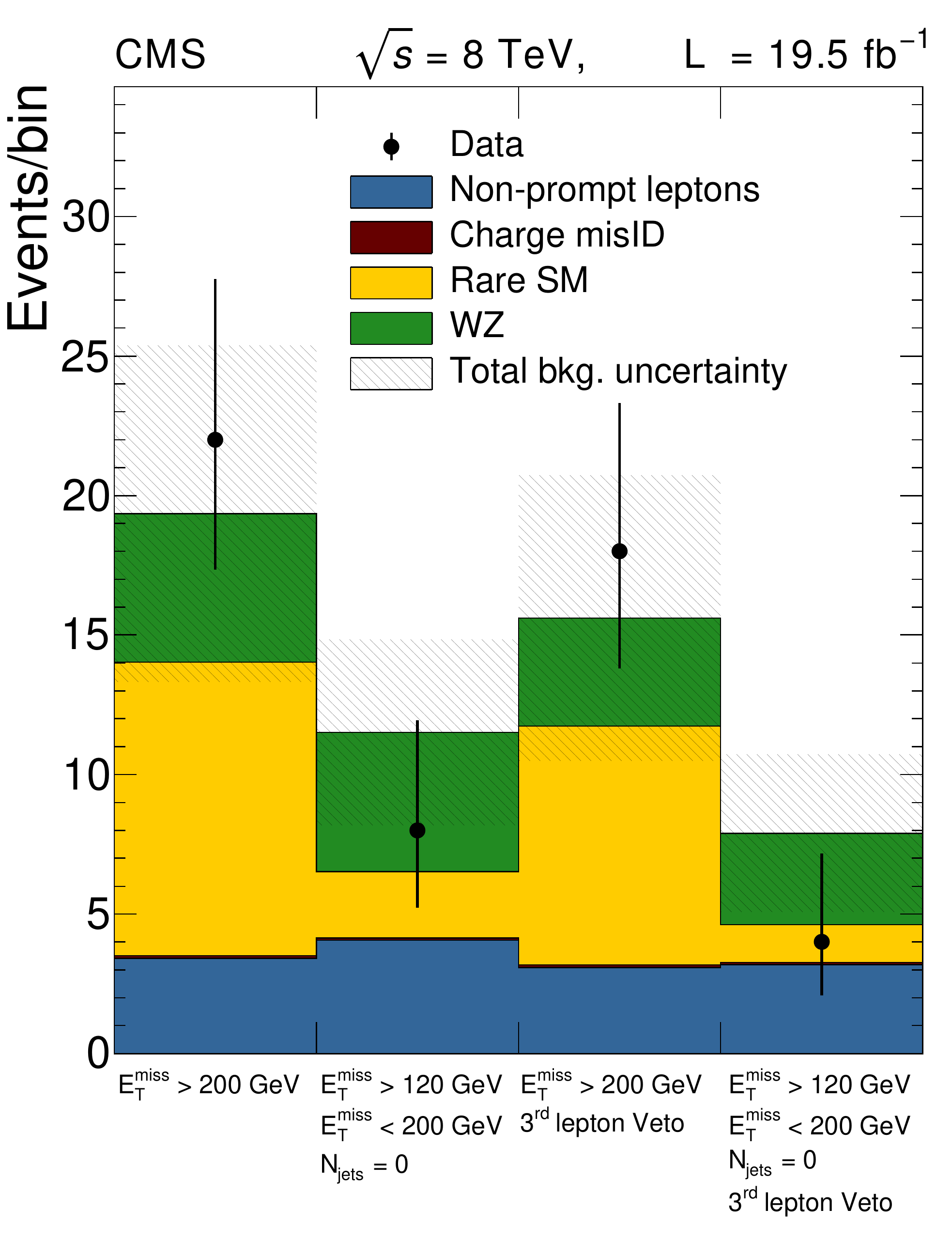}
\caption{(\cmsLeft) \MET\ distribution for same-sign dilepton candidates in comparison with the SM
expectations. The bottom panel shows the ratio and corresponding uncertainty of the observed
and total SM expected distributions. The third lepton veto is not applied.
The distributions of example signal scenarios are overlaid.
(\cmsRight) Observed yields and expected backgrounds for the different search regions.
In both plots, events with $\MET>120\GeV$ are displayed, and the hashed band shows the combined
statistical and systematic uncertainties of the total background.
\label{fig:SSMET}
}
\end{figure}

The observed yields and corresponding SM expectations are given in Table~\ref{tab:SSYields}.
Results are presented both with and without the veto of events with a third selected lepton.
The distribution of \MET\ in comparison with the SM expectation is shown in Fig.~\ref{fig:SSMET}, along with the observations
and expectations in each search region. The interpretation, presented in  Section~\ref{sec:interpretation},
is based on the two signal regions defined above, and includes  the third lepton veto in order to simplify
combination with the results of the three-lepton search.

\begin{table*}[htb]
 \begin{center}
	 \topcaption{
Observed yields and SM expectations for the
same-sign dilepton search, with and without a veto on the
presence of a third lepton.  The uncertainties include both
the statistical and systematic components. The $\njets$ variable refers to the number of jets with $\pt > 40\GeV$ and $\abs{\eta} < 2.5$.
}
   \label{tab:SSYields}
   \cmsRow{\begin{tabular}{l|c|c|c|c}
    \hline\hline
     \multirow{3}{*}{Sample}           & $\MET>200$\GeV & $\MET$ 120--200\GeV & $\MET>200$\GeV & $\MET$ 120--200\GeV    \\
                        &   & $\njets = 0$ &   & $\njets = 0$    \\
                        &   &   & $3^{\text{rd}}$ lepton veto  & $3^{\text{rd}}$ lepton veto     \\
    \hline
 Non-prompt leptons     &  3.4 $\pm$ 1.9 &  4.1 $\pm$ 2.2 &  3.1 $\pm$ 1.7 &  3.2 $\pm$ 1.7    \\
 Charge misidentification           &  0.09 $\pm$ 0.01 &  0.08 $\pm$ 0.01 &  0.09 $\pm$ 0.01 &  0.07 $\pm$ 0.01    \\
 Rare SM                &  10.5 $\pm$ 5.7 &  2.4 $\pm$ 2.4 &  8.6 $\pm$ 4.8 &  1.4 $\pm$ 2.1    \\
 WZ                     &  5.3 $\pm$ 0.8 &  5.0 $\pm$ 0.8 &  3.9 $\pm$ 0.6 &  3.3 $\pm$ 0.5    \\
    \hline
 Total background       &  19.4 $\pm$ 6.0 &  11.5 $\pm$ 3.3 &  15.6 $\pm$ 5.1 &  7.9 $\pm$ 2.8    \\
    \hline
 Data                   &   22    &   8    &   18    &   4       \\
    \hline
    \hline
    \end{tabular}}
 \end{center}
\end{table*}

\section{Search in the
\texorpdfstring{$\W\Z/\Z\Z + \MET$}{WZ/ZZ + MET} final state with two
leptons and two jets}
\label{diboson}
The three- and four-lepton searches described above
are sensitive not only to the processes of Fig.~\ref{fig:charginos-slep},
but also to those of Fig.~\ref{fig:charginos-wz},
with on-shell or off-shell vector
  bosons. In this section, we describe a search for events with two leptons consistent with a \Z boson
  and at least two jets ($\Z+{dijet}$), which extends the sensitivity to some of the processes of
  Fig.~\ref{fig:charginos-wz}. Specifically,
  we select events in which an on-shell $\Z$ boson decays to either an $\Pe^+\Pe^-$ or $\mu^+\mu^-$
  pair, while an on-shell $\W$ boson or another on-shell $\Z$ boson decays to two jets. The
  object selection and background determination procedures are based on
  those presented in
Ref.~\cite{SUS-11-021-paper}:
both leptons must have $\pt>20\GeV$ and
the dilepton invariant mass must be consistent with the $\Z$ boson mass to within $10\GeV$.
At least two jets with $\pt>30\GeV$ are required.
  Events with a third lepton are rejected in order to reduce the
  background from $\W\Z$ production.

  Following the lepton and jet selection, the dominant background is
from \zjets\ events. This background is strongly suppressed
by requiring large values of \MET ,
leaving \ttbar\ production as the dominant background.
The \ttbar\ background is reduced by a factor of $\sim10$ by applying the veto
on events with \bjetsnohyphen mentioned in Section~\ref{detector}.
Background from
  $\ttbar$ and $\zjets$ events is reduced further by requiring the dijet
  mass $\mjj$ formed from the two highest $\pt$ jets to be consistent with a $\W$ or $\Z$ boson,
namely $70 < \mjj < 110\GeV$.

  For the remaining background from $\zjets$ events, significant \MET\
  arises primarily because of the mismeasurement of jet \pt.  We evaluate
  this background using a sample of $\gjets$ events as described in Ref.~\cite{SUS-11-021-paper},
accounting for the different kinematic
  properties of the events in the control and signal samples.

The remaining background other than that from $\zjets$ events
  is dominated by \ttbar production, but includes events with $\W\W$, single-top-quark, and
  $\tau\tau$ production. This background is characterized by equal
  rates of $\Pe\Pe+\mu\mu$ versus $\Pe\mu$ events and so is denoted ``flavor symmetric''
  (FS).  To evaluate the FS background, we use an
  $\Pe\mu$ control sample, and correct for the different electron vs. muon selection efficiencies.
  The SM backgrounds from events with $\W\Z$ and $\Z\Z$ production are
  estimated from simulation and assigned uncertainties based on comparisons with data
in control samples with exactly three leptons ($\W\Z$ control sample)
and exactly four leptons ($\Z\Z$ control sample), and at least two jets.
 Background from rare SM processes with ${\ttbar}\Z$, $\Z\Z\Z, \Z\Z\W$, and $Z\W\W$ production
is determined from simulation with an assigned uncertainty of 50\%.
The background estimation methodology is validated in a signal-depleted control region, defined
by $\mjj > 110\GeV$, which is orthogonal to the search region. The observed yields
are found to be consistent with the expected backgrounds in this control region.

\begin{figure}[htbp]
\centering
\includegraphics[width=0.49\textwidth]{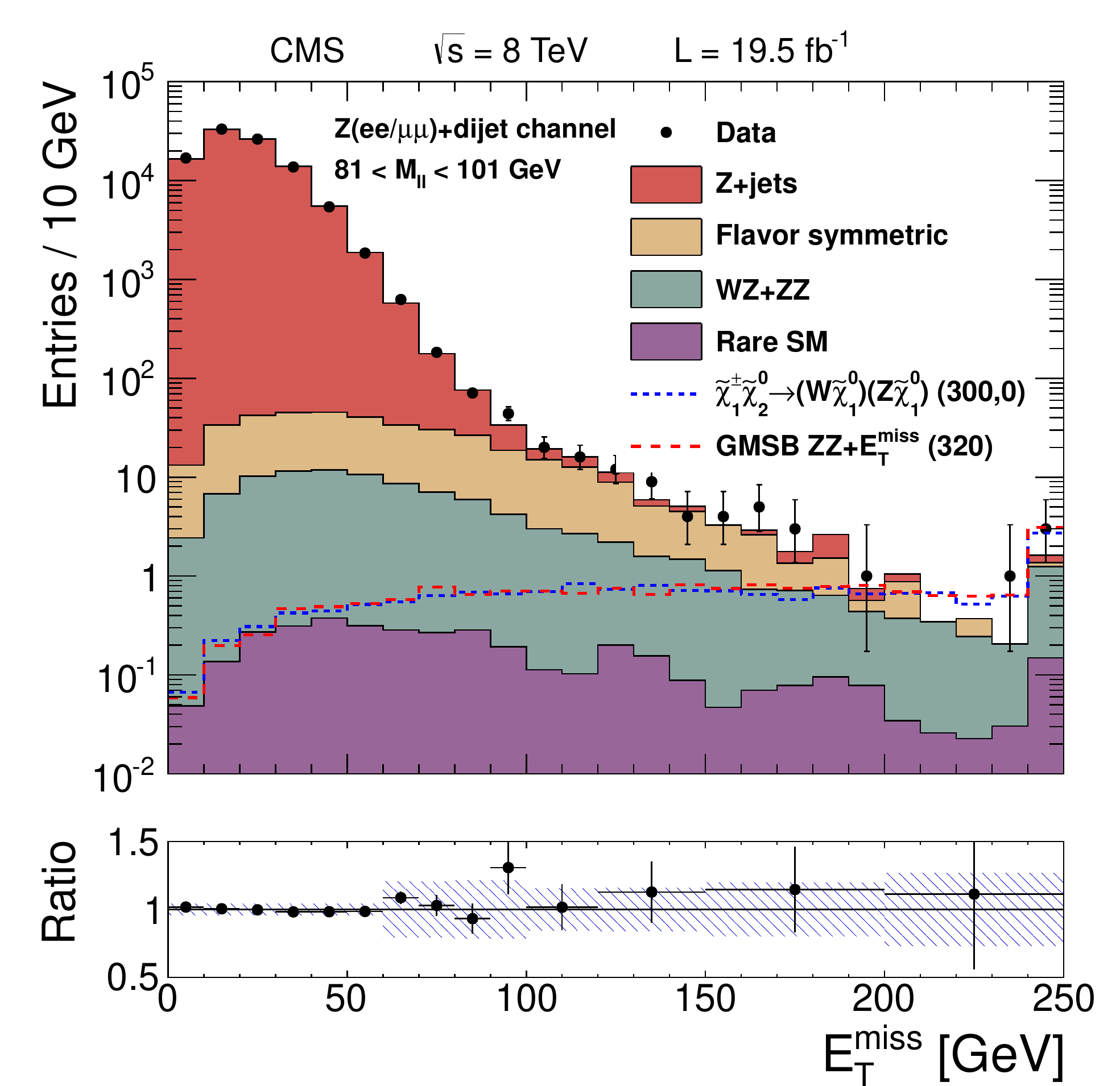}
\includegraphics[width=0.49\textwidth]{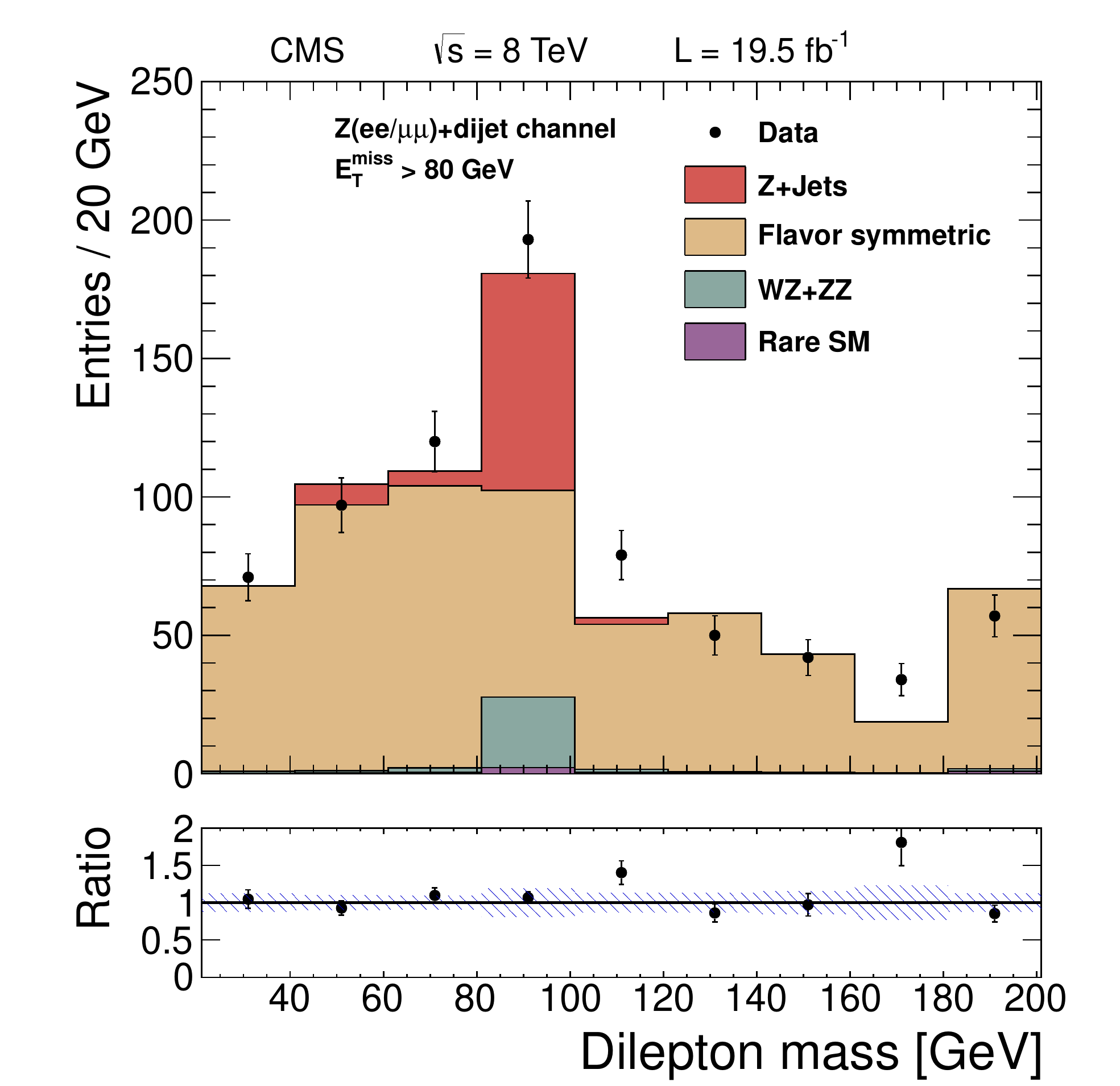}
\caption{
Distributions for $\Z+\text{dijet}$ events in comparison with SM expectations: (\cmsLeft) \MET distribution for events with the
dilepton invariant mass satisfying $81 < \mdil < 101$\GeV; expected results for two signal scenarios are overlaid,
(\cmsRight) $\mdil$ distribution for $\MET>80$\GeV.
The ratio of the observed to predicted yields in each bin is shown in the lower panels. The error bars indicate the
statistical uncertainties of the data and the shaded band the total background uncertainty.
\label{fig:pfmet_eemm}
}
\end{figure}

The results are presented in Table~\ref{tab:results_targ}. The five exclusive intervals
with $\MET>80$\GeV are treated as signal regions in the interpretations
presented in Section~\ref{sec:interpretation}.
Figure~\ref{fig:pfmet_eemm} displays the observed \MET\ and dilepton mass distributions
compared with the sum of the expected backgrounds.

\begin{table*}[hbt]
\centering
\topcaption{\label{tab:results_targ}%\footnotesize
Observed yields and SM expectations, in bins of \MET, for the $\Z+\text{dijet}$ analysis.
The total background is the sum of the \zjets\ background, the flavor-symmetric (FS)
background, and the $\W\Z, \Z\Z$, and rare SM backgrounds.
All uncertainties include both
the statistical and systematic components.
The expected yields for the \wzmet model with $\mchi=300$\GeV and $\mlsp=0$\GeV, and the GMSB \zzmet model
with $\mu=320$\GeV (see Section~\ref{app:combo}) are also indicated.
}
\cmsRow{
\begin{tabular}{l|c|c|c|c}
\hline
\hline
Sample                     &   \MET\ 0--30\GeV   &  \MET\ 30--60\GeV   &  \MET\ 60--80\GeV   & \MET\ 80--100\GeV     \\
\hline
\hline
        \zjets\ bkg   &  75839 $\pm$ 3042\phantom{0}   &   21234 $\pm$ 859\phantom{00}   &     690 $\pm$ 154   &   65 $\pm$ 22     \\
             FS bkg   &   70 $\pm$ 12   &   97 $\pm$ 16   &    48.3 $\pm$ 8.3\phantom{0}   &    35.2 $\pm$ 6.2\phantom{0}     \\
             WZ bkg   &    16.1 $\pm$ 8.1\phantom{0}   &   27 $\pm$ 14   &    11.8 $\pm$ 5.9\phantom{0}   &     6.8 $\pm$ 3.4     \\
             ZZ bkg   &     2.9 $\pm$ 1.4   &     6.0 $\pm$ 3.0   &     3.3 $\pm$ 1.7   &     2.8 $\pm$ 1.4     \\
        Rare SM bkg   &     0.5 $\pm$ 0.2   &     1.0 $\pm$ 0.5   &     0.6 $\pm$ 0.3   &     0.5 $\pm$ 0.2     \\
\hline
          Total bkg   &  75929 $\pm$ 3042   &   21364 $\pm$ 859   &     754 $\pm$ 154   &      110 $\pm$ 23     \\
               Data   &             76302   &             20991   &               809   &               115     \\
\hline
\wzmet (300/0)       &     0.6 $\pm$ 0.1   &     1.4 $\pm$ 0.1   &     1.2 $\pm$ 0.1   &     1.3 $\pm$ 0.1     \\
  GMSB  (320)  &     0.5 $\pm$ 0.0   &     1.5 $\pm$ 0.1   &     1.4 $\pm$ 0.1   &     1.4 $\pm$ 0.1     \\
\hline
\hline
Sample                    &\MET\ 100--120\GeV   &\MET\ 120--150\GeV   &\MET\ 150--200\GeV   & $\MET> 200\GeV$  \\
\hline
\hline
        \zjets\ bkg   &     7.8 $\pm$ 3.1   &     3.7 $\pm$ 1.6   &     2.0 $\pm$ 1.0   &     0.4 $\pm$ 0.3  \\
             FS bkg   &    21.9 $\pm$ 4.0\phantom{0}   &    13.2 $\pm$ 2.5\phantom{0}   &     5.7 $\pm$ 1.6   &     0.8 $\pm$ 0.4  \\
             WZ bkg   &     3.7 $\pm$ 1.9   &     2.9 $\pm$ 1.5   &     1.9 $\pm$ 0.9   &     0.9 $\pm$ 0.4  \\
             ZZ bkg   &     1.8 $\pm$ 0.9   &     1.9 $\pm$ 0.9   &     1.4 $\pm$ 0.7   &     1.3 $\pm$ 0.7  \\
        Rare SM bkg   &     0.2 $\pm$ 0.1   &     0.4 $\pm$ 0.2   &     0.4 $\pm$ 0.2   &     0.3 $\pm$ 0.1  \\
\hline
          Total bkg   &    35.4 $\pm$ 5.5   &    22.2 $\pm$ 3.5   &    11.3 $\pm$ 2.2   &     3.6 $\pm$ 1.0  \\
               Data   &                36   &                25   &                13   &                 4  \\
\hline
\wzmet\ (300/0)   &         1.5 $\pm$ 0.1   &     2.3 $\pm$ 0.1   &     3.4 $\pm$ 0.1   &     5.2 $\pm$ 0.2  \\
GMSB  (320)&         1.4 $\pm$ 0.1   &     2.2 $\pm$ 0.1   &     3.9 $\pm$ 0.1   &     5.7 $\pm$ 0.2  \\
\hline
\hline
\end{tabular}
}
\end{table*}

\section{Searches in the \texorpdfstring{$\W\PH + \MET$}{WH + MET} final state}
\label{sec:wh}

The recent observation of a Higgs boson~\cite{Aad:2012tfa,Chatrchyan:2012ufa,Chatrchyan:2013lba} offers the novel
possibility to perform beyond-the-SM searches by exploiting the measured properties
of this particle. In particular, the heavy neutralinos
are expected to decay predominantly via a Higgs boson
in large regions of SUSY parameter space, and in this section we report searches
for such decays.

Three exclusive final states sensitive to the process of Fig.~\ref{fig:charginos-wz}(center) are considered here.
In all searches, the \PW\ boson is required to decay leptonically.
A search in the single-lepton final state provides
sensitivity to events in which the Higgs boson decays to a \bbbar pair.
A search in the same-sign dilepton
final state targets events with the decay $\PH\to\PW^+\PW^-$ in which one of the \PW\ bosons
decays leptonically and the other hadronically. The results of the CMS inclusive multilepton
search~\cite{SUS13002}
are reinterpreted, covering final states with at least three leptons.
It is used to target the decays $\PH\to\PWp\PWm$, $\PH\to\Z\Z$, and
$\PH\to\tau^+\tau^-$, where the \PW\ and \Z\ bosons, and the $\tau$ lepton, decay leptonically.
The results from these searches are combined to place limits on the production of the \whmet\ final state.

\subsection{Search in the single-lepton final state}
\label{sec:singlelepton}

\subsubsection{Overview of the search}
\label{sec:singlelepton_overview}

In this section we report the results from a search for
$\chipmo\chitn\to(\PW\lsp)(\PH\lsp) \to \ell\nu \bbbar+\MET$ events.
Previous searches involving the $\PH \to \bbbar$ decay mode,
  corresponding to the largest SM branching fraction (56\%)~\cite{Heinemeyer:2013tqa}, have targeted
the associated production with a leptonically decaying \PW\ boson~\cite{CMS-PAS-HIG-13-012}.
In the present search, we impose additional kinematic requirements on \MET\ and related quantities. These requirements
strongly suppress both the SM backgrounds and the
SM production of a Higgs boson while retaining efficiency for the SUSY signal.
This search is an extension of a search for direct top-squark pair production~\cite{SUS13011},
which targets events with a single lepton, at least four jets,
and \MET, with similar object selection and analysis methodology.
The final state considered here is similar, except that we expect only two jets.

Events are required to contain a single lepton, exactly two b jets, and \MET. The largest background arises from \ttbar\ production, due both to semileptonic \ttbar\ events and to events where both top quarks decay leptonically but one lepton is not identified.
Events with \wjets\ production also constitute an important source of background.
The SM backgrounds are suppressed using several kinematic requirements based on large values of \MET.
Signal regions are defined by successively tighter requirements on \MET.
The signal is expected to produce a peak in the dijet mass spectrum at $M_{\bbbar}=m_{\PH}$.

\subsubsection{Event selection}
\label{sec:singlelepton_eventselection}

Events are required to contain exactly one electron (muon) with $\pt>30$ (25)\GeV and $\abs{\eta}<1.4442\,(2.1)$.
Electrons are restricted to the central region of the detector for consistency with the search for top-squarks~\cite{SUS13011}.
There must be exactly two jets with $\abs{\eta}<2.4$ and no jets  with $2.4 < \abs{\eta} < 4.7$.
This latter requirement substantially reduces the \ttsl\ background, which typically has four jets.
The two selected jets must satisfy the CSVM b-tagging criteria and have $\pt>30$\GeV.
We require $\mt > 100$\GeV, which primarily rejects backgrounds with a
single $\PW \to \ell\nu$ decay and no additional \MET, such as \ttsl, \wjets,
and SM \whbb\ events, and single-top-quark events in the
  $t$ and $s$ channels.
To suppress the dilepton \ttbar\ backgrounds, events with an isolated high-\pt\ track or $\tauh$ candidate are rejected.

Further suppression of the \ttbar\ backgrounds is achieved by using the \mtbl\ variable~\cite{mt2w}, which is defined as the
minimum ``mother'' particle mass compatible with the four-momentum of the lepton, b-tagged jets, and \MET.  It has an endpoint at
the top-quark mass for \ttbar\ events without mismeasurement effects, while signal events may have larger values.  We require $\mtbl > 200\GeV$.

The dijet mass \mbb\ formed from the two selected jets is required to satisfy $100< M_{\bbbar} <150\GeV$. This requirement has an efficiency of about 80\% for signal events.

\subsubsection{Backgrounds and their estimation methodology}
\label{sec:singlelepton_bkg}

Backgrounds are grouped into six categories.
The largest background arises from \ttbar\ events and from single-top-quark production in the tW channel,
in which both \PW\ bosons decay leptonically (dilepton top-quark background).
Backgrounds from \ttbar\ and single-top-quark production with one leptonically decaying \PW\ boson are referred
to as the single-lepton top-quark background.
Backgrounds from \PW\Z\ production, where the \PW\ boson decays leptonically and the \Z\ boson
decays to a \bbbar\ pair, are referred to as the \wzbb\ background.
Backgrounds from \PW\ bosons produced in associated production with a \bbbar\ pair
are referred to as the \wbb\ background,
while production of \PW\ bosons with other partons constitutes the \wl\ background.
Finally, the ``rare background'' category consists of processes with two top quarks and a W, Z or Higgs boson,
as well as diboson, triboson, \zjets, and SM \whbb\ events. The \zjets\ process has a large cross section but
is included in the rare background category because its contribution is very small after the signal-region requirements
are imposed.  With the exception of the \wl\ background, the background estimation is based on simulation.

The simulation is validated in three data control regions (CR) that are enriched in different background components.
A data sample enriched in \wl\ is obtained by vetoing events with b-tagged jets (CR-0b).
A data sample enriched in the dilepton top-quark background is obtained by requiring either exactly
two leptons satisfying the lepton selection criteria, or one such lepton and an isolated high-\pt\ track (CR-2$\ell$).
Finally, the \mbb\ requirement is inverted to obtain a data sample (CR-\mbb) consisting of a
mixture of backgrounds with similar composition as the signal region.

The agreement between the data and the simulation in the three data control regions is used to determine
scale factors and uncertainties for the background predictions. In CR-2$\ell$, the data are found to agree with the predictions from simulation, which are dominated by the dilepton top-quark background.
A 40\% uncertainty is assessed on the dilepton top-quark background, based on the limited statistical precision of the event sample after
applying all the kinematical requirements. Correction factors of $0.8\pm0.3$, $1.2\pm0.5$, and $1.0\pm0.6$ are evaluated
for the \wzbb, \wbb, and single-lepton top-quark backgrounds, respectively, based on studies of the CR-\mbb\ and CR-0b samples.
The rare backgrounds are taken from simulation with a 50\% systematic uncertainty.

The \wl\ background prediction is evaluated using the CR-0b sample, using the b-tagging misidentification rate for
   light flavor jets predicted by simulation. This rate includes all flavors except b quarks.
The uncertainty is 40\%, due to uncertainties in the b-tagging misidentification rate
and its variation with jet $\pt$.

\subsubsection{Results}
\label{sec:singlelepton_results}

\begin{figure*}[htbp]
\centering
\includegraphics[width=0.48\textwidth]{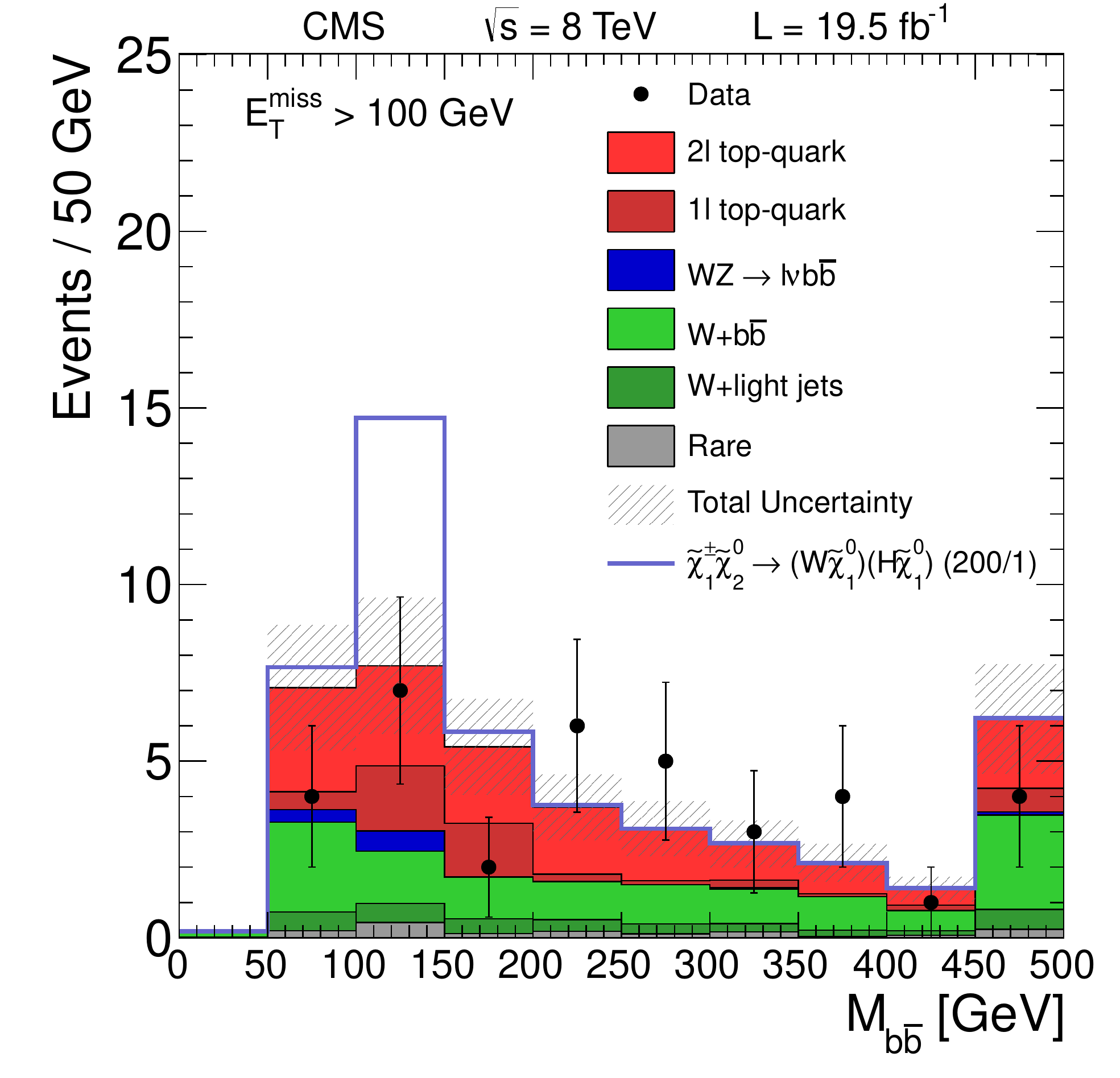}
\includegraphics[width=0.48\textwidth]{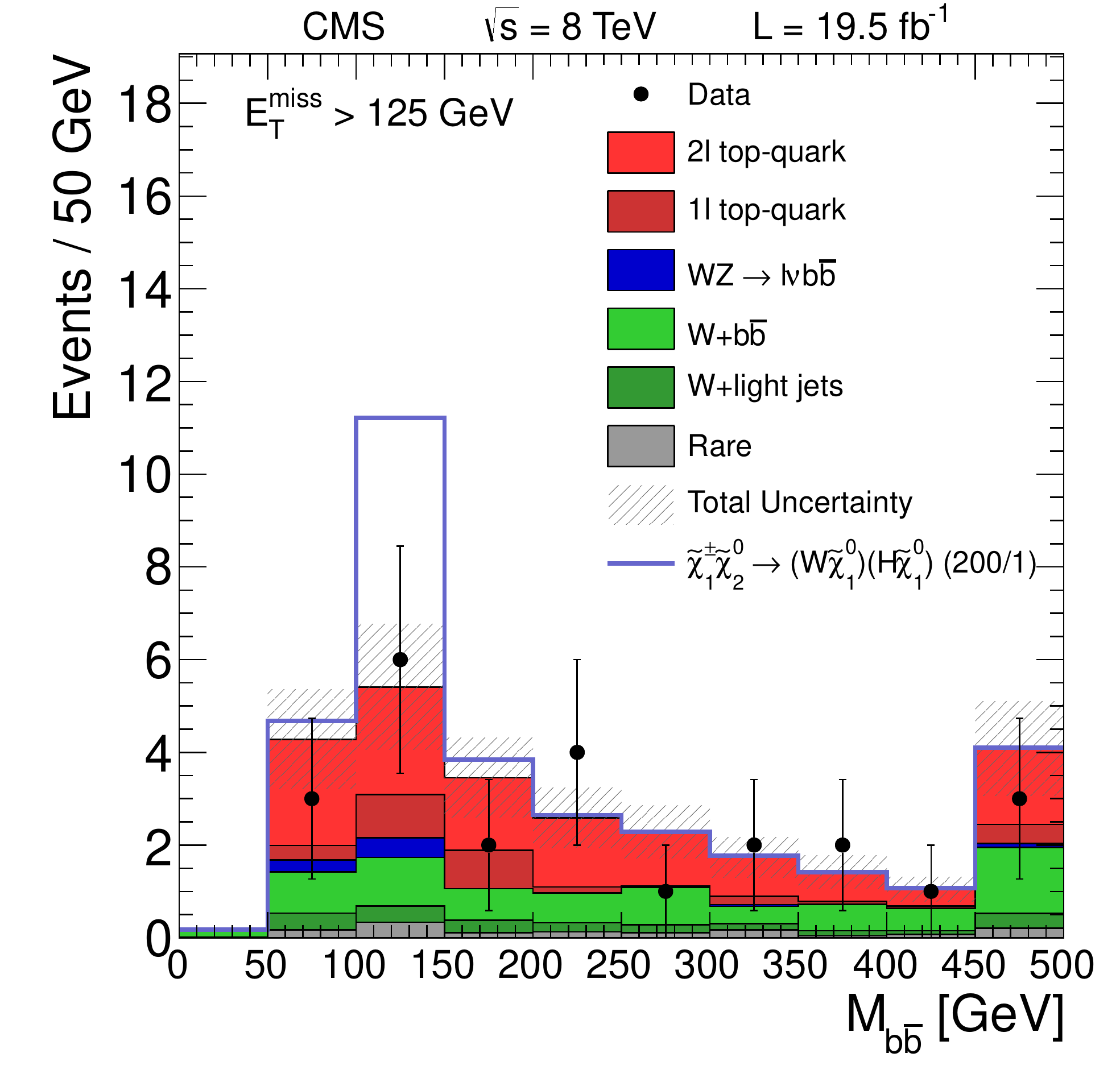}
\includegraphics[width=0.48\textwidth]{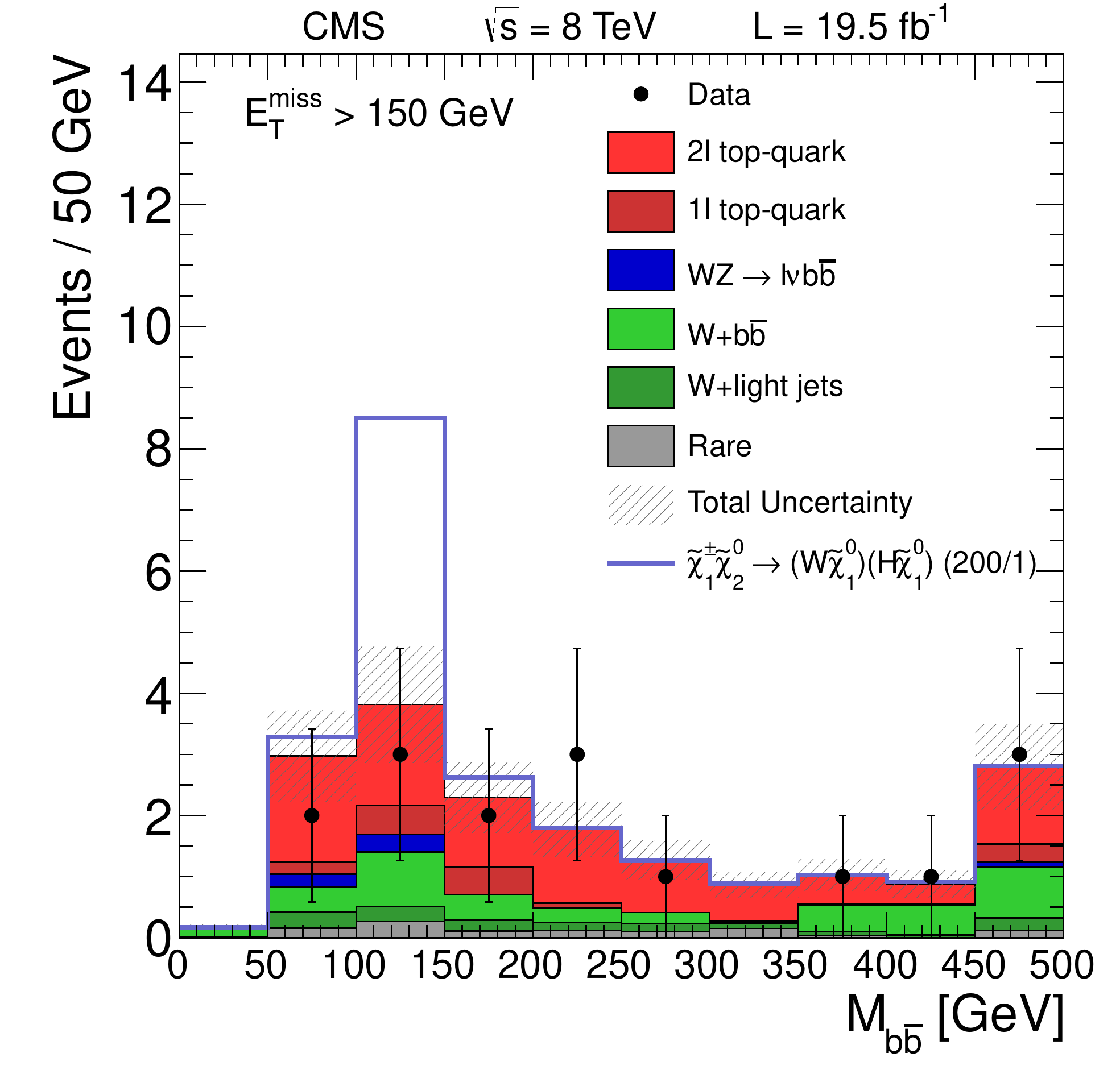}
\includegraphics[width=0.48\textwidth]{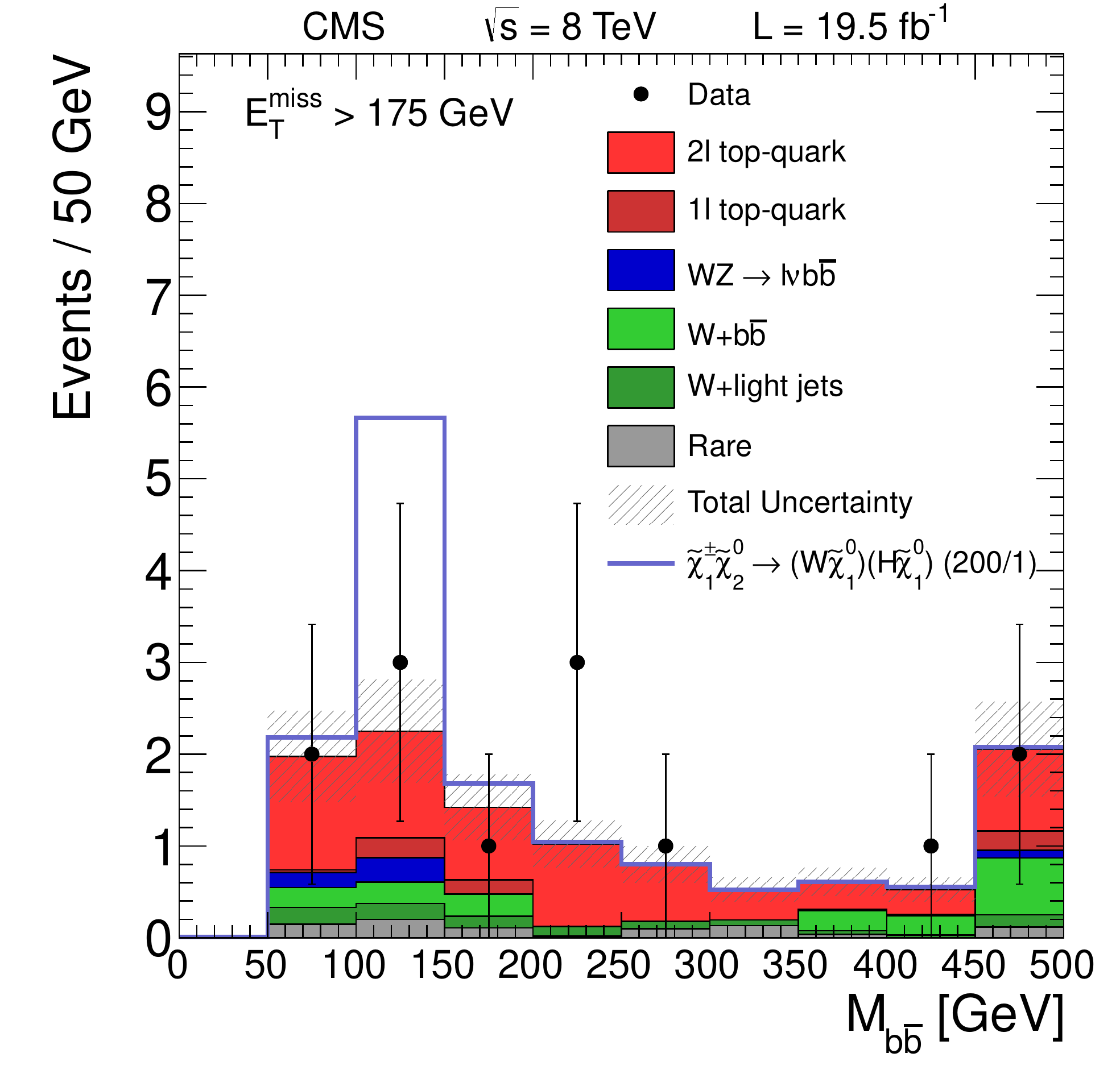}
\caption{ Distributions of \mbb\ for the single-lepton \whmet\ analysis for
(upper left) $\MET>100$\GeV, (upper right) $\MET>125$\GeV, (lower left) $\MET>150$\GeV, and (lower right) $\MET>175$\GeV after all signal region requirements have been applied except for that on \mbb.
The data are compared to the sum of the expected backgrounds.
The labels ``2$\ell$ top'' and ``1$\ell$ top'' refer to the dilepton top-quark and single-lepton top-quark backgrounds, respectively.
The band indicates the total uncertainty of the background prediction.
Results from an example signal scenario are shown, stacked on top of the SM background.
\label{fig:singlelepton}
}
\end{figure*}

Four overlapping signal regions are defined by the requirements $\MET>100,125,150$, and 175\GeV.
In general, signal regions with tighter \MET\ requirements are more sensitive to signal
scenarios with larger mass differences $\mchi - \mlsp$.
The results for these signal regions are summarized in Table~\ref{tab:singlelepton}.
The data are seen to agree with the background predictions to within the uncertainties.
The expected yields for several signal scenarios are indicated, including systematic
uncertainties that are discussed in Section ~\ref{sec:interpretation}.
The distributions of \mbb\ are displayed in Fig.~\ref{fig:singlelepton}.
No evidence for a peak at $\mbb=m_{\PH}$ is observed.

\begin{table*}[htbp]
\topcaption{\label{tab:singlelepton}
Observed yields and SM expectations, in several bins of \MET, for the single-lepton \whmet\ analysis.
The expectations from several signal scenarios are shown;
the first number indicates \mchi\ and the second \mlsp (\GeVns).
The uncertainties include both the statistical and systematic components.
}
\centering
\footnotesize
\begin{tabular}{l|c|c|c|c}
\hline
\hline
                  Sample    &     $\MET > 100\GeV$    &     $\MET > 125\GeV$    &     $\MET > 150\GeV$    &     $\MET > 175\GeV$   \\
\hline
            Dilepton top-quark    &         2.8 $\pm$ 1.2    &         2.3 $\pm$ 1.0    &         1.7 $\pm$ 0.7    &         1.2 $\pm$ 0.5   \\
       Single-lepton top-quark    &         1.8 $\pm$ 1.1    &         0.9 $\pm$ 0.6    &         0.5 $\pm$ 0.3    &         0.2 $\pm$ 0.2   \\
                         \wzbb    &         0.6 $\pm$ 0.2    &         0.4 $\pm$ 0.2    &         0.3 $\pm$ 0.1    &         0.3 $\pm$ 0.1   \\
                          \wbb    &         1.5 $\pm$ 0.9    &         1.0 $\pm$ 0.7    &         0.9 $\pm$ 0.6    &         0.2 $\pm$ 0.3   \\
                           \wl    &         0.5 $\pm$ 0.2    &         0.3 $\pm$ 0.1    &         0.2 $\pm$ 0.1    &         0.2 $\pm$ 0.1   \\
                          Rare    &         0.4 $\pm$ 0.2    &         0.3 $\pm$ 0.2    &         0.3 $\pm$ 0.2    &         0.2 $\pm$ 0.1   \\
\hline
                Total background  &         7.7 $\pm$ 1.9    &         5.4 $\pm$ 1.3    &         3.8 $\pm$ 1.0    &         2.3 $\pm$ 0.6   \\
                    Data    &                     7    &                     6    &                     3    &                     3   \\
\hline
\footnotesize \signal\ (130/1)  &        9.0 $\pm$ 1.2    &         7.5 $\pm$ 1.0    &         6.0 $\pm$ 0.8    &         4.5 $\pm$ 0.6     \\
\footnotesize \signal\ (150/1)  &        7.2 $\pm$ 1.0    &         6.1 $\pm$ 0.9    &         5.0 $\pm$ 0.7    &         3.5 $\pm$ 0.5     \\
\footnotesize \signal\ (200/1)  &        7.0 $\pm$ 0.9    &         5.8 $\pm$ 0.8    &         4.7 $\pm$ 0.7    &         3.4 $\pm$ 0.5     \\
\footnotesize \signal\ (300/1)  &        5.2 $\pm$ 0.7    &         4.9 $\pm$ 0.7    &         4.4 $\pm$ 0.6    &         3.9 $\pm$ 0.5     \\
\footnotesize \signal\ (400/1)  &        3.2 $\pm$ 0.4    &         3.0 $\pm$ 0.4    &         2.8 $\pm$ 0.4    &         2.5 $\pm$ 0.3     \\
\hline
\hline
\end{tabular}
\end{table*}

\subsection{Search in the same-sign dilepton final state}
\label{sec:samesign}

\begin{figure}[htbp]
\centering
\includegraphics[width=\cmsFigWidth]{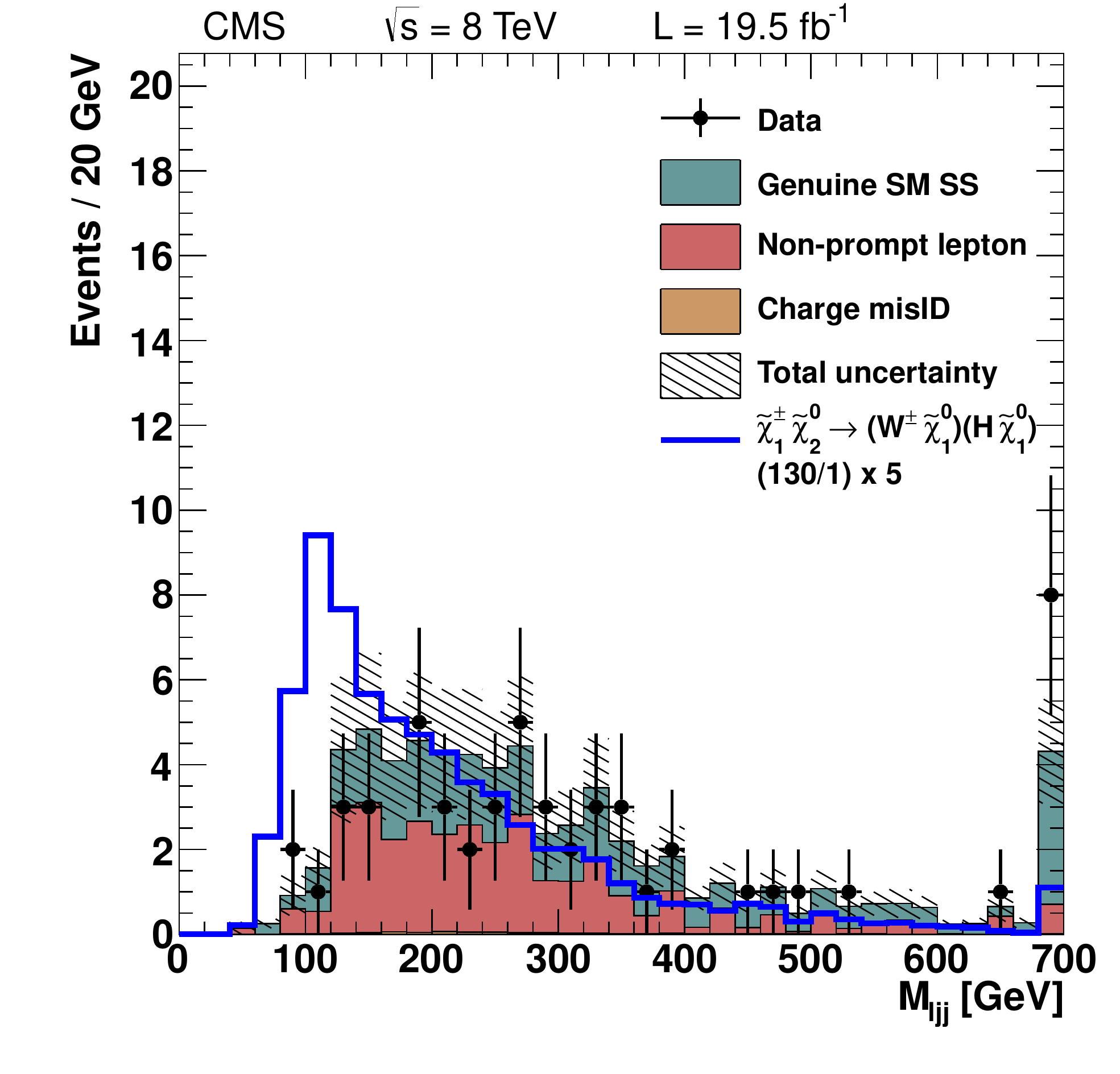}
\caption{\mljj\ distribution for the same-sign dilepton \whmet\ analysis, compared to the expected backgrounds,
after all selection requirements have been applied except for that on \mljj. An example signal scenario with
$\mchi = 130$\GeV and $\mlsp = 1$\GeV is overlaid.  For better visibility, the signal normalization has been
  increased by a factor of five relative to the theory prediction.
\label{fig:mljj}
}
\end{figure}

The object selection and background
estimation methodology for the SS dilepton search follow those presented in Section~\ref{dilepton}.
We define the quantity \mljj\ as the three-body invariant mass of the system obtained by combining the two highest \pt\ jets in an event with the lepton closest to the  dijet axis.  Signal events peak below  $m_{\PH}$, due to the undetected neutrino, as shown in Fig.~\ref{fig:mljj}. Background events generally have larger values of \mljj.  Events are required to satisfy $\mljj< 120$\GeV.

We require the presence of exactly two SS leptons  (\Pe\Pe, $\mu\mu$, or \Pe$\mu$), each with $\pt > 20\GeV$, and of either exactly two or exactly three jets, each with $\pt> 30\GeV$.   The \MET value must exceed 40\GeV.
To suppress \ttbar background, events with a ``tight'' CSV b jet or with two or more ``loose'' CSV b jets are rejected,
where the tight (loose) CSV working point corresponds to an efficiency of about 55\% (83\%) for b jets,
and a misidentification probability for light-parton jets of about 0.1\% (10\%)~\cite{btag}.
Events with  an additional electron or muon or with a \tauh candidate are rejected in order to suppress background from SM processes with multiple electroweak bosons.

The transverse mass \mt\ is computed for each of the selected leptons, and at least one lepton must satisfy $\mt > 110$\GeV.
This requirement suppresses processes containing a single leptonically decaying W boson.
We additionally require a separation $\Delta \eta(\ell_1,\ell_2)<1.6$ in order to reduce background with non-prompt leptons as well as SM events with two W bosons.

To suppress \ttbar\ events in which the decays of a W boson and a b quark lead to an SS lepton pair, we calculate the
quantity $\mtj$~\cite{ref:MT2J},
which is the minimum mass of a mother particle compatible with the four-momenta of the two leptons,  jets, and \MET.  For events with three jets, $\mtj$ is calculated with the two jets that minimize the result. We require $\mtj > 100\GeV$.

The background estimation methodology (Section~\ref{dilepton})  is validated
   using a signal-depleted data control region defined by inverting
   the \mljj\ requirement.  We observe 51 events in this control region,
   consistent with the background estimate of $62 \pm 22$ events.

The results are summarized in Table~\ref{tab:SS}.
No evidence for a peak in the \mljj\ distribution is observed, as seen from Fig.~\ref{fig:mljj}.
In the signal region $\mljj< 120\GeV$, we observe 3 events whereas $2.9 \pm 1.2$ SM background events are expected.

\begin{table*}[htb]
\centering
\topcaption{\label{tab:SS} Observed yields and SM expectations for the same-sign dilepton \whmet\ analysis. The expectations from several signal scenarios are shown;
the first number indicates \mchi\ and the second \mlsp (\GeVns).
The uncertainties include both the statistical and systematic components. }
\begin{tabular}{l|cccc} \hline\hline
Sample & $\Pe\Pe$ & $\mu\mu$ & $\Pe\mu$ & Total \\
\hline
                        Non-prompt leptons   &  0.3 $\pm$  0.3 &  0.2 $\pm$  0.2 &  0.8 $\pm$  0.5 &  1.3 $\pm$  0.8 \\
                  Charge misidentification   &  $<$0.01       &  $<$0.01       &  $<$0.03       &  $<$0.03 \\
                   Genuine SM SS dileptons   &  0.4 $\pm$  0.4 &  0.4 $\pm$  0.4 &  0.8 $\pm$  0.6 &  1.6 $\pm$  0.9 \\
\hline
                   Total background  &  0.7 $\pm$  0.5 &  0.6 $\pm$  0.5 &  1.6 $\pm$  0.7 & 2.9 $\pm$  1.2 \\
                               Data &     1 &     1 &     1 &     3 \\
\hline%\hline
\footnotesize \signal\ (130/1) & 0.7$\pm$0.1 &       0.9$\pm$0.1 &       1.8$\pm$0.2 &       3.4$\pm$0.5 \\
\footnotesize \signal\ (150/1) & 0.5$\pm$0.1 &       0.6$\pm$0.1 &       1.2$\pm$0.2 &       2.3$\pm$0.3 \\
\footnotesize \signal\ (200/1) & 0.19$\pm$0.03 &       0.35$\pm$0.05 &       0.52$\pm$0.07 &       1.1$\pm$0.1 \\
\footnotesize \signal\ (300/1) & 0.06$\pm$0.01 &       0.10$\pm$0.02 &       0.17$\pm$0.03 &       0.33$\pm$0.05 \\
\footnotesize \signal\ (400/1) & 0.02$\pm$0.00 &       0.03$\pm$0.00 &       0.05$\pm$0.01 &       0.10$\pm$0.01 \\
\hline\hline
\end{tabular}
\end{table*}

\subsection{Search in the multilepton final state}
\label{sec:multilepton}

For the multilepton search presented in Ref.~\cite{SUS13002}, events with at least three leptons are selected, including up to one \tauh candidate.
These events are categorized into multiple exclusive signal regions based on the number and flavor of
the leptons, the presence or absence of an OSSF pair, the invariant mass of the OSSF pair (if present),
the presence or absence of a tagged b jet, and the \MET and $\Ht$ values. The most sensitive signal regions for this search are those with exactly three leptons,
no tagged b jets (using the CSVM criteria), and a low $\Ht$ value.

Backgrounds from dilepton \ttbar\ events with non-prompt leptons are evaluated from simulation,
while other backgrounds with non-prompt leptons are determined using data control samples.
Backgrounds from \PW\Z and $\Z\Z$ diboson processes are estimated from simulation, with a correction
to the \MET\ resolution based on comparisons to data in control regions.

The data yields in the signal regions are found to be consistent with the expected
SM backgrounds.
The observed data yields, expected SM backgrounds, and expected signal yields
for the five most sensitive signal regions for the $\mchi=130$\GeV,
$\mlsp=1$\GeV scenario, where the multilepton analysis has the best sensitivity,
are shown in Table~\ref{tab:multileptonResults130}.
Additional signal-depleted regions are used to constrain the backgrounds and associated uncertainties.
Similar tables for other scenarios are presented in Appendix~\ref{app:multilepton}.

\begin{table}[htb]
\centering
\topcaption{\label{tab:multileptonResults130}
Observed yields and SM expectations for the multilepton \whmet\ search for the five signal regions with best sensitivity for the
$\mchi=130\GeV$, $\mlsp=1\GeV$ scenario. All five signal regions require exactly three leptons, no \tauh candidate, no tagged b jet,
and $\Ht < 200\GeV$.  The ``Below Z'' entries indicate the requirement of an OSSF lepton pair with $M_{\ell\ell}< 75\GeV$.
}
\begin{tabular}{c|c|c|c|c}
\hline
\hline
OSSF pair & $\MET$ [\GeVns{}] & Data & Total SM & Signal\\
\hline
 Below \Z & 50--100  & 142 & 125 $\pm$ 28   & 24.4 $\pm$ 4.4  \\
 Below \Z & 100--150 & 16  & 21.3 $\pm$ 8.0 & 6.8 $\pm$ 1.2   \\
 None     & 0--50    & 53  & 52 $\pm$ 12    & 8.7 $\pm$ 1.7   \\
 None     & 50--100  & 35  & 38 $\pm$ 15    & 10.8 $\pm$ 2.0  \\
 None     & 100--150 & 7   & 9.3 $\pm$ 4.3  & 3.37 $\pm$ 0.54 \\
\hline
\hline
\end{tabular}
\end{table}

\section{Searches in the final state with a non-resonant opposite-sign dilepton pair}
\label{osdilepton}
Finally, we present a search for events with an oppositely charged
\Pe\Pe, \Pe$\mu$, or $\mu\mu$
pair in which the lepton pair is inconsistent
with \Z boson decay.  The search is sensitive to the processes shown
in Fig.~\ref{fig:charginos-ll}.

Both leptons are required to have $\pt>20\GeV$.  The $\Pe\Pe$ or $\mu\mu$ invariant mass
must differ from the \Z boson mass by at least 15\GeV.  Events must have $\MET>60\GeV$ and no tagged b jet defined with the CSVM criteria.
The remaining background is mostly composed of events with
$\ttbar$ and $\W\W$ production and is
reduced using the \mctp\ variable, which is defined in Ref.~\cite{Matchev:2009ad}.

  The \mctp\ variable is designed to identify events with two  boosted massive particles that each decay into a visible particle  and an invisible one.  For events with two W bosons that each  decay leptonically, and for perfect event reconstruction,  \mctp\ has an endpoint at the W boson mass.  In practice,  because of imperfect event reconstruction, background events can  appear at larger values of \mctp.  However, for SM events, the distribution of \mctp\ falls rapidly for $\mctp>m_\PW$. In contrast, for the signal scenario, the \mctp\ distribution can extend
to much higher values.

The background evaluation for this search is based on templates that describe the shape
of the \mctp\ distribution for each of the major background categories.
The templates are obtained either from data control samples or simulation.
The template shapes are fit to data to determine their respective normalizations. Because backgrounds
from Z and ZZ processes contribute predominantly to the ee and $\mu\mu$ final states,
separate templates are derived for same-flavor and opposite-flavor events.

A top-quark control sample is selected by inverting the b-jet veto. The corresponding template
accounts for backgrounds with \ttbar\ events (with or without accompanying vector bosons)
and single-top-quark events produced with W bosons. We verify with simulation that the
corresponding \mctp\ template accurately models the shape of the targeted event sample in the signal region.

A template derived from simulation accounts for events with diboson production and for rare events, where by 'rare' we in this case mean events from  Higgs and triboson production.
 The simulation is validated using control regions. A first control region is selected by requiring
the dilepton mass to be consistent with the Z boson mass. A second control region is selected
by requiring a third isolated electron or muon. The two control regions are dominated by events with ZZ and
WZ production, respectively. The \mctp\ distribution is found to be well described by the simulation
for both control regions.

The simulation of events with WW production is validated using the three-lepton WZ-dominated
 control sample. One lepton is removed from the event, and its four-momentum is added to
the \MET\ vector. Rescaling the \mctp\ value of each event by $m_\PW/m_\Z$  yields a distribution
with very similar properties to events with WW production, as verified with simulation.
The number of events in the control sample is small, and we assign a systematic
 uncertainty to each \mctp\ bin defined by the difference between the yield in the data control
sample and the WW event simulation, or else the statistical uncertainty of the data control sample,
 whichever is larger.

Similarly, a template distribution for backgrounds with two leptons from an off-shell Z boson, with
\MET\ from misreconstructed jets, is obtained from simulation. We weight the simulated
events such that the \MET distribution agrees  with data in the on-Z
($\abs{M_{\ell\ell} - M_\Z} < 15\GeV$) control region. We then examine the \mctp\
distribution in the $\mctp < 100\GeV$, on-Z control region,
where this background is expected to dominate, to validate the simulation after all
corrections have been applied. We assign a bin-by-bin systematic uncertainty
given by the fractional difference between the data and template in this control region
(around 25\% for each bin).

We construct a template describing backgrounds with a leptonically decaying W boson and a non-prompt lepton
from a data control sample, obtained by selecting events with two same-charge leptons, one
of which has a relative isolation in a sideband defined by $0.2<\Irel<0.3$.
All other selection
   requirements are the same as for the nominal analysis.
Due to the small number of events
in the control sample, we assign a 30\% systematic uncertainty to each
bin.

\begin{figure}[htbp]
\centering
\includegraphics[width=.48\textwidth]{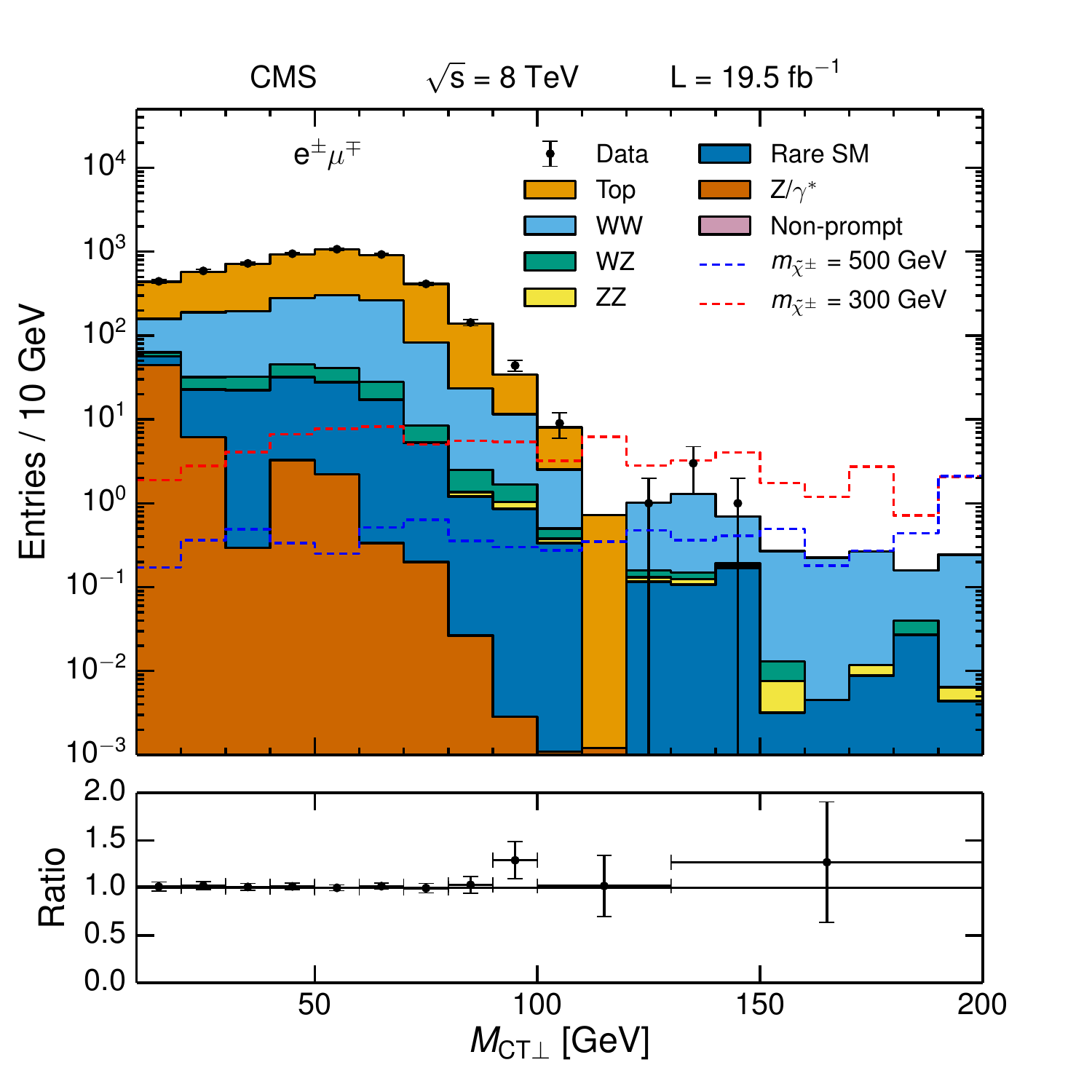}
\includegraphics[width=.48\textwidth]{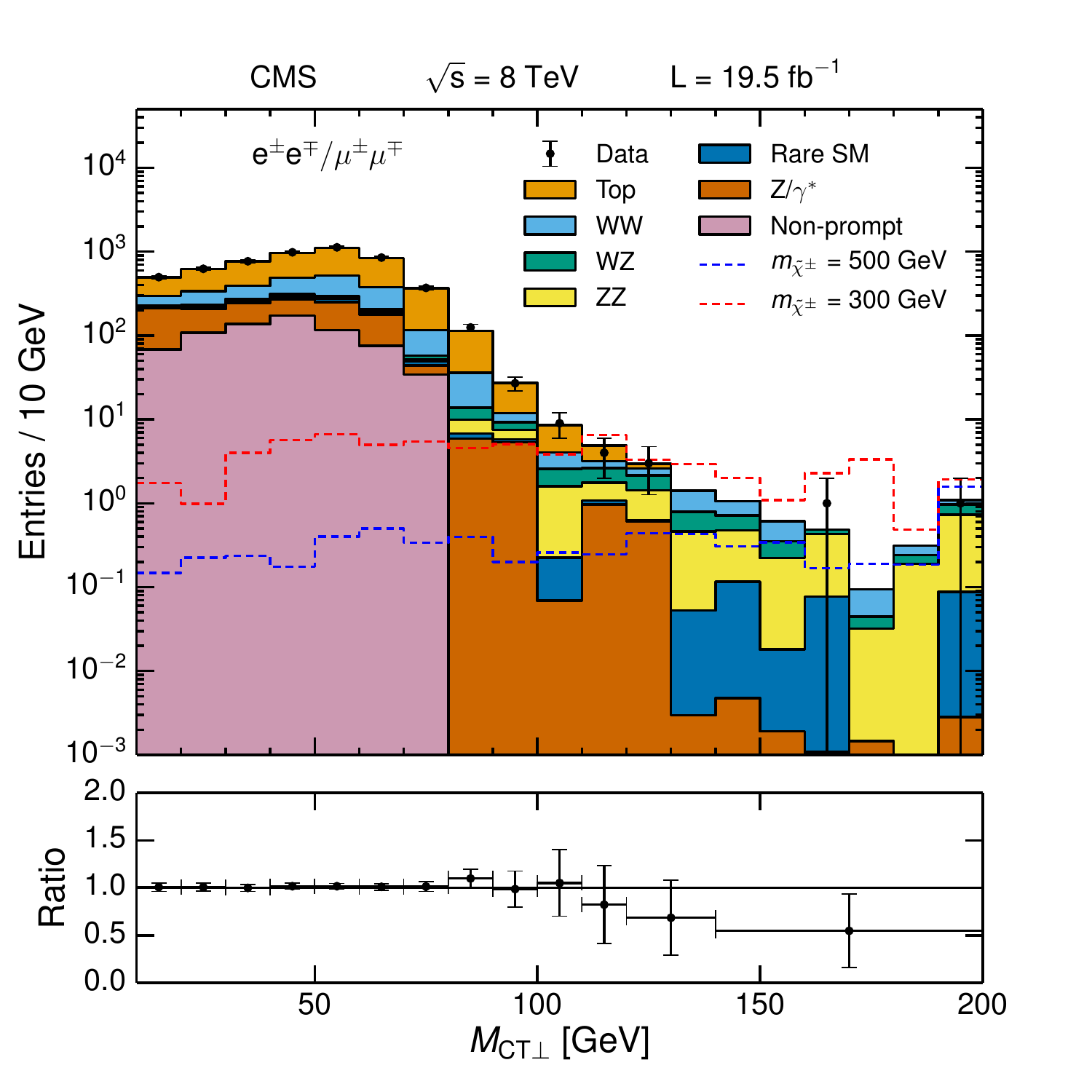}
\caption{
\mctp\ distribution for the non-resonant opposite-sign dilepton analysis compared to the background prediction
for the
(\cmsLeft) opposite-flavor and
       (\cmsRight) same-flavor channels. The background prediction is based on a fit of
        templates derived from control samples or simulation.
The signal distributions with two different chargino mass values for the SUSY
   scenario shown in Fig.~\ref{fig:charginos-slep}(\cmsLeft) are also shown, with the LSP mass set to zero.
   The ratio of the data to the fitted distribution is shown in the lower panels.
\label{fig:mctdist}}
\end{figure}

A binned maximum likelihood fit of the \mctp\ distribution is performed
  for $\mctp >10\GeV$ in order to determine the normalizations of the templates.
  The fit assumes the SM-only hypothesis.  The fitting procedure is
  validated using simulation to verify that it behaves as expected
  both with and without injected signal. The results of the fit are
presented in Table~\ref{t:osdilresults} and Fig.~\ref{fig:mctdist}.
We use a binned Anderson-Darling test~\cite{anderson1952asymptotic} to verify that the fit
  results are consistent with the SM, finding a $p$ value of 0.41
  with respect to SM-only pseudo-experiments.
\begin{table*}
\centering
    \topcaption{Results from a maximum likelihood fit of the background-only hypothesis to the \mctp\ distribution in data
    for $\mctp> 10\GeV$ for the non-resonant opposite-sign dilepton analysis.
    The corresponding results from simulation are also shown.}
    \label{t:osdilresults}
    \begin{tabular}{l|c|c|c|c}
    \hline\hline
         \multirow{2}{*}{Sample} & \multicolumn{2}{c|}{Opposite flavor} & \multicolumn{2}{c}{Same flavor}\\ \cline{2-5}
        & Fit & Simulation & Fit & Simulation \\
         \hline
Top quark & $3750\pm750$ & 3360 & $2780\pm420$ & 2472\\
Diboson and rare SM & $1460\pm210$ & 1433 & $1170\pm180$ & 1211\\
\Z/$\Pgg^*$ & $57\pm50$ & 106 & $710\pm420$ & 917\\
Non-prompt & $<$96 & 477 & $710\pm520$ & 156\\
\hline \hline
    \end{tabular}
\end{table*}

We can recast the analysis as a comparison of event counts in a high-\mctp\ signal region. To do this, we use the same templates, but fit the background normalizations in the $10 < \mctp < 120\GeV$ region, where signal contributions are expected to be negligible. We then use
these fitted normalizations to extrapolate to the $\mctp > 120\GeV$ region. Since the \ttbar\ and diboson background
shapes are similar in the low-\mctp\ region, we constrain the ratio of the \ttbar\ to diboson yields to the value obtained from simulation, assigning a 10\% uncertainty.

The results are given in Table~\ref{t:mct_cutcount}.
The sum of the yields from the low- and extrapolated high-\mctp\ regions agree
with the yields in Table~\ref{t:osdilresults}  to within the uncertainties. Note that the extra constraint on the ratio of the \ttbar\ to diboson yields leads to smaller uncertainties than those in Table~\ref{t:osdilresults}.
The numbers of observed events in the high-\mctp\ regions are found to be consistent with the background estimates, for both the opposite- and same-flavor channels.

\begin{table*}[htb]
\centering
\topcaption{
Results from a maximum likelihood fit of the background-only hypothesis to the \mctp\ distribution in data,
performed for events with $10<\mctp<120$\GeV and extrapolated to the $\mctp>120$\GeV region, for the non-resonant opposite-sign dilepton analysis.
Where the predicted value is zero, the one standard deviation upper limit is given.}
\label{t:mct_cutcount}
\cmsRow{
    \begin{tabular}{l|c|c|c|c}
    \hline\hline
       \multirow{2}{*}{Sample}  & \multicolumn{2}{c|}{Opposite flavor} & \multicolumn{2}{c}{Same flavor}\\ \cline{2-5}
        & $\mctp$ 10--120\GeV & $\mctp > 120$\GeV & $\mctp$ 10--120\GeV & $\mctp > 120$\GeV \\
         \hline
Top quark &  $3770\pm90$ & $< 0.4$ & $2770\pm110$ & $0.35\pm0.10$\\
Diboson and rare SM & $1430\pm110$ & $4\pm3$ & $1240\pm90$ & $9\pm3$\\
Z/$\Pgg^*$ & $57\pm25$ & $< 0.01$ & $700\pm240$ & $0.6\pm0.3$\\
Non-prompt & $<81$ & $< 0.01$ & $659\pm77$ & $< 0.5$\\
Total & $5260\pm130$ & $4\pm3$ & $5370\pm100$ & $10\pm3$\\ \hline
    Data       & 5309         & 5           & 5388         & 5          \\\hline \hline
    \end{tabular}
    }
\end{table*}

For slepton pair production (Fig.~\ref{fig:charginos-ll}(\cmsRight)), in which only same-flavor lepton pairs are produced,  we also consider a more  focused approach in which events with opposite-flavor dilepton pairs  provide a data control sample.  We use the \mctp\ distribution of the opposite-flavor dilepton events to define a template for  the flavor-symmetric background. The flavor-symmetric background includes top-quark and WW events, as well as WZ events in which one selected lepton comes from the W boson and the other from the Z boson.  By using a single template to account for several different processes, we reduce the number of free parameters, thereby increasing the statistical precision of the search.
To accommodate the new template, the diboson template is modified slightly so that it accounts only for non-flavor-symmetric diboson processes: WZ events where both selected leptons come from a Z boson, and ZZ events.
The Z$/\gamma^*$ and non-prompt templates remain unchanged.

\begin{figure}[htb]
\centering
\includegraphics[width=\cmsFigWidth]{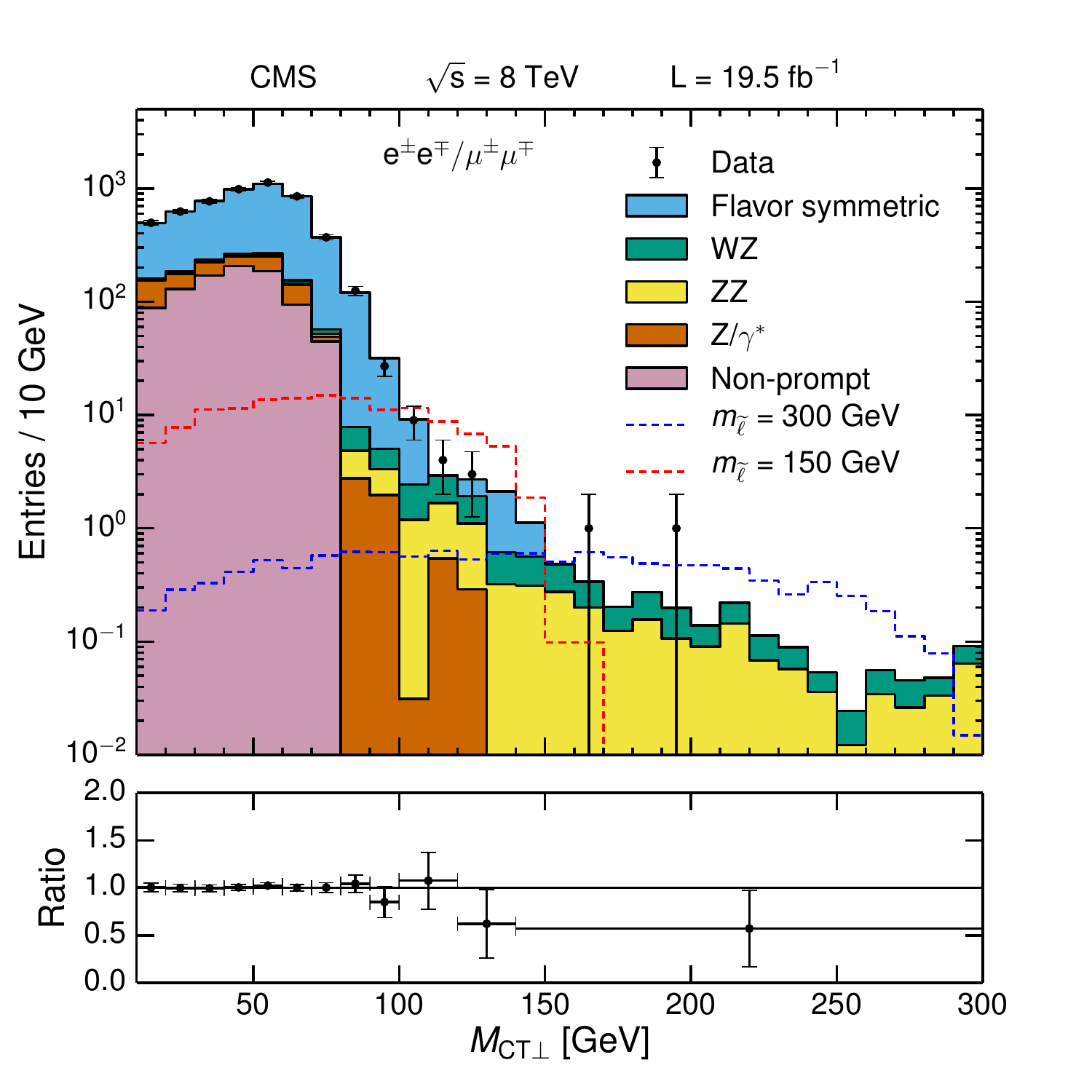}
        \caption{
\mctp\ distribution compared to the background prediction
for the same-flavor channel of the non-resonant opposite-sign dilepton analysis,
   where the background prediction is
   derived from an alternative template method that uses opposite-flavor
dilepton events as a control sample (see text).
The signal distributions with two different slepton mass values for the SUSY
scenario shown in Fig.~\ref{fig:charginos-ll}(\cmsRight) are also shown, with the LSP mass set to zero.
The ratio of the data to the fitted distribution is shown in the lower panel.
}
        \label{fig:mctdist2}

\end{figure}

We perform a maximum likelihood fit of these templates to the  measured same-flavor \mctp\ distribution under the SM-only hypothesis.
  The results are presented in Fig.~\ref{fig:mctdist2} and Table~\ref{t:osdilresults2}.  The resulting Anderson-Darling $p$ value is 0.22, implying consistency of the data with the SM.

\begin{table}[htb]
\centering
\topcaption{
Results from a maximum likelihood fit of the background-only hypothesis  to the \mctp\ distribution of the same-flavor channel
with $\mctp> 10\GeV$, for the non-resonant opposite-sign dilepton analysis, where the  background prediction is derived from
an alternative template method  that uses opposite-flavor dilepton events as a control sample (see text). For comparison, the
SM expected yields based on simulation are also indicated.
}
    \label{t:osdilresults2}
    \begin{tabular}{l|c|c}
    \hline\hline
      \multirow{2}{*}{Sample}    & \multicolumn{2}{c}{Same flavor} \\ \cline{2-3}
        & Fit & Simulation \\
         \hline
Flavor symmetric & $4040\pm490$ & 3620\\
Non-FS diboson & $98\pm50$ & 60\\
$\Z/\Pgg^*$ & $330^{+560}_{-330}$ & 917\\
Non-prompt & $920\pm840$ & 156\\
\hline\hline
    \end{tabular}
\end{table}

\section{Interpretations of the searches}
\label{sec:interpretation}

We now present the interpretation of our results in the context of models for the direct electroweak
pair production of charginos, neutralinos, and sleptons.
We compute 95\% confidence level (CL) upper limits on the new-physics cross sections using the \cls
method~\cite{Junk:1999kv,Read:2002hq,ATLAS:1379837}, incorporating the uncertainties in the signal
efficiency and acceptance described below and the uncertainties of the expected background ($\sigma_{\text{experiment}}$).
For each point in the signal parameter space we arrange the search regions according to their expected
sensitivity, and compute limits using the results from simultaneous counting experiments in the
most sensitive search regions.
For the $\PW\PH$ search we use  the search regions that contribute to 90\% of the total signal acceptance.
For the other searches, we use the ten most sensitive search regions.
The NLO+NLL cross sections from Refs.~\cite{Fuks:2012qx,Fuks:2013vua,Fuks:2013lya}
are used to place constraints on the masses of the charginos, neutralinos, and sleptons.

In setting limits, we account for the following sources of systematic uncertainty
associated with the signal event acceptance and efficiency. The
uncertainty of the integrated luminosity determination is 2.6\%~\cite{LUMIPAS}.
Samples of $\Z \to \ell\ell$ events are used to measure the lepton efficiencies, and the
corresponding uncertainties (3\% per lepton) are propagated to the signal event acceptance and efficiency.
The uncertainty of the trigger efficiency is 5\% for the dilepton and single-lepton triggers used.
The uncertainty of the b-jet tagging efficiency results in
an uncertainty for the acceptance that depends on the model details but is typically less than 5\%.
The energy scale of hadronic jets is known to 1--4\%, depending on
$\eta$ and \pt, yielding an uncertainty of 1--5\% for the signal event
selection efficiency. The larger uncertainties correspond to models for
which the difference $\Delta$M between the masses \mchi\ and \mlsp\ is small.
The experimental acceptance for signal events depends on the level of initial-state radiation activity,
especially in the small $\Delta$M region where an initial-state boost may be required for an event
to satisfy the selection requirements, including those on \MET\ and \MT.
We use the method of Ref.~\cite{SUS13011} to correct for
an observed overestimation in simulation (of up to 20\%) of the fraction of events with a large initial-state
boost, and to assign corresponding systematic uncertainties.
The signal cross sections are varied by their uncertainties~\cite{PDF4LHC} of approximately 5\%
to determine the $\pm 1$ standard deviation ($\sigma_{\mathrm{theory}}$) excluded regions.

\subsection{Limits on chargino-neutralino production with slepton-mediated decays}
\label{tri-ss-combine}

We first place limits on the production of chargino-neutralino pairs in models
with light sleptons, depicted in Fig.~\ref{fig:charginos-slep}, using the results of the three-lepton
(Section~\ref{trilepton}) and same-sign dilepton (Section~\ref{dilepton}) searches.
Three different scenarios are considered, with different assumptions about
the nature of the sleptons, which affect the number of $\tau$ leptons in the final state.
These interpretations depend on whether the sleptons are the SUSY partners $\slep_L$ or
$\slep_R$ of left-handed or right-handed leptons.  We consider two limiting cases.
 In one case, $\slep_R$ does not participate while $\slep_L$ and $\snu$ do: then
both diagrams of Fig.~\ref{fig:charginos-slep} exist, and the chargino
and neutralino decay to all three lepton flavors with equal
probability.  Furthermore, two additional diagrams in which the decay
$\chiz_2\to\ell\,\slep\to\ell\,\ell\,\chiz_1$ is replaced
by $\chiz_2\to\snu\,\nu\to\nu\,\nu\,\chiz_1$ reduce
the fraction of three-lepton final states by 50\%.  In the second
case, in which $\slep_R$ participates while $\slep_L$ and $\snu$ do
not, only the diagram of Fig.~\ref{fig:charginos-slep}(\cmsRight) exists, and
there is no reduction in the three-lepton final states.  Because the
$\slep_R$ couples to the chargino via its higgsino component, chargino
decays to $\slep_R$ strongly favor production of a $\tau$  lepton.  We thus consider three flavor scenarios:
\begin{itemize}
\item the ``flavor-democratic'' scenario: the chargino ($\chipm_1$) and
neutralino ($\chiz_2$) both decay with equal probability into all
three lepton flavors, as expected for $\slep_L$;
\item the ``$\tau$-enriched'' scenario: the chargino decays exclusively to a
$\tau$ lepton as expected for $\slep_R$, while the neutralino decays
democratically;
\item the ``$\tau$-dominated'' scenario: the chargino and neutralino both decay only to $\tau$ leptons.
\end{itemize}

\begin{figure}[htbp]
\centering
\includegraphics[width=\cmsFigWidth]{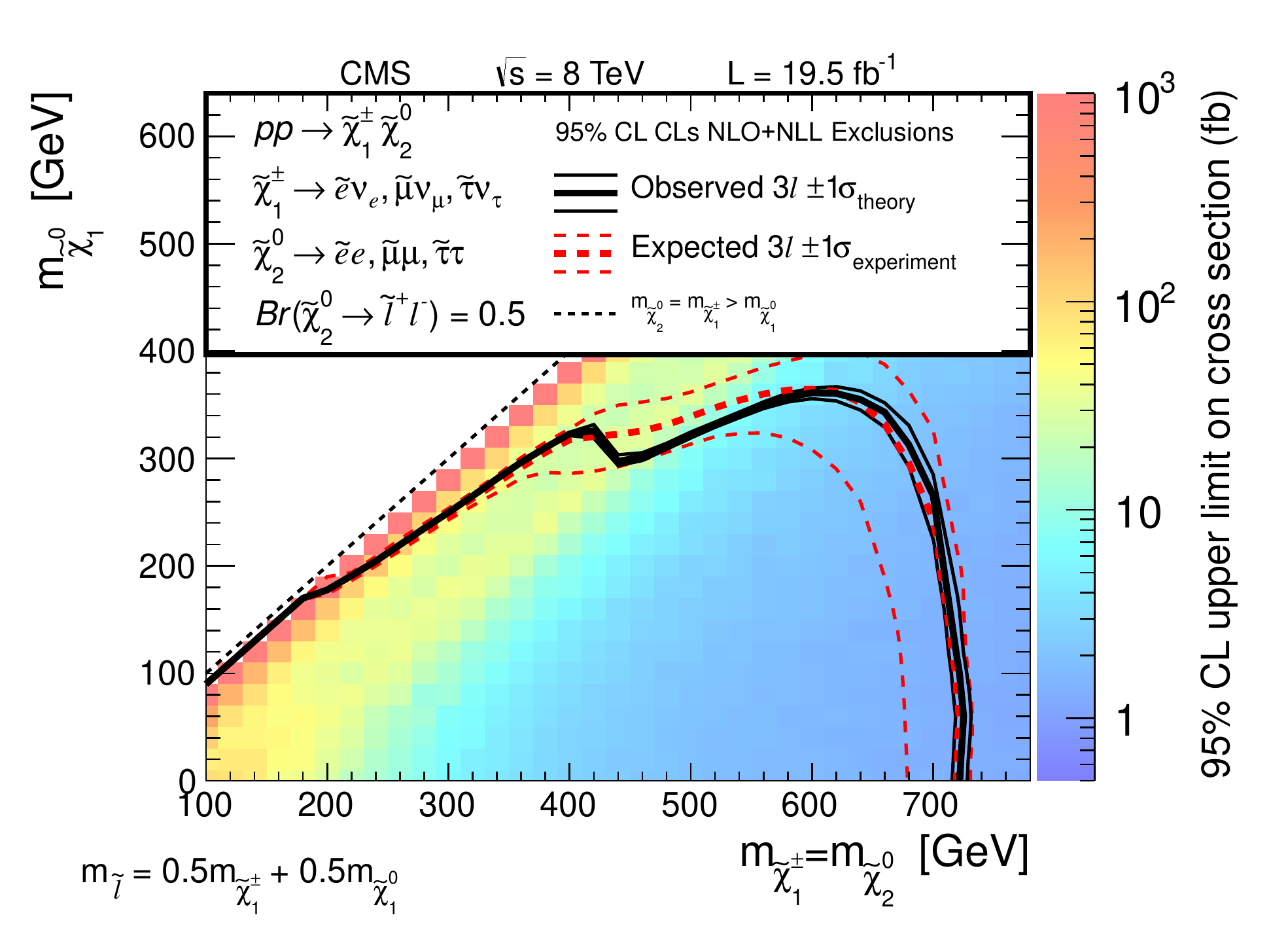}
\caption{
Interpretation of the results of the three-lepton search in the flavor-democratic
signal model with slepton mass parameter $\xslep=0.5$.
The shading in the $m_{\chiz_1}$ versus $m_{\chiz_2}$
($=m_{\chipm_1}$) plane indicates the 95\% CL upper limit on the
chargino-neutralino production cross section times branching fraction.
The contours bound the mass regions excluded at 95\%
CL assuming the NLO+NLL cross sections for a branching fraction of 50\%, as appropriate for the visible
decay products in this scenario.
The observed, ${\pm}1\sigma_{\text{theory}}$ observed, median expected, and $\pm1\sigma_{\text{experiment}}$ expected bounds are shown.
\label{fig:ULtriA}
}
\end{figure}

Figure~\ref{fig:ULtriA} displays the results from the three-lepton search,
interpreted in the flavor-democratic scenario.
The figure depicts the 95\% CL upper limit
on the cross section times branching fraction in the $m_{\chiz_1}$
versus $m_{\chiz_2}$ (${=}m_{\chipm_1}$) plane.
The 50\% branching fraction to three leptons is taken into account.
The upper limit on the cross section times branching fraction generally becomes
more stringent with the increasing mass difference between the chargino or
heavy neutralino and the LSP. A drop in sensitivity is observed in the region where
this mass difference leads to dilepton pairs with invariant masses close to that of the \Z boson,
and is caused by a higher rate for the WZ background.

\begin{figure}[htbp]
\centering
\includegraphics[width=0.49\textwidth]{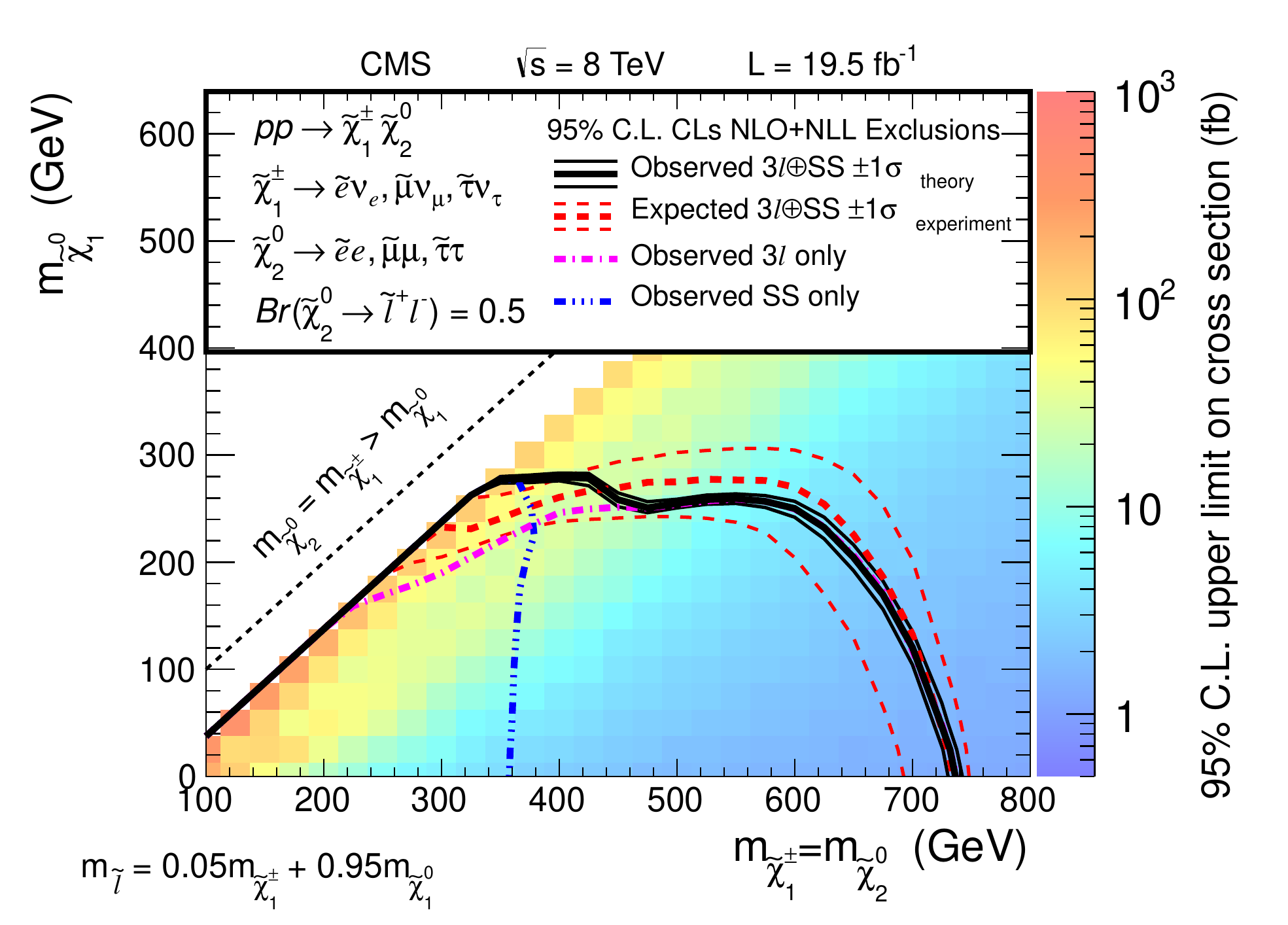}
\includegraphics[width=0.49\textwidth]{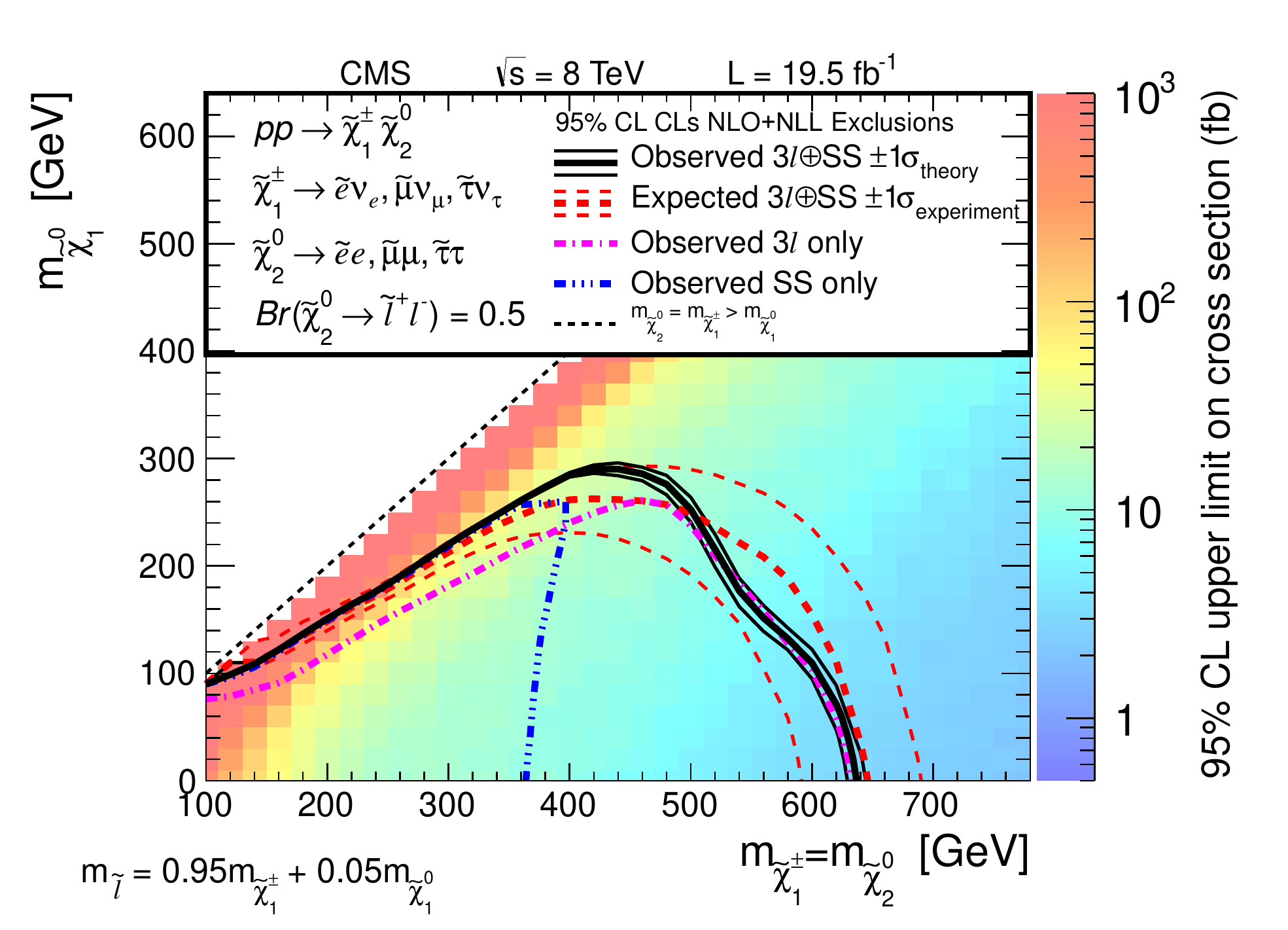}
\caption{
Interpretation of the results of the three-lepton search, the same-sign dilepton search, and their combination,
in the flavor-democratic signal model with two different values of the slepton mass parameter: (\cmsLeft) $\xslep=0.05$, (\cmsRight) $\xslep=0.95$.
The shading indicates the 95\% CL upper limits on the cross section times branching fraction,
and the contours the excluded regions assuming the NLO+NLL signal cross sections.
}
\label{fig:ULtriAss}
\end{figure}

The corresponding results for the combination of the SS dilepton and three-lepton searches are shown in
Fig.~\ref{fig:ULtriAss} for two values of $\xslep$ (0.05 and 0.95).

\begin{figure}[tbhp]
\begin{center}
\includegraphics[width=0.49\textwidth]{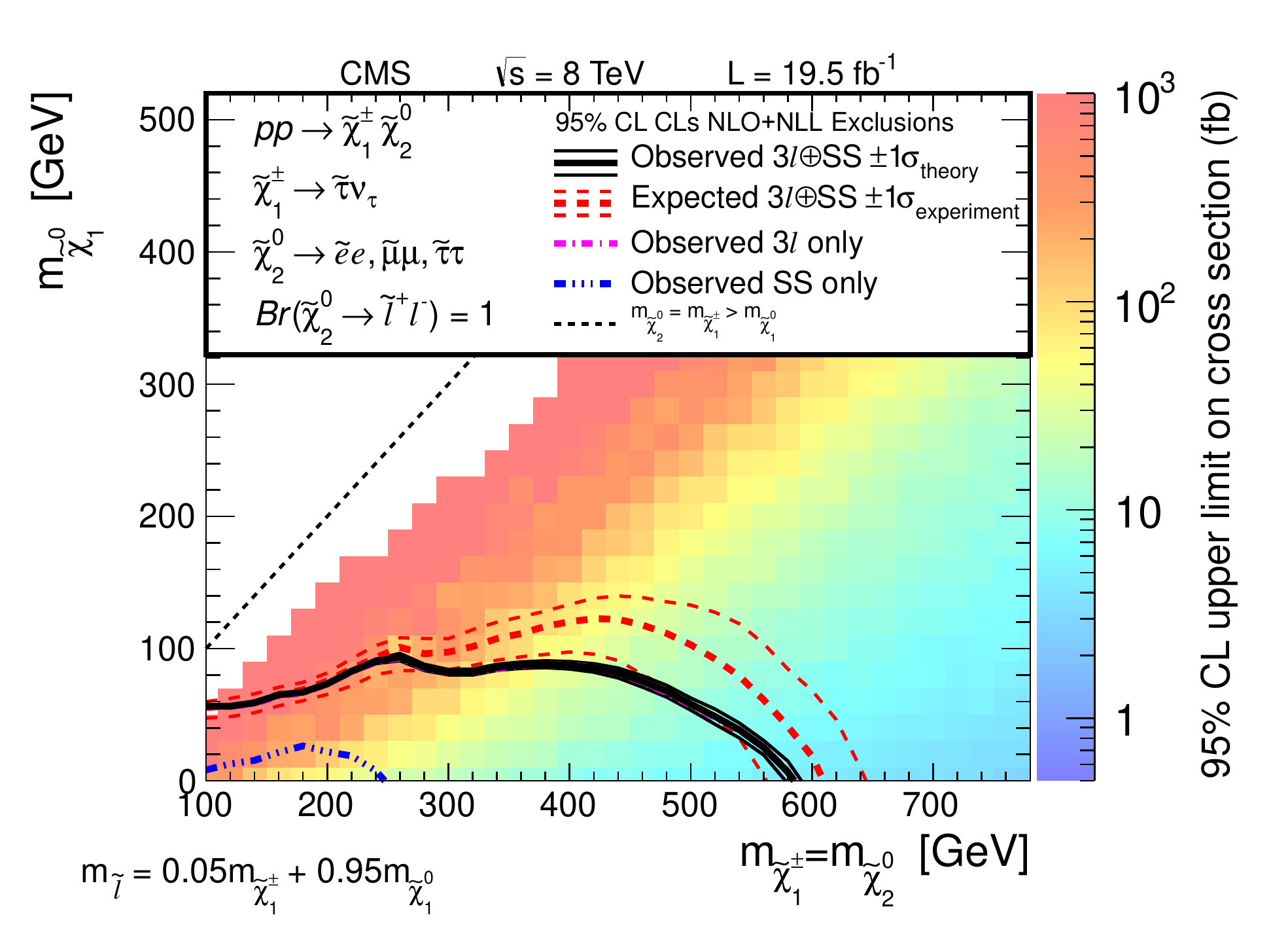}
\includegraphics[width=0.49\textwidth]{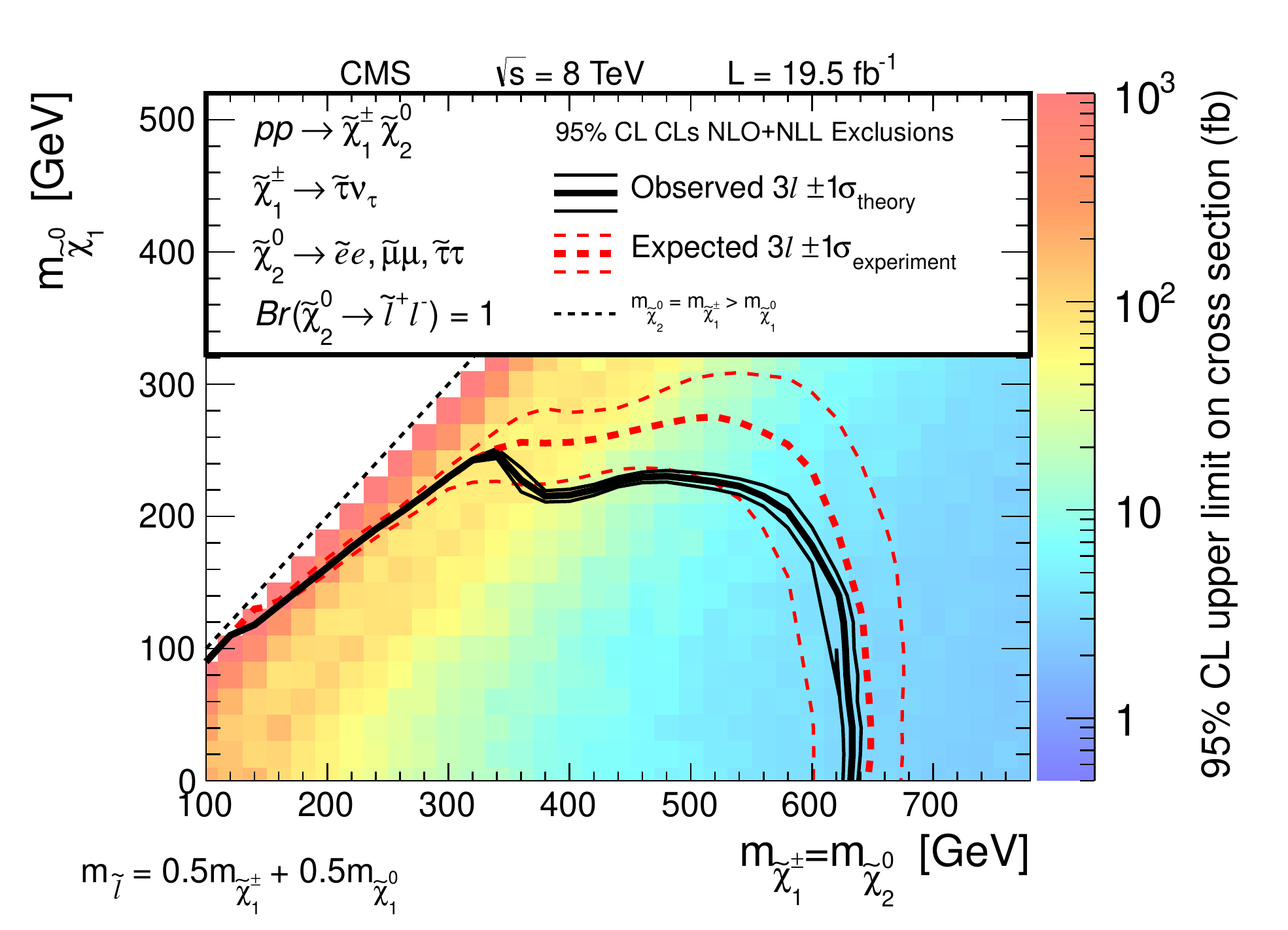}
\includegraphics[width=0.49\textwidth]{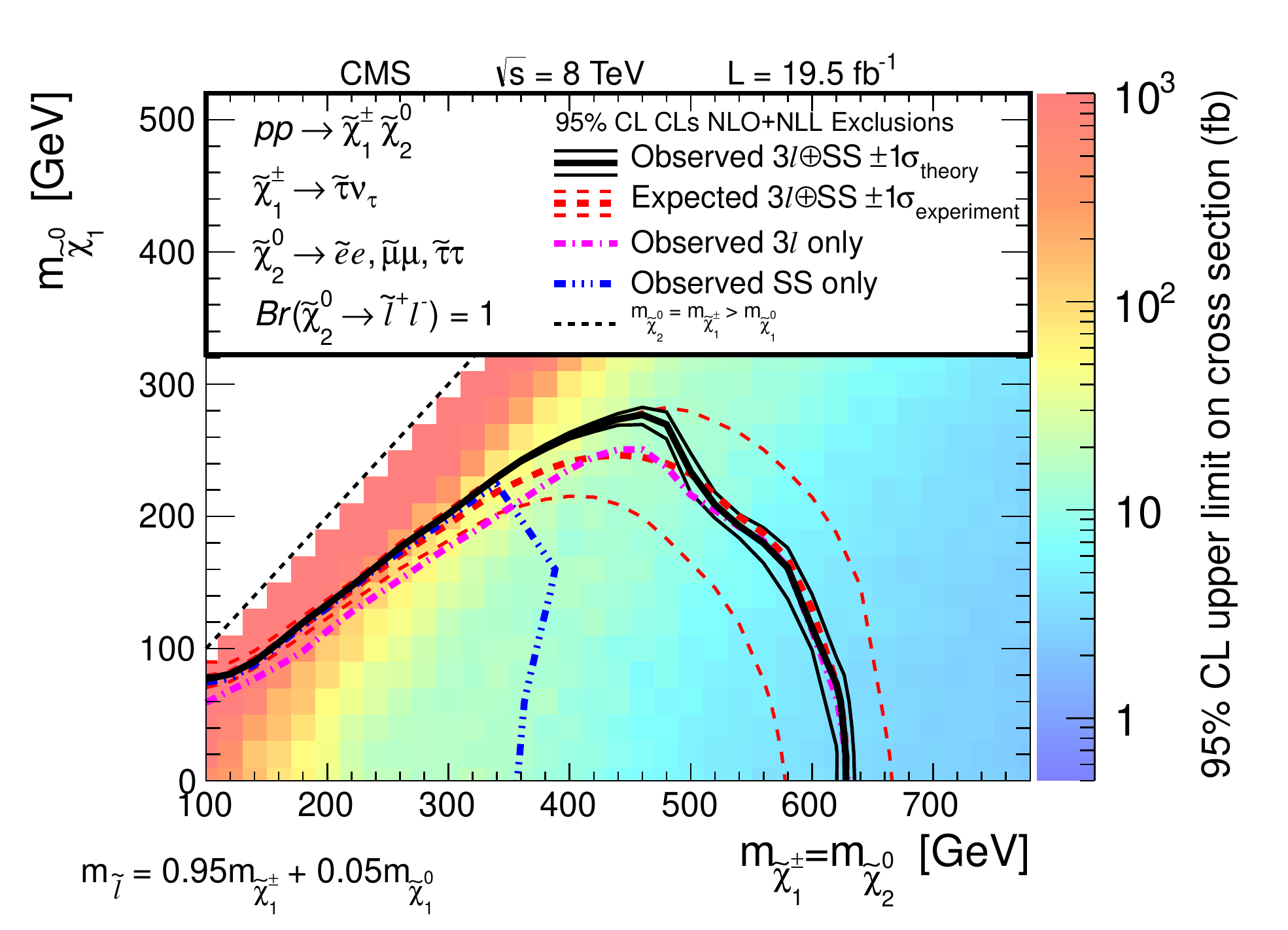}
\caption{
Interpretation of the results of the three-lepton search, the same-sign dilepton search, and their combination, for the $\tau$-enriched signal model with
 (\cmsTLeft) $\xslep=0.05$ and (bottom) $\xslep=0.95$; (\cmsTCenter) interpretation of the three-lepton search for the $\tau$-enriched signal model with $\xslep=0.5$.
The shading indicates the 95\% CL upper limits on the cross section times branching fraction,
and the contours the excluded regions assuming the NLO+NLL signal cross sections.
}
\label{fig:ULtriB}
\end{center}
\end{figure}
\begin{figure}[htbp]
\begin{center}
\includegraphics[width=\cmsFigWidth]{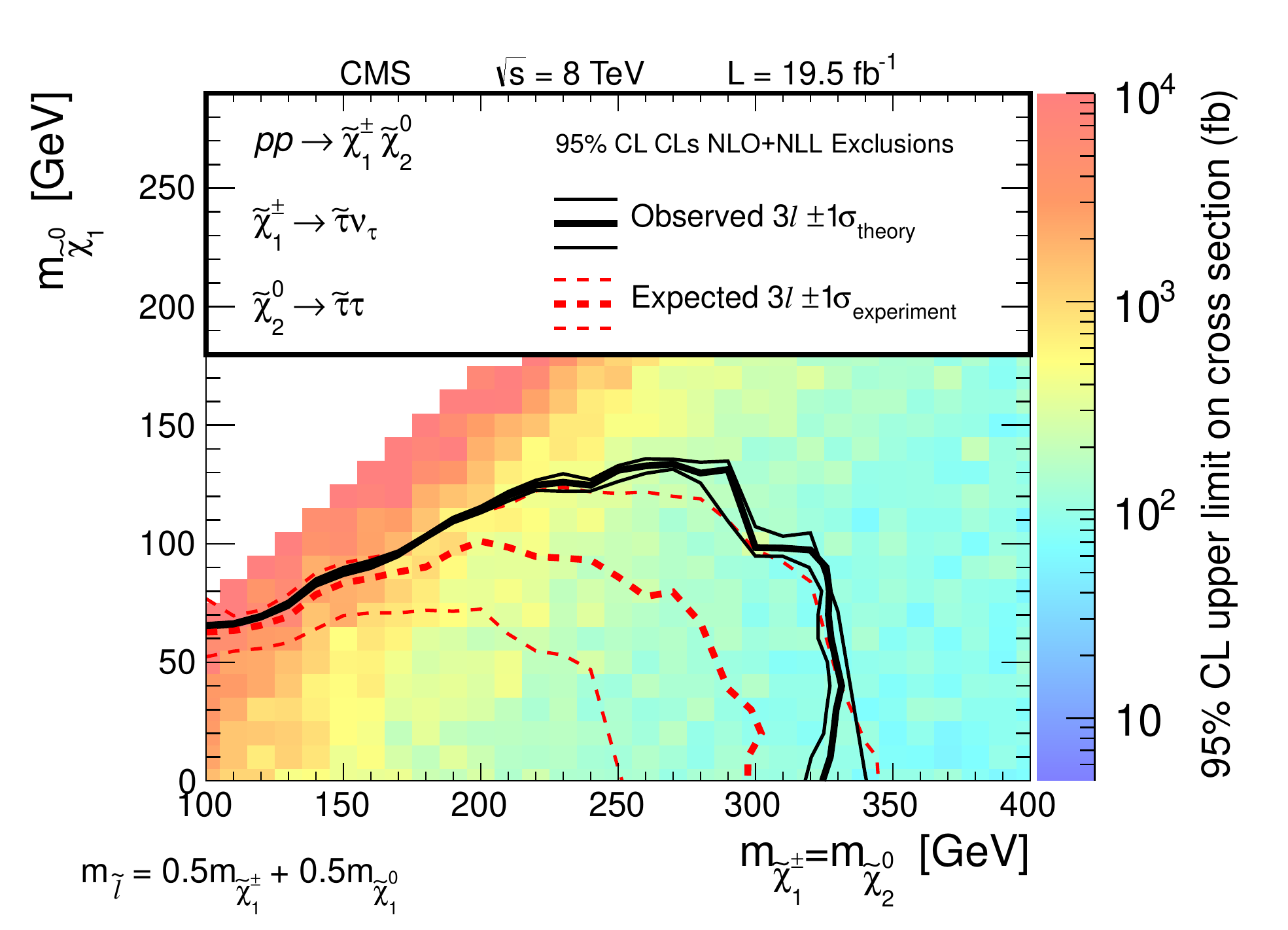}
\caption{
Interpretation of the results of the three-lepton search in the $\tau$-dominated signal model.
The shading indicates the 95\% CL upper limits on the cross section times branching fraction,
and the contours the excluded regions assuming the NLO+NLL signal cross sections.
}
\label{fig:ULtriC}
\end{center}
\end{figure}
Figure~\ref{fig:ULtriB} presents the corresponding limits for the $\tau$-enriched scenario
and Fig.~\ref{fig:ULtriC} for the $\tau$-dominated scenario.
For the $\xslep=0.50$ scenario, all three leptons are produced with significant values of \pt.
As a consequence, the trilepton analysis
is more sensitive than the SS dilepton search, for which the limit contours are omitted in
Figs.~\ref{fig:ULtriA},~\ref{fig:ULtriB}(\cmsTCenter), and~\ref{fig:ULtriC}.
For the other limit curves in Figs.~\ref{fig:ULtriAss}-\ref{fig:ULtriC}, the increase in the
combined mass limit due to incorporation of the SS dilepton search occurs in the experimentally challenging
region where the two neutralinos have similar masses.

For the models with $\xslep = 0.05$ (Figs.~\ref{fig:ULtriAss}(\cmsLeft) and
\ref{fig:ULtriB}(\cmsTLeft)), the decay $\sTau\to\tau\lsp$ is not kinematically
allowed for signal scenarios with $m_{\chipmo}-m_{\lsp} < 20 m_\tau$.
Therefore, in this region, the decay $\chipmo\to\sTau\nu_\tau$ is
suppressed.  Similarly, in the models with $\xslep = 0.95$
(Figs.~\ref{fig:ULtriAss}(\cmsRight) and \ref{fig:ULtriB}(\cmsTRight)), the decay
$\chitn\to\sTau\tau$ is not kinematically allowed in the region with
$m_{\chitn}-m_{\lsp} < 20 m_\tau$.

\subsection{Limits on chargino-neutralino production without light sleptons}

We next place limits on chargino-neutralino production under the assumption that
the sleptons are too heavy to participate, as depicted in Fig.~\ref{fig:charginos-wz}.
The chargino is assumed to always decay to a \PW\ boson
and the $\chiz_1$ LSP. The $\chiz_2$ is expected to decay to a $\chiz_1$ LSP and either
a \Z boson or the Higgs boson. The relative branching fraction ($\mathcal{B}$) for these two decays
is in general model-dependent~\cite{Howe:2012xe}. We thus consider two limiting cases, in which
either $\mathcal{B}(\chiz_2\to\Z\chiz_1)=1$ (Section~\ref{sec:wzmet}), or
$\mathcal{B}(\chiz_2\to\PH\chiz_1)=1$ (Section~\ref{sec:whmet}). The sensitivity in a generic
model lies between these two extremes.

\subsubsection{Limits on chargino-neutralino production in the \wzmet\ final state}
\label{sec:wzmet}

To evaluate upper limits on the process of Fig.~\ref{fig:charginos-wz}(\cmsLeft), we
use the results of the \wzzmet analysis (Section~\ref{diboson})
together with the three-lepton analysis (Section~\ref{trilepton}).
Figure~\ref{fig:WZetmiss}(\cmsLeft) displays the observed limits for the individual studies
and their combination. The sensitivities of the three-lepton
and \wzzmet analyses are complementary, with the three-lepton results dominating
the sensitivity in the region where the difference between the neutralino masses is small,
and the \wzzmet results dominating the sensitivity in the region where \mchi\ is large.
A significant degradation in sensitivity is present in the region of parameter space in which
$\Delta M\approx M_Z$, causing the chargino and neutralino decay products to be produced with low momentum
in the rest frame of their mother particles.
The observed limits are less stringent than the expected limits because
the data lie above the expected background in the three-lepton ee and $\mu\mu$ OSSF search regions with $\MT>160\gev$ and  $75<Mll<105\gev$ (see Fig.~\ref{fig:OSSFMET} and Table~\ref{tab:L3OSSF}).

\subsubsection{Limits on chargino-neutralino production in the \whmet\ final state}
\label{sec:whmet}

\begin{figure}[htbp]
\centering
\includegraphics[width=0.49\textwidth]{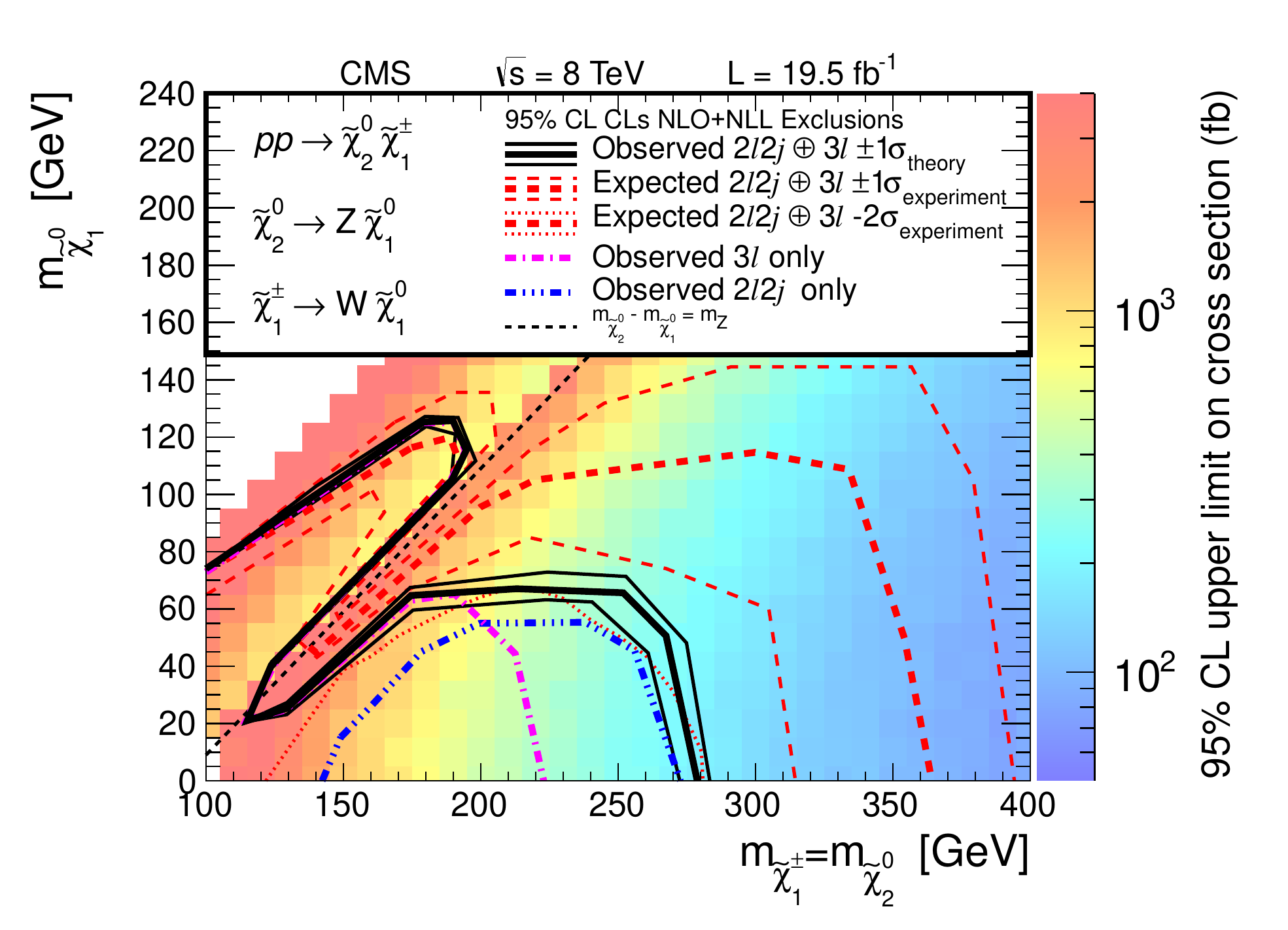}
\includegraphics[width=0.49\textwidth]{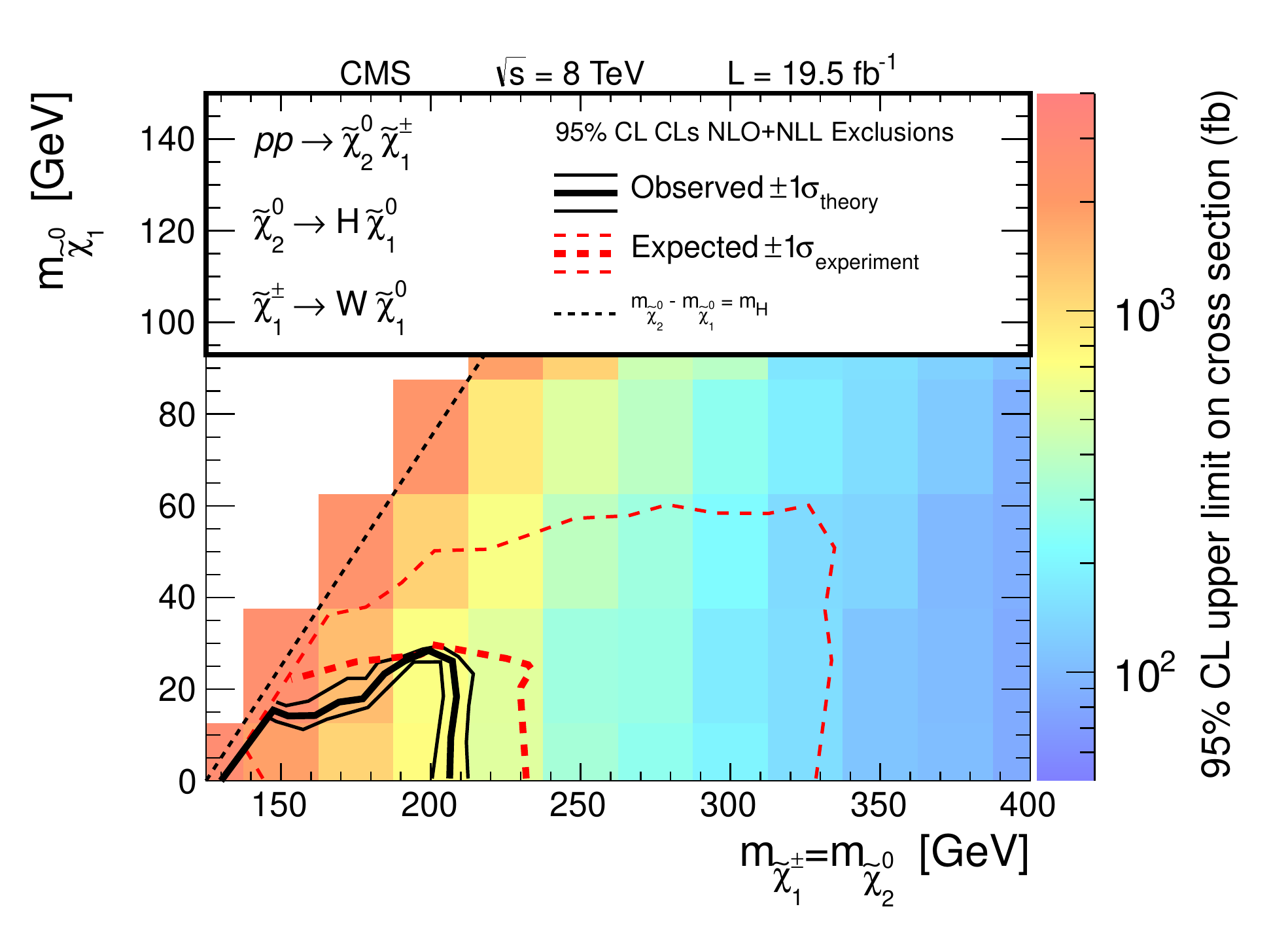}
\caption{ (\cmsLeft) Interpretation of the results of the \zdijet\ search,
the three-lepton search, and their combination, in the \wzmet\ model.
(\cmsRight) Interpretation of the combined results of the single-lepton, same-sign dilepton, and multilepton search regions, in the \whmet\ model.
The shading indicates the 95\% CL upper limits on the cross section times branching fraction,
and the contours the excluded regions assuming the NLO+NLL signal cross sections.
\label{fig:WZetmiss}}
\end{figure}

To evaluate upper limits for the process of Fig.~\ref{fig:charginos-wz}(center), we
combine the results of the single-lepton, SS dilepton, and multilepton searches
described in Section~\ref{sec:wh}.
Figure~\ref{fig:WZetmiss}(\cmsRight) displays the observed limits for the
combination of these analyses.
The multilepton search provides the best sensitivity at low \mchi, while the single-lepton search dominates
at high \mchi.  The same-sign dilepton search contributes to the combination at low \mchi.
In Appendix \ref{app:1Dplots} the observed and expected results for the \whmet\ final state are presented as a function of \mchi, for a fixed mass $\mlsp=1$\GeV, for each of the three search regions and their combination.

\subsection{Limits on a \texorpdfstring{$\Z$}{Z}-boson enriched GMSB model
}

\label{app:combo}

\begin{figure}[htbp]
\centering
    \includegraphics[width=\cmsFigWidth]{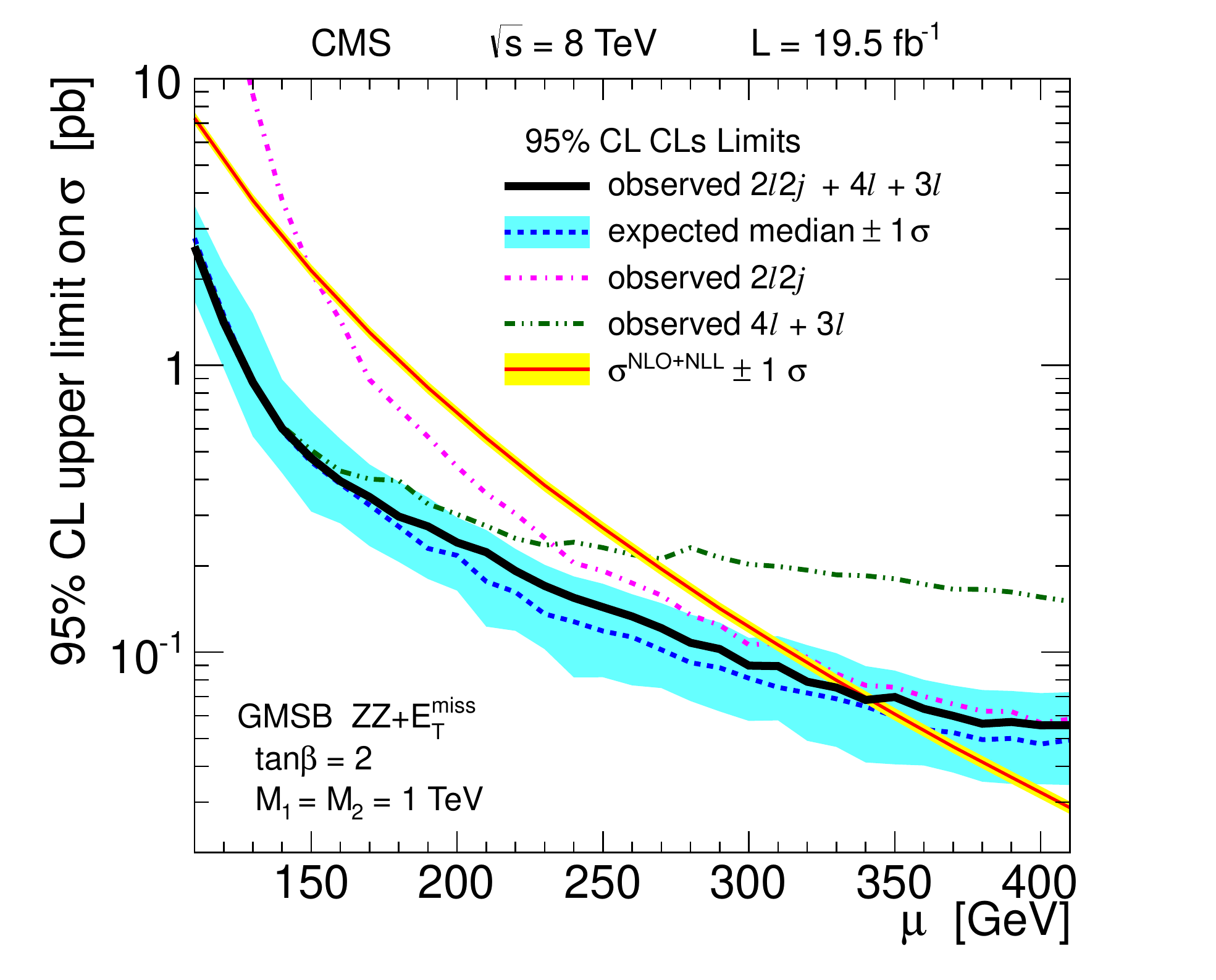}
\caption{
\label{fig:combo}
Interpretation of the results of the $\Z+\text{dijet}$ search, the three- and four-lepton searches,
and their combination, in the GMSB scenario discussed in the text.
The observed and expected 95\% CL upper limits on the cross section are indicated as a function of the higgsino mass
parameter $\mu$, and are compared to the theoretical cross section.
}
\end{figure}

We also consider a gauge-mediated symmetry breaking
(GMSB) \Z-boson enriched higgsino
model which predicts an enhanced branching fraction to the \zzmet\ final state. The
LSP in this model is an almost massless gravitino ($\sGra$), the
next-to-lightest SUSY particle is a higgsino $\chiz_1$,
and the $\chipm_1$ and $\chiz_2$ particles are nearly mass degenerate with the $\chiz_{1}$.  We
set the gaugino mass parameters to $M_1=M_2=1\TeV$ and
the ratio of Higgs bosons vacuum expectation values to $\tan\beta=2$.
The results are presented as a function of the higgsino mass parameter $\mu$, where
$m_{\chiz_1} \approx m_{\chiz_2} \approx m_{\chipm_1} \approx \mu$ to within typical mass differences
of a few\GeV.
The branching fraction
to the \zzmet\ final state
varies from 100\% at $\mu=130$\GeV to 85\% at
$\mu=420$\GeV.
We use the results of the three-lepton (Section~\ref{trilepton}),
four-lepton (Section~\ref{quadlepton}), and
\wzzmet\ (Section~\ref{diboson})
searches to constrain the GMSB scenario.
The results
are presented in
Fig.~\ref{fig:combo}. %The region $\mu < 330$\GeV is excluded at 95\% CL.

\subsection{Limits on chargino and slepton pair production}
\label{limchargino}

\begin{figure}[htbp]
\centering
\includegraphics[width=0.49\textwidth]{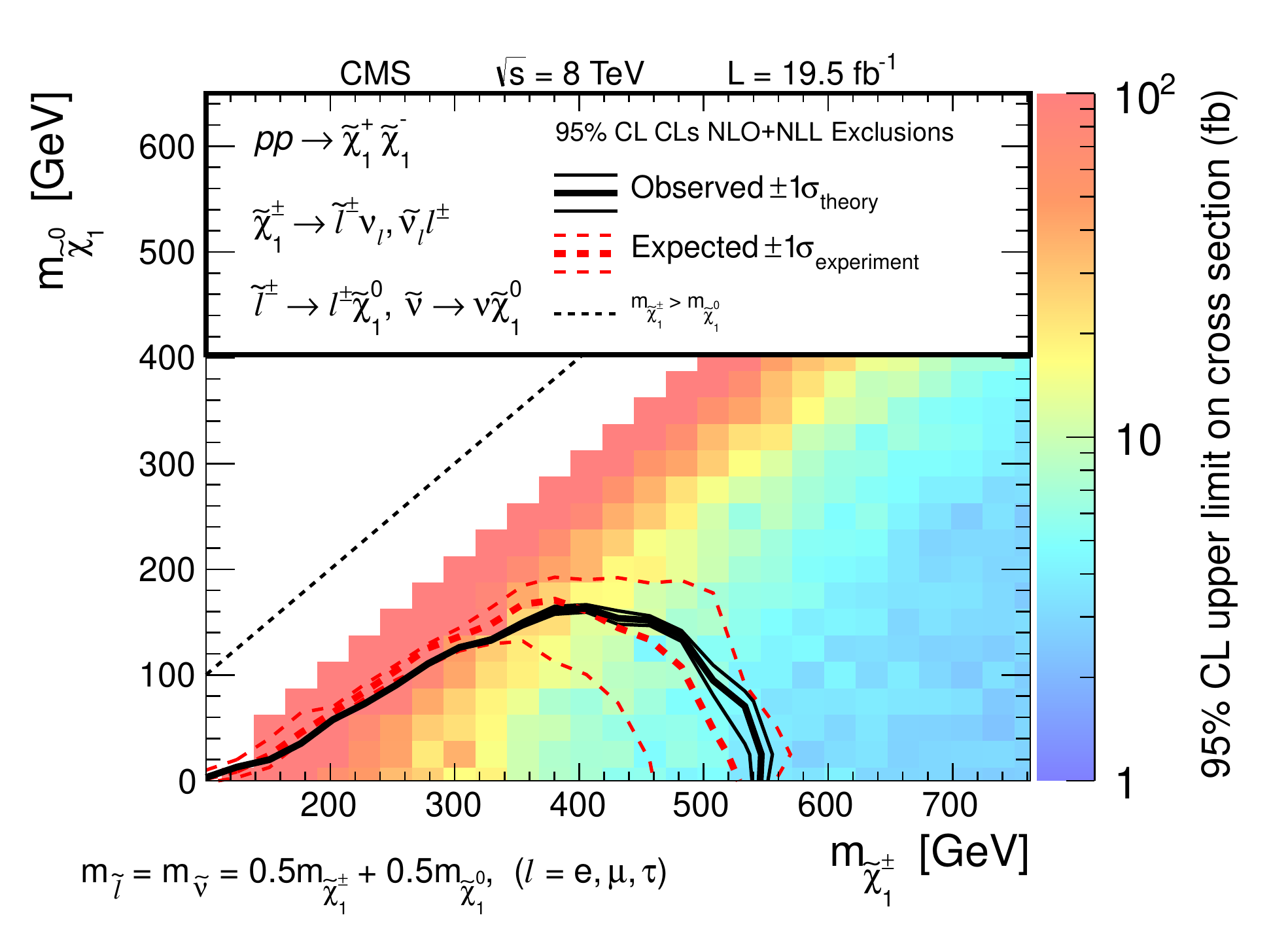}
\includegraphics[width=0.49\textwidth]{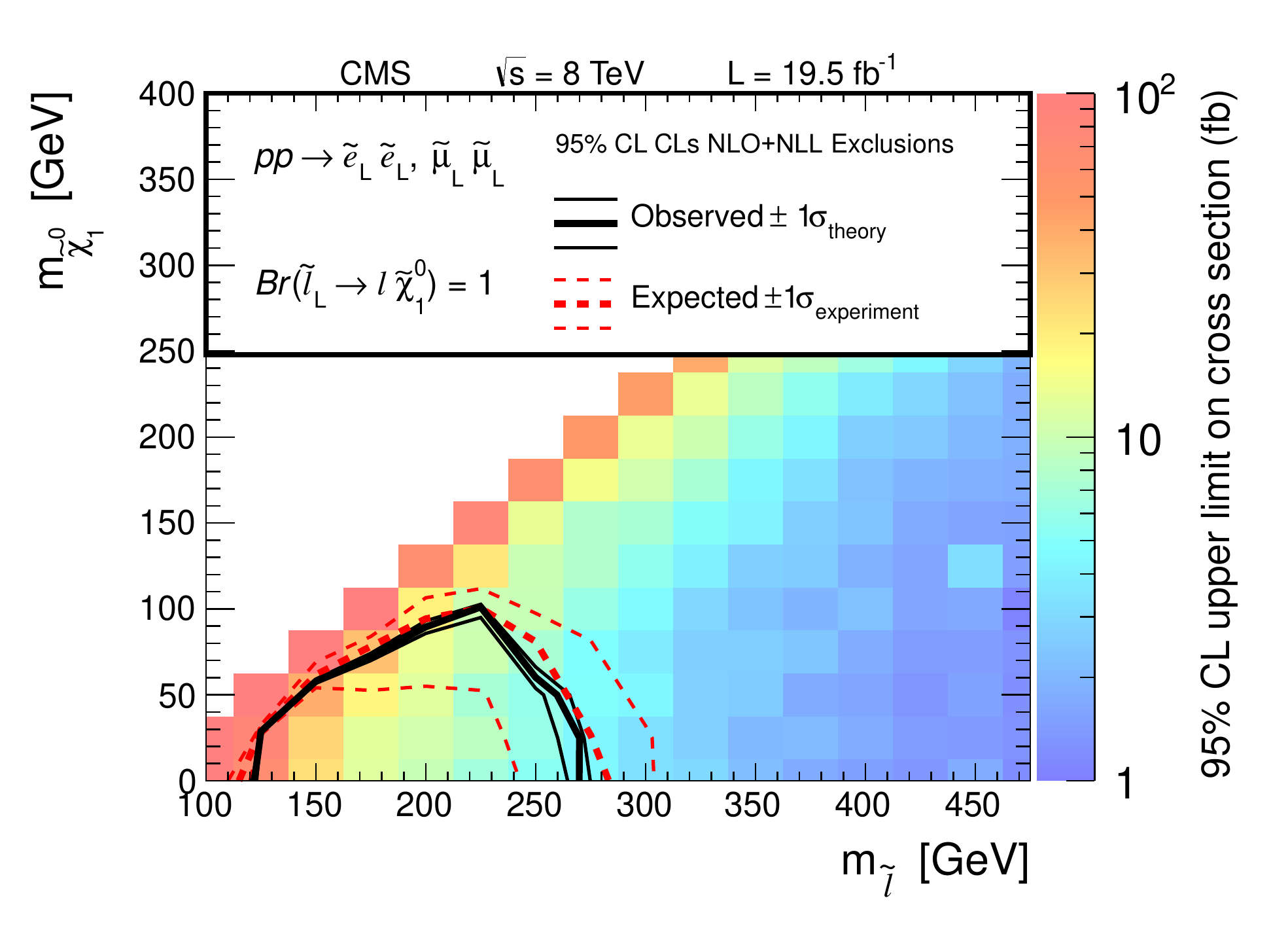}
\includegraphics[width=0.49\textwidth]{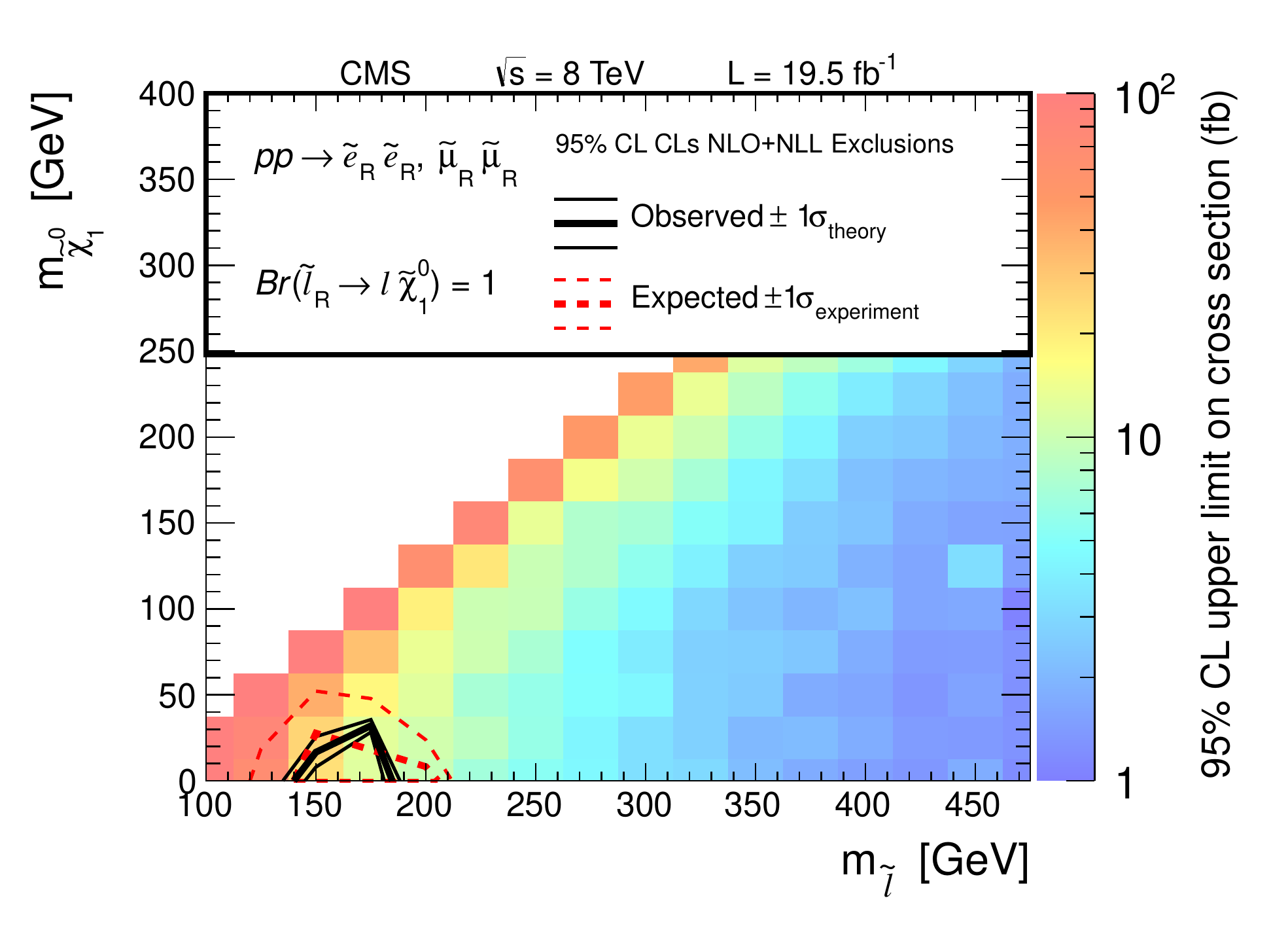}
\caption{
\label{fig:tchipmslepslep}
Interpretation of the results of the opposite-sign non-resonant dilepton search, in the models with
(\cmsTLeft) chargino pair production ($\chipm_1\PSGc^\mp_1$), (\cmsTCenter) left-handed slepton pair production ($\slep_L \slep_L$), and (\cmsTRight) right-handed slepton pair production ($\slep_R \slep_R$).
The shading indicates the 95\% CL upper limits on the cross section times branching fraction,
and the contours the excluded regions assuming the NLO+NLL signal cross sections.
}
\end{figure}

Figure~\ref{fig:tchipmslepslep} shows limits on  the chargino and
slepton pair-production cross section times branching fraction for the processes of
Fig.~\ref{fig:charginos-ll}. The limits for chargino pair production are
determined using both the opposite- and same-flavor dilepton search regions discussed in Section~\ref{osdilepton},
while the limits for slepton pair production are set using only the same-flavor dilepton search region.
The production cross sections for left-handed sleptons are larger than those for right-handed sleptons,
enhancing the sensitivity.

\section{Summary}
\label{sec:summary}

This paper presents searches for the direct electroweak pair production of supersymmetric charginos, neutralinos, and sleptons in a wide variety of signatures with
leptons, and \PW, \Z, and Higgs bosons. Results are based on a sample of proton-proton collision data collected at center-of-mass energy $\sqrt{s} = 8$ \TeV
with the CMS detector in 2012, corresponding to an integrated luminosity of 19.5\fbinv.

The direct electroweak production of SUSY particles may result in several different signal topologies with one or more leptons
and missing transverse energy (\MET).
The relative sensitivity of each signature depends on unknown parameters, including the SUSY particle masses. This situation, along with the
relatively small cross sections typical of electroweak SUSY production, motivates a strategy based on multiple dedicated search regions that target
each possible signal topology. In each of these search regions, the data are found to be in agreement with the standard model background expectations.
No significant evidence for a signal-like excess is observed.

\begin{figure}[htbp]
\centering
\includegraphics[width=0.49\textwidth]{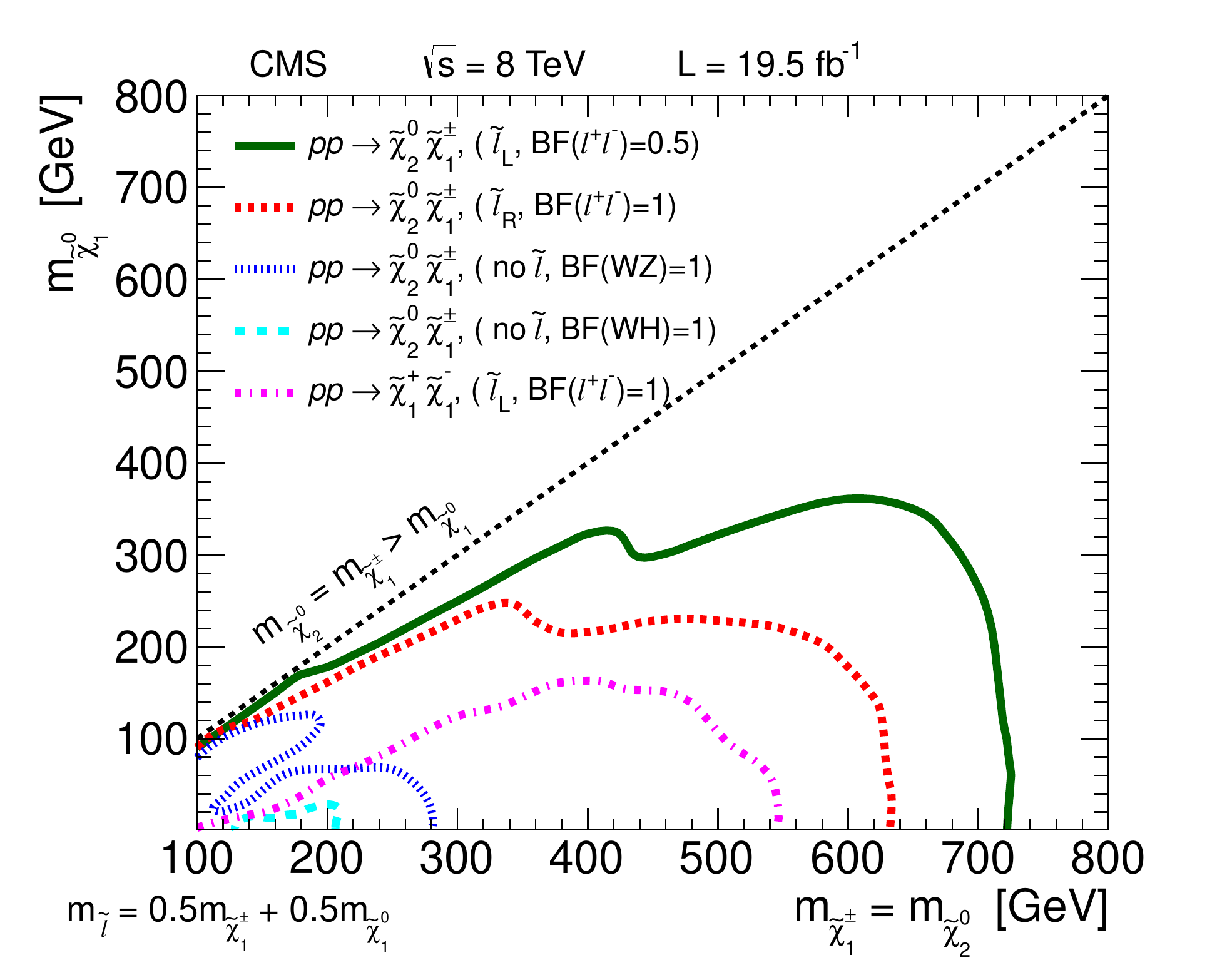}
\includegraphics[width=0.49\textwidth]{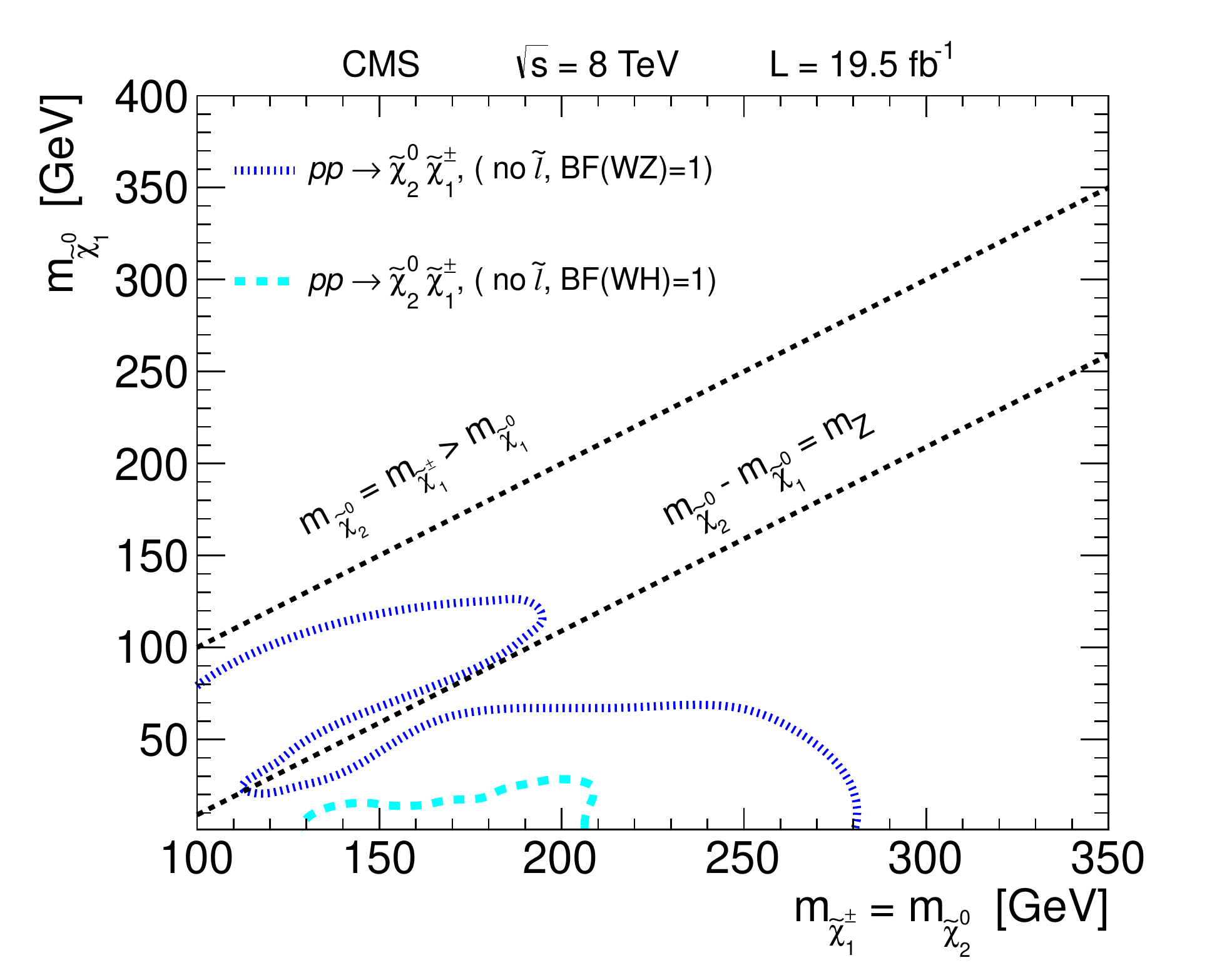}
\caption{
\label{fig:SummaryResult}
(\cmsLeft) Contours bounding the mass regions excluded at 95\% CL for
chargino-neutralino production with decays to left-handed sleptons, right-handed sleptons,
or direct decays to Higgs and vector bosons, and for chargino-pair production, based on NLO+NLL signal cross sections.
Where applicable, the $\xslep$ value used to calculate the slepton mass is 0.5.
(\cmsRight) Expanded view for chargino-neutralino production with decays to Higgs and vector bosons.
}
\end{figure}

The results are interpreted in the context of models dominated by direct electroweak SUSY production.
Several of the interpretation results are summarized in Fig.~\ref{fig:SummaryResult}.
We consider models with a wino-like chargino and neutralino pair
with degenerate mass $m_{\PSGc}$, and a bino-like lightest SUSY particle with mass $m_{\mathrm{LSP}}$. We also consider the presence of light sleptons,
either produced in the decays of charginos or neutralinos, or produced directly in pairs. The limits on the signal production cross sections are most stringent in the region of
parameter space with large $\Delta M \equiv m_{\PSGc}-m_{\mathrm{LSP}}$ (or, for direct slepton production, $\Delta M \equiv m_{\widetilde{\ell}}-m_{\mathrm{LSP}}$), and less stringent in the region of small $\Delta M$, where the final-state objects are less energetic.

The electroweak SUSY process with the largest cross section is chargino-neutralino pair production.
The resulting signal topologies depend on
the properties of the sleptons. Models with light sleptons enhance the branching fraction to final states
with three leptons.
Depending on the left/right mixing and flavor of these sleptons, our results probe charginos and neutralinos with masses up
to 320\GeV, 620\GeV, and 720\GeV, for the $\tau$-dominated, $\tau$-enriched, and flavor-democratic scenarios, respectively.
In such models, searches in the same-sign dilepton final state enhance the
sensitivity in the experimentally challenging region with small $\Delta M$.

Models without light sleptons lead to \wzmet\ or \whmet signatures, with model-dependent branching fractions. To probe the \wzmet\ signature,
searches in the three-lepton and \Z boson plus jets (with leptonic \Z\ decay) final states are performed. To probe the \whmet\ signature, searches
are performed in the single-lepton final state with $\PH\to\bbbar$, in the same-sign dilepton final state with $\PH \to \PW(\ell\nu)\PW(\text{jj})$, where j denotes a jet,
and in final states with three or more leptons with $\PH\to\PWp\PWm$, \Z\Z, or \TT. If the \wzmet\ (\whmet) branching fraction is assumed to be 100\%,
our results probe charginos and neutralinos with masses up to 270\GeV (200\GeV).
The \wzmet\ search is particularly important in the region with small $\Delta M$,
where we probe charginos and neutralinos with masses up to 200\GeV. We also consider
a specific model based on gauge-mediated SUSY breaking that predicts an enhancement in the
\zzmet\ production rate.
Our results probe higgsinos with masses
up to 330\GeV in this scenario.

Following chargino-neutralino pair production, the electroweak SUSY process with the largest cross section is chargino pair production, which leads to a final state consisting of an opposite-sign lepton pair and \MET.
Our results probe chargino masses up to 540\GeV in a scenario with light sleptons. The direct pair production of sleptons leads to a similar
signature, with a lower cross section. For left-handed (right-handed) sleptons, our results probe sleptons with masses up to 260 (180)\GeV.

\section*{Acknowledgements}
{\tolerance=1200
\hyphenation{Bundes-ministerium Forschungs-gemeinschaft Forschungs-zentren} We congratulate our colleagues in the CERN accelerator departments for the excellent performance of the LHC and thank the technical and administrative staffs at CERN and at other CMS institutes for their contributions to the success of the CMS effort. In addition, we gratefully acknowledge the computing centres and personnel of the Worldwide LHC Computing Grid for delivering so effectively the computing infrastructure essential to our analyses. Finally, we acknowledge the enduring support for the construction and operation of the LHC and the CMS detector provided by the following funding agencies: the Austrian Federal Ministry of Science, Research and Economy and the Austrian Science Fund; the Belgian Fonds de la Recherche Scientifique, and Fonds voor Wetenschappelijk Onderzoek; the Brazilian Funding Agencies (CNPq, CAPES, FAPERJ, and FAPESP); the Bulgarian Ministry of Education and Science; CERN; the Chinese Academy of Sciences, Ministry of Science and Technology, and National Natural Science Foundation of China; the Colombian Funding Agency (COLCIENCIAS); the Croatian Ministry of Science, Education and Sport, and the Croatian Science Foundation; the Research Promotion Foundation, Cyprus; the Ministry of Education and Research, Estonian Research Council via IUT23-4 and IUT23-6 and European Regional Development Fund, Estonia; the Academy of Finland, Finnish Ministry of Education and Culture, and Helsinki Institute of Physics; the Institut National de Physique Nucl\'eaire et de Physique des Particules~/~CNRS, and Commissariat \`a l'\'Energie Atomique et aux \'Energies Alternatives~/~CEA, France; the Bundesministerium f\"ur Bildung und Forschung, Deutsche Forschungsgemeinschaft, and Helmholtz-Gemeinschaft Deutscher Forschungszentren, Germany; the General Secretariat for Research and Technology, Greece; the National Scientific Research Foundation, and National Innovation Office, Hungary; the Department of Atomic Energy and the Department of Science and Technology, India; the Institute for Studies in Theoretical Physics and Mathematics, Iran; the Science Foundation, Ireland; the Istituto Nazionale di Fisica Nucleare, Italy; the Korean Ministry of Education, Science and Technology and the World Class University program of NRF, Republic of Korea; the Lithuanian Academy of Sciences; the Ministry of Education, and University of Malaya (Malaysia); the Mexican Funding Agencies (CINVESTAV, CONACYT, SEP, and UASLP-FAI); the Ministry of Business, Innovation and Employment, New Zealand; the Pakistan Atomic Energy Commission; the Ministry of Science and Higher Education and the National Science Centre, Poland; the Funda\c{c}\~ao para a Ci\^encia e a Tecnologia, Portugal; JINR, Dubna; the Ministry of Education and Science of the Russian Federation, the Federal Agency of Atomic Energy of the Russian Federation, Russian Academy of Sciences, and the Russian Foundation for Basic Research; the Ministry of Education, Science and Technological Development of Serbia; the Secretar\'{\i}a de Estado de Investigaci\'on, Desarrollo e Innovaci\'on and Programa Consolider-Ingenio 2010, Spain; the Swiss Funding Agencies (ETH Board, ETH Zurich, PSI, SNF, UniZH, Canton Zurich, and SER); the Ministry of Science and Technology, Taipei; the Thailand Center of Excellence in Physics, the Institute for the Promotion of Teaching Science and Technology of Thailand, Special Task Force for Activating Research and the National Science and Technology Development Agency of Thailand; the Scientific and Technical Research Council of Turkey, and Turkish Atomic Energy Authority; the National Academy of Sciences of Ukraine, and State Fund for Fundamental Researches, Ukraine; the Science and Technology Facilities Council, UK; the US Department of Energy, and the US National Science Foundation.

Individuals have received support from the Marie-Curie programme and the European Research Council and EPLANET (European Union); the Leventis Foundation; the A. P. Sloan Foundation; the Alexander von Humboldt Foundation; the Belgian Federal Science Policy Office; the Fonds pour la Formation \`a la Recherche dans l'Industrie et dans l'Agriculture (FRIA-Belgium); the Agentschap voor Innovatie door Wetenschap en Technologie (IWT-Belgium); the Ministry of Education, Youth and Sports (MEYS) of the Czech Republic; the Council of Science and Industrial Research, India; the HOMING PLUS programme of Foundation for Polish Science, cofinanced from European Union, Regional Development Fund; the Compagnia di San Paolo (Torino); and the Thalis and Aristeia programmes cofinanced by EU-ESF and the Greek NSRF.
\par}

\bibliography{auto_generated}   % will be created by the tdr script.

\appendix
\clearpage
\section{Additional plots for the three-lepton and four-lepton searches}
\label{app:3lplots}
\begin{figure}[htb]
\centering
\includegraphics[width=\cmsFigWidth]{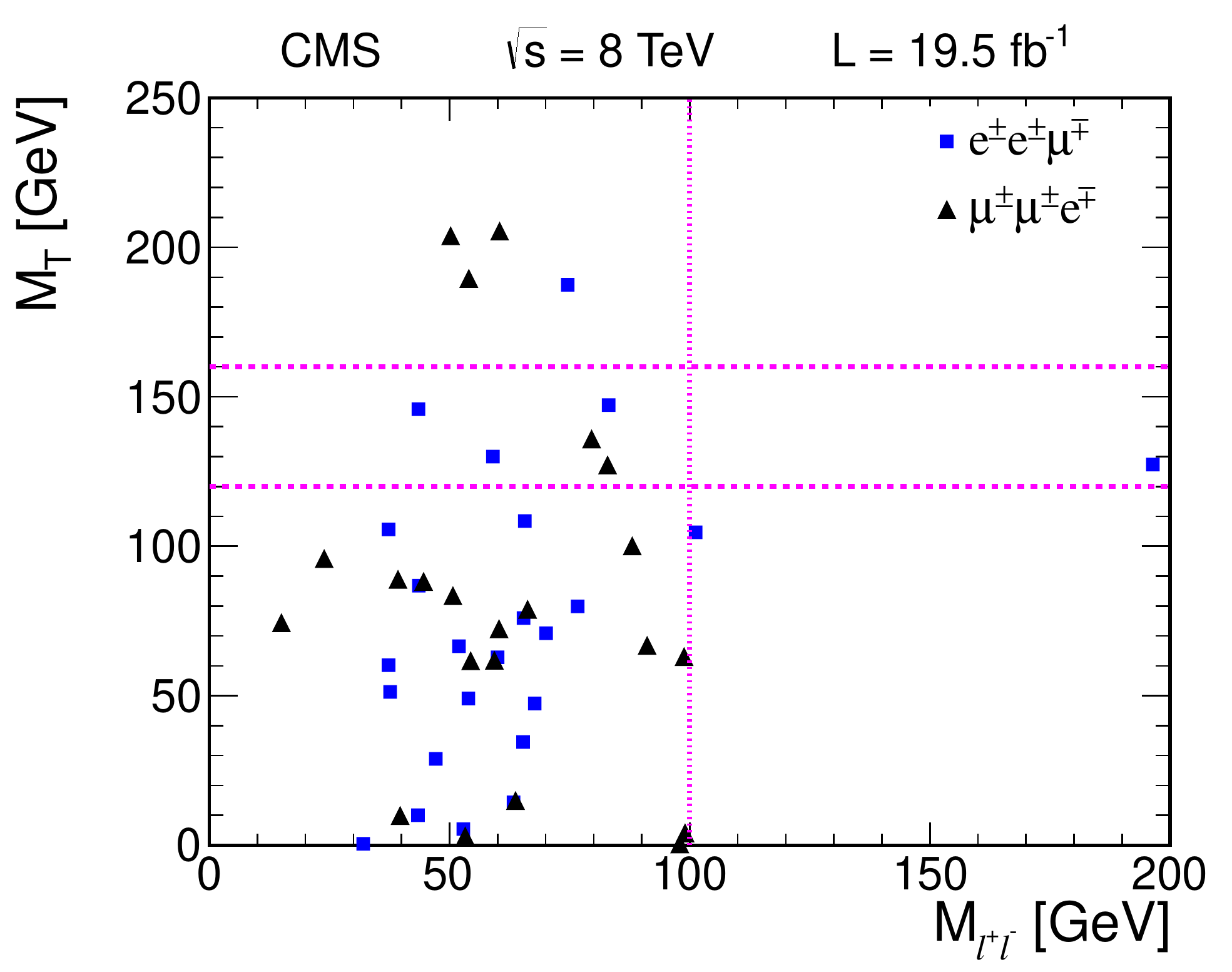}
\caption{Distribution of $\MT$ versus $\mdil$ for three-lepton $\Pe\Pe\mu$ and $\Pe\mu\mu$ events without an OSSF pair. $\mdil$ is calculated by combining opposite-sign leptons and choosing
the pair closest to the corresponding dilepton mass determined from $\Z \to \tau \tau$ simulation.   $\MT$ is calculated using the remaining lepton.
\label{fig:noOSSFscatter}
}
\end{figure}

\begin{figure*}[htb]
\centering
\includegraphics[width=0.7\textwidth]{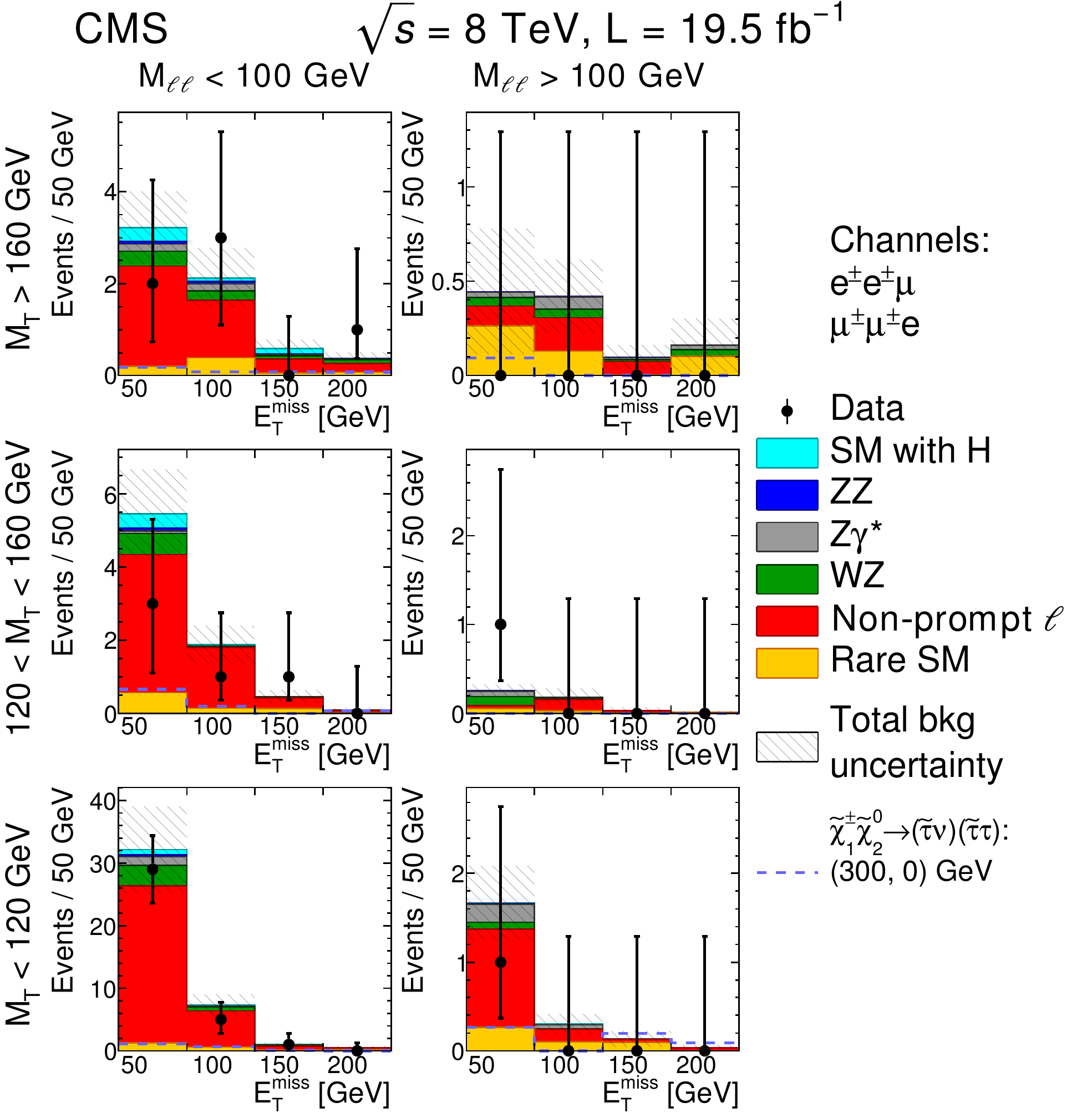} \\
\caption{
The $\MET$ distributions for three-lepton $\Pe\Pe\mu$ and $\Pe\mu\mu$ events without an OSSF pair.
The SM expectations are also shown. The $\ETmiss$ distributions for an example signal scenario is overlaid.
The first (second) number in parentheses indicates the value of \mchi\ (\mlsp).
\label{fig:noOSSFMET}
}
\end{figure*}

\begin{figure}[htbp]
\centering
\includegraphics[width=\cmsFigWidth]{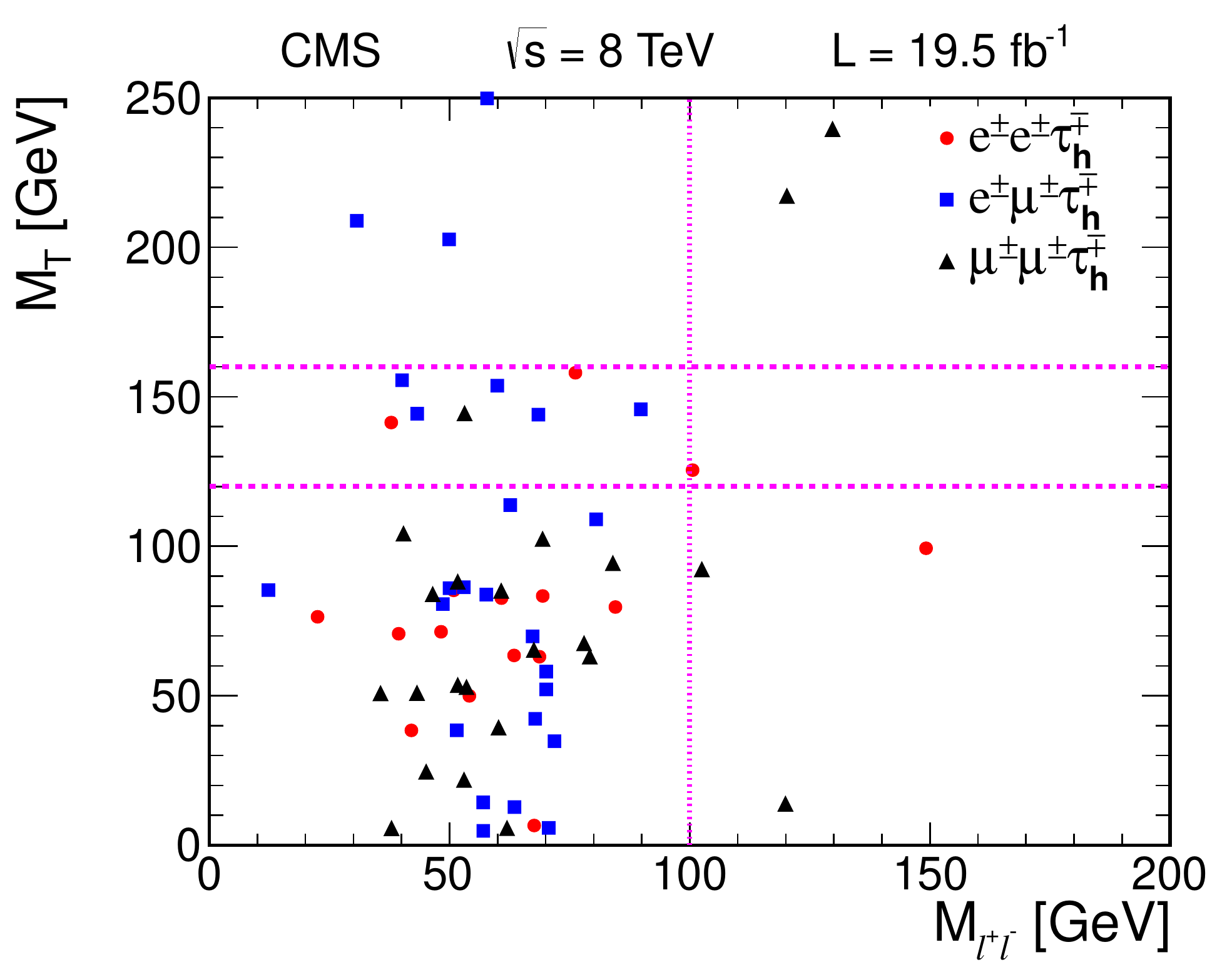}
\caption{Distribution of $\MT$ versus $\mdil$
for three-lepton events with
a same-sign $\Pe\Pe$, $\Pe\mu$, or $\mu\mu$ pair and one \tauh. $\mdil$ is calculated by combining opposite-sign leptons and choosing
the pair closest to the corresponding dilepton mass determined from $\Z \to \tau \tau$ simulation.   $\MT$ is calculated using the remaining lepton.
\label{fig:SStauscatter}
}
\end{figure}

\begin{figure*}[htb]
\centering
\includegraphics[width=0.7\textwidth]{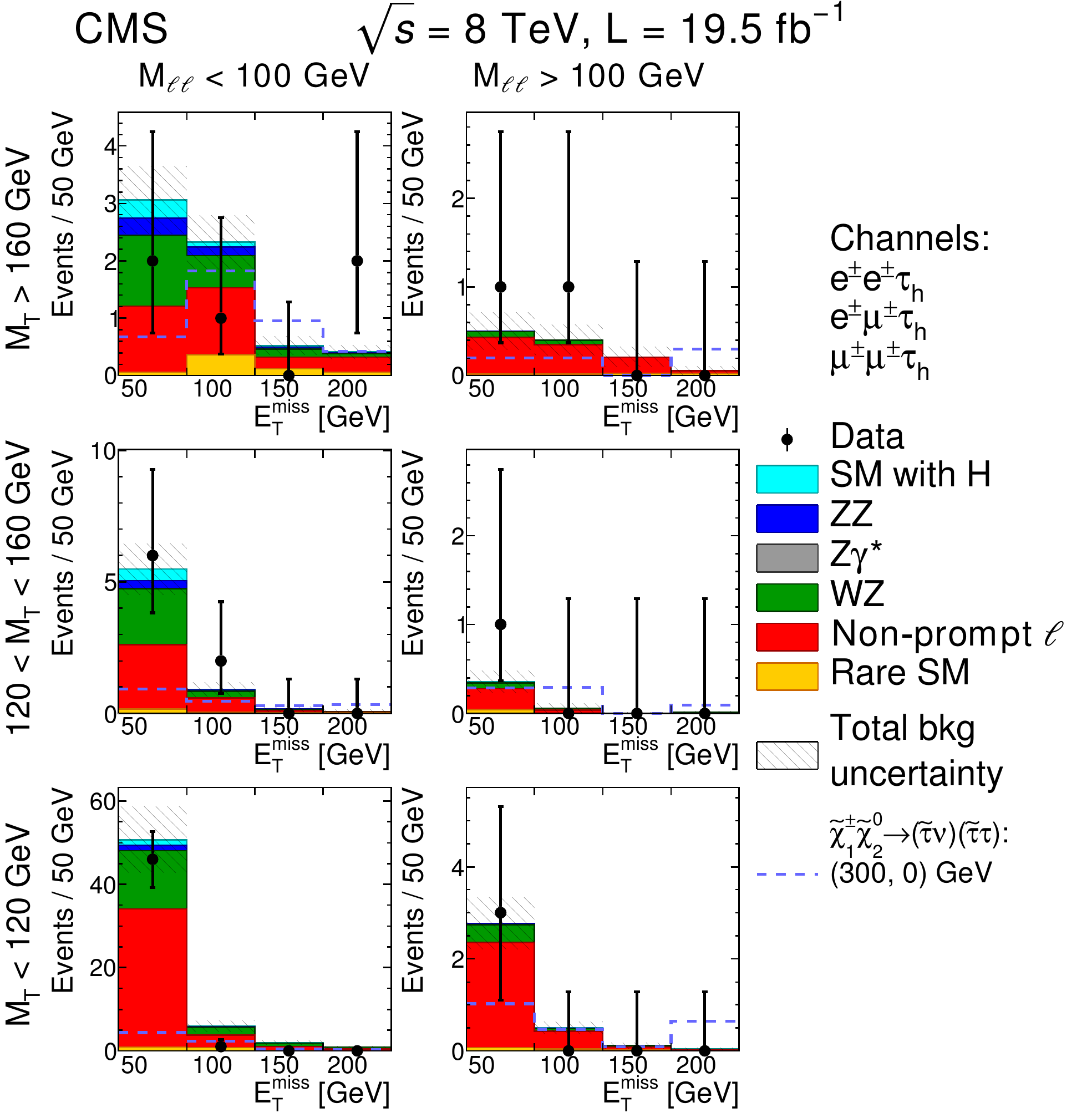} \\
\caption{The $\MET$ distributions for three-lepton events with
a same-sign $\Pe\Pe$, $\Pe\mu$, or $\mu\mu$ pair and one \tauh.
The SM expectations are also shown.  The $\ETmiss$ distributions for an example signal scenario is overlaid.
The first (second) number in parentheses indicates the value of \mchi\ (\mlsp).
\label{fig:noOSSFtau1MET}
}
\end{figure*}

\begin{figure}[htb]
\centering
\includegraphics[width=\cmsFigWidth]{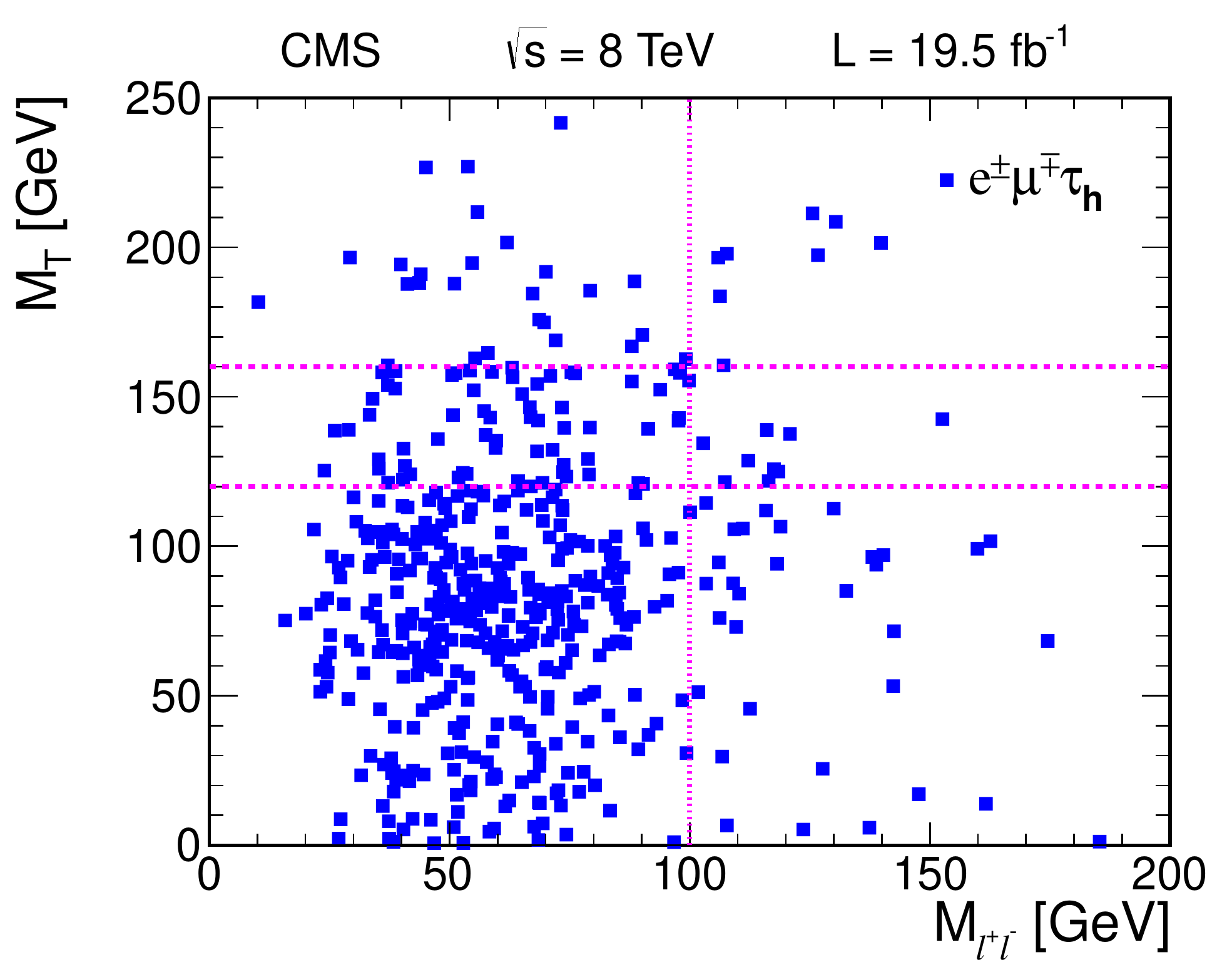}
\caption{Distribution of $\MT$ versus $\mdil$
for three-lepton events with
an opposite-sign $\Pe\mu$ pair and one \tauh. $\mdil$ is calculated by combining opposite-sign leptons and choosing
the pair closest to the corresponding dilepton mass determined from $\Z \to \tau \tau$ simulation.   $\MT$ is calculated using the remaining lepton.
\label{fig:OSOFtauscatter}
}
\end{figure}

\begin{figure*}[htb]
\centering
\includegraphics[width=0.7\textwidth]{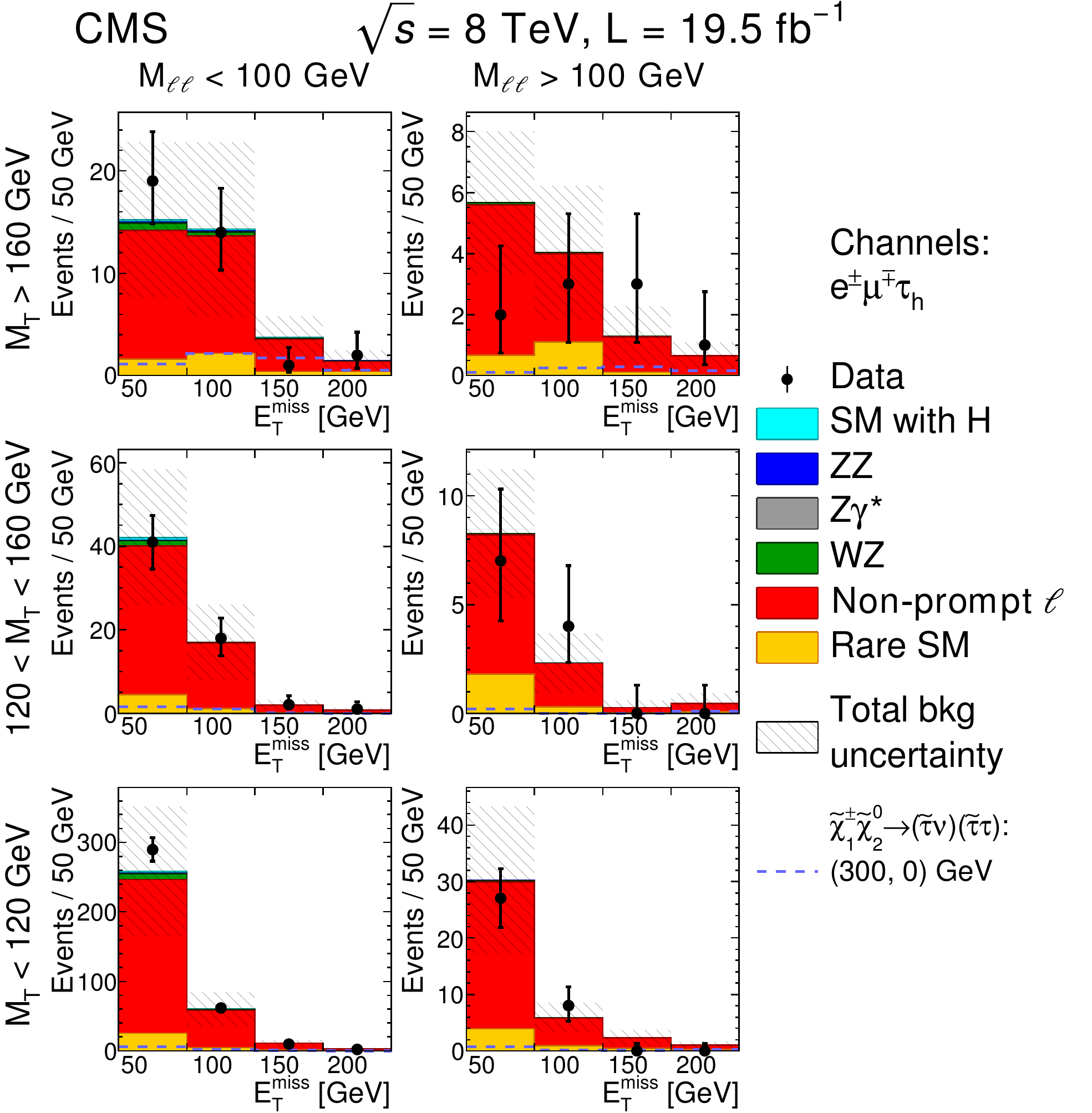}
\caption{The $\MET$ distributions for three-lepton events with
an opposite-sign $\Pe\mu$ pair and one \tauh.
The SM expectations are also shown. The $\ETmiss$ distributions for an example signal scenario is overlaid.
The first (second) number in parentheses indicates the value of \mchi\ (\mlsp).
\label{fig:OSOFtau1MET}
}
\end{figure*}

\begin{figure}[htb]
\begin{center}
\includegraphics[width=\cmsFigWidth]{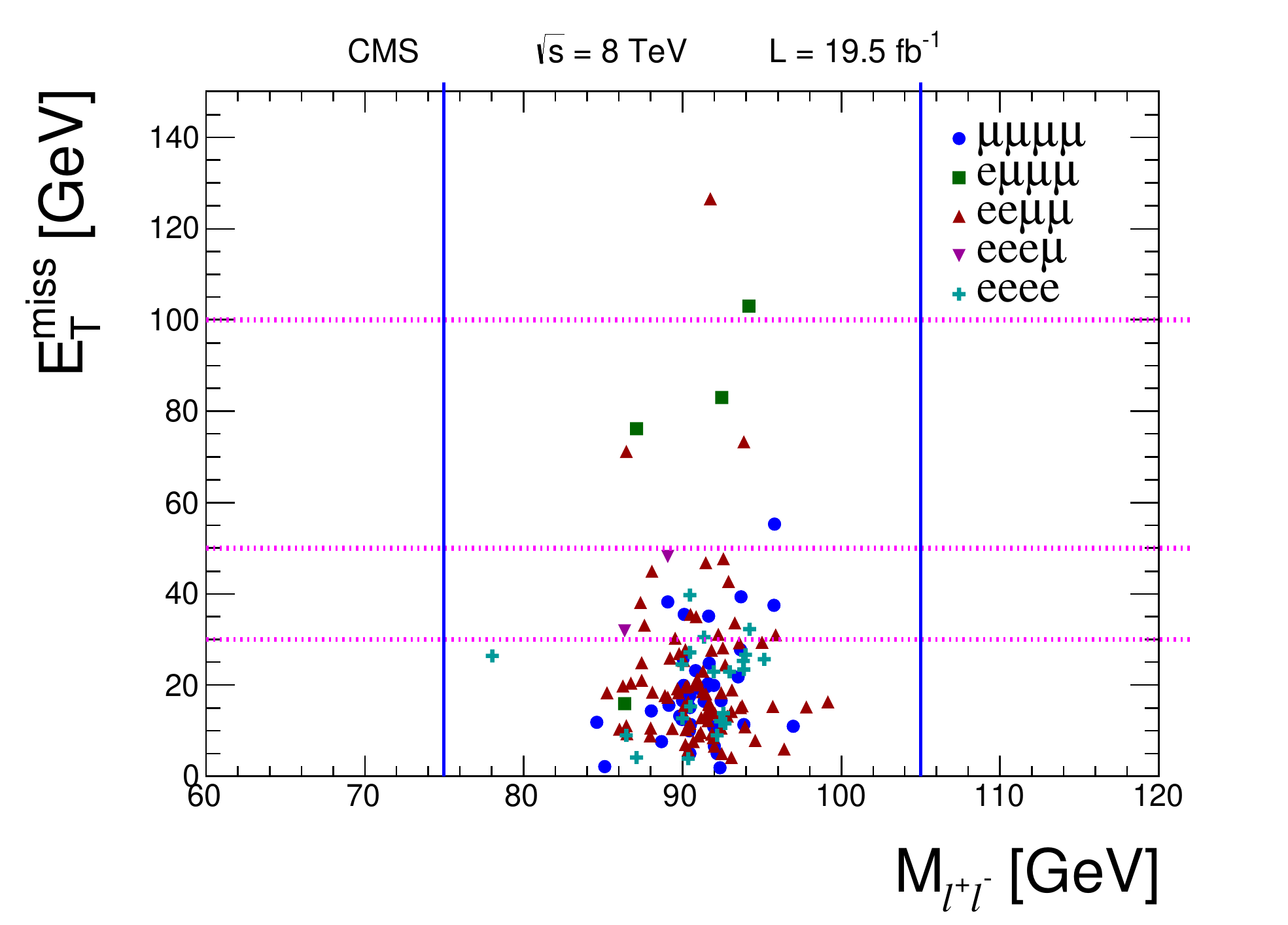} \\
\caption{$\MET$ versus $\mdil$
for four-lepton events with an on-Z OSSF pair and no \tauh.
The legend indicates the flavor breakdown of events. For events with two OSSF pairs,
we choose the pair with mass closest to the $\Z$ boson mass.
\label{fig:quadscatter}
}
\end{center}
\end{figure}

This appendix presents additional results from the three-lepton and four-lepton
searches. The distributions of $\MT$ versus $\mdil$ for three-lepton
events are presented in Figs.~\ref{fig:noOSSFscatter}-\ref{fig:OSOFtau1MET}.
The corresponding numerical results are presented in Tables ~\ref{tab:L3noOSSF}-\ref{tab:L3OSOFtau}.
The distribution of \MET versus $\mdil$ for four-lepton events is presented in Fig.~\ref{fig:quadscatter}.

\begin{table*}[htbp]
\centering
\topcaption{
\label{tab:L3noOSSF}
Observed yields and SM expectations for three-lepton $\Pe\Pe\mu$ and $\Pe\mu\mu$ events without an OSSF pair.
The uncertainties include both the statistical and systematic components.
}
\begin{tabular}{ l | l | c c |  c  c  }\hline\hline
\multirow{2}{*}{$\MT$ (\GeVns)} &\multirow{2}{*}{$\ETm$ (\GeVns)}  & \multicolumn{2}{c|}{$M_{\ell\ell} < 100$\GeV} & \multicolumn{2}{c}{$M_{\ell\ell} > 100$\GeV}\\\cline{3-6}
&& Total bkg & Observed & Total bkg & Observed \\
\hline\hline
\multirow{4}{*}{$>$160}& 50--100&3.2 $\pm$ 0.8&2&0.44 $\pm$ 0.33&0\\
& 100--150&2.1 $\pm$ 0.7&3&0.42 $\pm$ 0.19&0\\
& 150--200&0.59 $\pm$ 0.18&0&0.10 $\pm$ 0.06&0\\
& $>$200&0.37 $\pm$ 0.13&1&0.16 $\pm$ 0.14&0\\
\hline\hline
\multirow{4}{*}{120--160}& 50--100&5.5 $\pm$ 1.2&3&0.25 $\pm$ 0.07&1\\
& 100--150&1.9 $\pm$ 0.5&1&0.19 $\pm$ 0.10&0\\
& 150--200&0.46 $\pm$ 0.18&1&0.03 $\pm$ 0.03&0\\
& $>$200&0.10 $\pm$ 0.05&0&0.008 $\pm$ 0.010&0\\
\hline\hline
\multirow{4}{*}{0--120}& 50--100&32 $\pm$ 7&29&1.7 $\pm$ 0.4&1\\
& 100--150&7.3 $\pm$ 1.7&5&0.30 $\pm$ 0.11&0\\
& 150--200&1.0 $\pm$ 0.3&1&0.14 $\pm$ 0.09&0\\
& $>$200&0.53 $\pm$ 0.24&0&0.03 $\pm$ 0.03&0\\
\hline\hline
\end{tabular}
\end{table*}

\begin{table*}[htb]
\centering
\topcaption{
\label{tab:L3SStau}
Observed yields and SM expectations for events with a same-sign $\Pe\Pe$ , $\Pe\mu$, or $\mu\mu$ pair and one \tauh.
The uncertainties include both the statistical and systematic components.
}
\begin{tabular}{ l | l | c c | c  c  }\hline\hline
\multirow{2}{*}{$\MT$ (\GeVns{})} &\multirow{2}{*}{$\ETm$ (\GeVns{})}  & \multicolumn{2}{c|}{$M_{\ell\ell} < 100$\GeV} & \multicolumn{2}{c}{$M_{\ell\ell} > 100$\GeV}\\\cline{3-6} && Total bkg & Observed & Total bkg & Observed \\
\hline\hline
\multirow{4}{*}{$>$160}& 50--100&3.1 $\pm$ 0.6&2&0.5 $\pm$ 0.2&1\\
& 100--150&2.3 $\pm$ 0.5&1&0.4 $\pm$ 0.2&1\\
& 150--200&0.5 $\pm$ 0.2&0&0.2 $\pm$ 0.1&0\\
& $>$200&0.4 $\pm$ 0.1&2&0.06 $\pm$ 0.05&0\\
\hline\hline
\multirow{4}{*}{120--160}& 50--100&6 $\pm$ 1&6&0.4 $\pm$ 0.1&1\\
& 100--150&0.9 $\pm$ 0.3&2&0.06 $\pm$ 0.05&0\\
& 150--200&0.3 $\pm$ 0.1&0&0.00 $\pm$ 0.01&0\\
& $>$200&0.06 $\pm$ 0.08&0&0.01 $\pm$ 0.01&0\\
\hline\hline
\multirow{4}{*}{0--120}& 50--100&51 $\pm$ 8&46&2.8 $\pm$ 0.6&3\\ & 100--150&6 $\pm$ 1&1&0.5 $\pm$ 0.1&0\\
& 150--200&2.0 $\pm$ 0.4&0&0.11 $\pm$ 0.07&0\\
& $>$200&0.9 $\pm$ 0.2&0&0.04 $\pm$ 0.02&0\\
\hline\hline
\end{tabular}
\end{table*}

\begin{table*}[htb]
\centering
\topcaption{
\label{tab:L3OSOFtau}
Observed yields and SM expectations for events with an opposite-sign $\Pe\mu$ pair and \tauh.
The uncertainties include both the statistical and systematic components.
}
\begin{tabular}{ l | l | c c | c  c  }\hline\hline
\multirow{2}{*}{$\MT$ (\GeVns)} &\multirow{2}{*}{$\ETm$ (\GeVns)}  & \multicolumn{2}{c|}{$M_{\ell\ell} < 100$\GeV} & \multicolumn{2}{c}{$M_{\ell\ell} > 100$\GeV}\\\cline{3-6}
&& Total bkg & Observed & Total bkg & Observed \\
\hline\hline
\multirow{4}{*}{$>$160}& 50--100&15 $\pm$ 8&19&5.7 $\pm$ 2.3&2\\
& 100--150&14 $\pm$ 9&14&4.0 $\pm$ 2.2&3\\
& 150--200&3.7 $\pm$ 2.1&1&1.3 $\pm$ 1.0&3\\
& $>$200&1.5 $\pm$ 1.0&2&0.7 $\pm$ 0.4&1\\
\hline\hline
\multirow{4}{*}{120--160}& 50--100&42 $\pm$ 16&41&8.3 $\pm$ 2.9&7\\
& 100--150&17 $\pm$ 9&18&2.3 $\pm$ 1.3&4\\
& 150--200&2.0 $\pm$ 1.2&2&0.27 $\pm$ 0.32&0\\
& $>$200&0.8 $\pm$ 0.5&1&0.5 $\pm$ 0.4&0\\
\hline\hline
\multirow{4}{*}{0--120}& 50--100&259 $\pm$ 93&290&30 $\pm$ 13&27\\
& 100--150&60 $\pm$ 25&62&5.9 $\pm$ 2.6&8\\
& 150--200&11 $\pm$ 5&10&2.3 $\pm$ 1.4&0\\
& $>$200&2.9 $\pm$ 1.4&2&1.1 $\pm$ 0.6&0\\
\hline\hline
\end{tabular}
\end{table*}

\clearpage
\section{Additional results for the multilepton analysis}
\label{app:multilepton}

In this appendix, we present similar results as those presented in Table~\ref{tab:multileptonResults130}
for the multilepton analysis of Section~\ref{sec:multilepton} but for different values of $\mchi$.

\begin{table*}[htb]
\topcaption{
Multilepton results for the $\mchi=150$\GeV,$\mlsp=1$\GeV scenario. See Table~\ref{tab:multileptonResults130} for details.
\label{tab:multileptonResults150}}
\begin{center}
\begin{tabular}{c|c|c|c|c|c}
\hline
\hline
$N_{\tauh}$ & OSSF pair & $\MET$ (\GeVns) & Data & Total SM & Signal\\
\hline
0 & Below \Z & 50--100 & 142 & 125 $\pm$ 28 & 14.9 $\pm$ 2.8\\
0 & Below \Z & 100--150 & 16 & 21.3 $\pm$ 8.0 & 5.06 $\pm$ 0.86\\
0 & None & 0--50	 & 53 & 52 $\pm$ 12 & 4.61 $\pm$ 0.99\\
0 & None & 50--100 & 35 & 38 $\pm$ 15 & 6.5 $\pm$ 1.1\\
0 & None & 100--150 & 7 & 9.3 $\pm$ 4.3 & 2.32 $\pm$ 0.43\\

\hline
\hline
\end{tabular}
\end{center}
\end{table*}

\begin{table*}[htb]
\topcaption{
Multilepton results for the $\mchi=200$\GeV,$\mlsp=1$\GeV scenario. See Table~\ref{tab:multileptonResults130} for details.
\label{tab:multileptonResults200}}
\begin{center}
\begin{tabular}{c|c|c|c|c|c}
\hline
\hline
$N_{\tauh}$ & OSSF pair & $\MET$ (\GeVns) & Data & Total SM & Signal\\
\hline
0 & Below \Z & 50--100  & 142 & 125 $\pm$ 28 & 4.90 $\pm$ 0.91\\
0 & Below \Z & 100--150 &  16 & 21.3 $\pm$ 8.0 & 2.63 $\pm$ 0.43\\
0 & Below \Z & 150--200 &   5 & 2.9 $\pm$ 1.0 & 0.61 $\pm$ 0.16\\
0 & None     & 50--100  &  35 & 38 $\pm$ 15 & 2.31 $\pm$ 0.43\\
0 & None     & 100--150 &   7 & 9.3 $\pm$ 4.3 & 1.31 $\pm$ 0.26\\

\hline
\hline
\end{tabular}
\end{center}
\end{table*}

\begin{table*}[htb]
\topcaption{
Multilepton results for the $\mchi=300$\GeV,$\mlsp=1$\GeV scenario. See Table~\ref{tab:multileptonResults130} for details.
\label{tab:multileptonResults300}}
\begin{center}
\begin{tabular}{c|c|c|c|c|c}
\hline
\hline
$N_{\tauh}$ & OSSF pair & $\MET$ (\GeVns) & Data & Total SM & Signal\\
\hline
0 & Below \Z & 100--150 & 16 & 21.3 $\pm$ 8.0 & 0.70 $\pm$ 0.13\\
0 & Below \Z & 150--200 & 5 & 2.9 $\pm$ 1.0 & 0.348 $\pm$ 0.067\\
0 & Below \Z & $>$200  & 0 & 0.88 $\pm$ 0.31 & 0.218 $\pm$ 0.041\\
0 & Above \Z & 150--200 & 1 & 2.48 $\pm$ 0.68 & 0.180 $\pm$ 0.045\\
1 & None     & 150--200 & 8 & 15.1 $\pm$ 7.4 & 0.44 $\pm$ 0.12\\

\hline
\hline
\end{tabular}
\end{center}
\end{table*}

\begin{table*}[htb]
\topcaption{
Multilepton results for the $\mchi=400$\GeV,$\mlsp=1$\GeV scenario. See Table~\ref{tab:multileptonResults130} for details.
\label{tab:multileptonResults400}}
\begin{center}
\begin{tabular}{c|c|c|c|c|c}
\hline
\hline
$N_{\tauh}$ & OSSF pair & $\MET$ (\GeVns) & Data & Total SM & Signal\\
\hline
0 & Below \Z & 100--150 & 16 & 21.3 $\pm$ 8.0 & 0.167 $\pm$ 0.028\\
0 & Below \Z & 150--200 & 5 & 2.9 $\pm$ 1.0 & 0.138 $\pm$ 0.025\\
0 & Below \Z & $>$200 & 0 & 0.88 $\pm$ 0.31 & 0.137 $\pm$ 0.025\\
0 & None     & $>$200 & 0 & 0.42 $\pm$ 0.22 & 0.057 $\pm$ 0.011\\
1 & None     & $>$200 & 3 & 2.4 $\pm$ 1.1 & 0.152 $\pm$ 0.038\\

\hline
\hline
\end{tabular}
\end{center}
\end{table*}

\clearpage
\section{One-dimensional exclusion plots in the \texorpdfstring{$\W\PH + \MET$}{WH + MET} final state}
\label{app:1Dplots}

\begin{figure*}[htbp]
\centering
\includegraphics[width=0.49\textwidth]{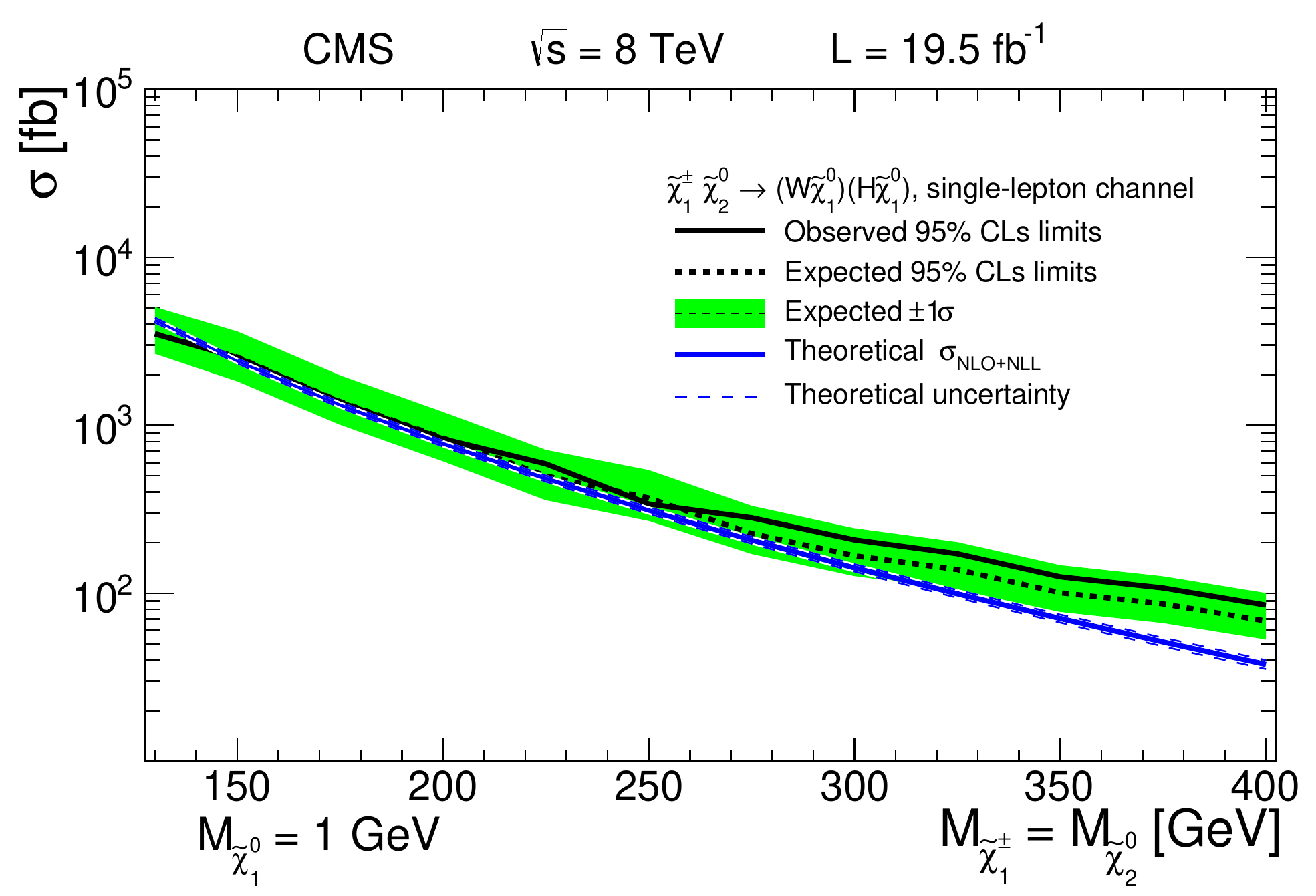}
\includegraphics[width=0.49\textwidth]{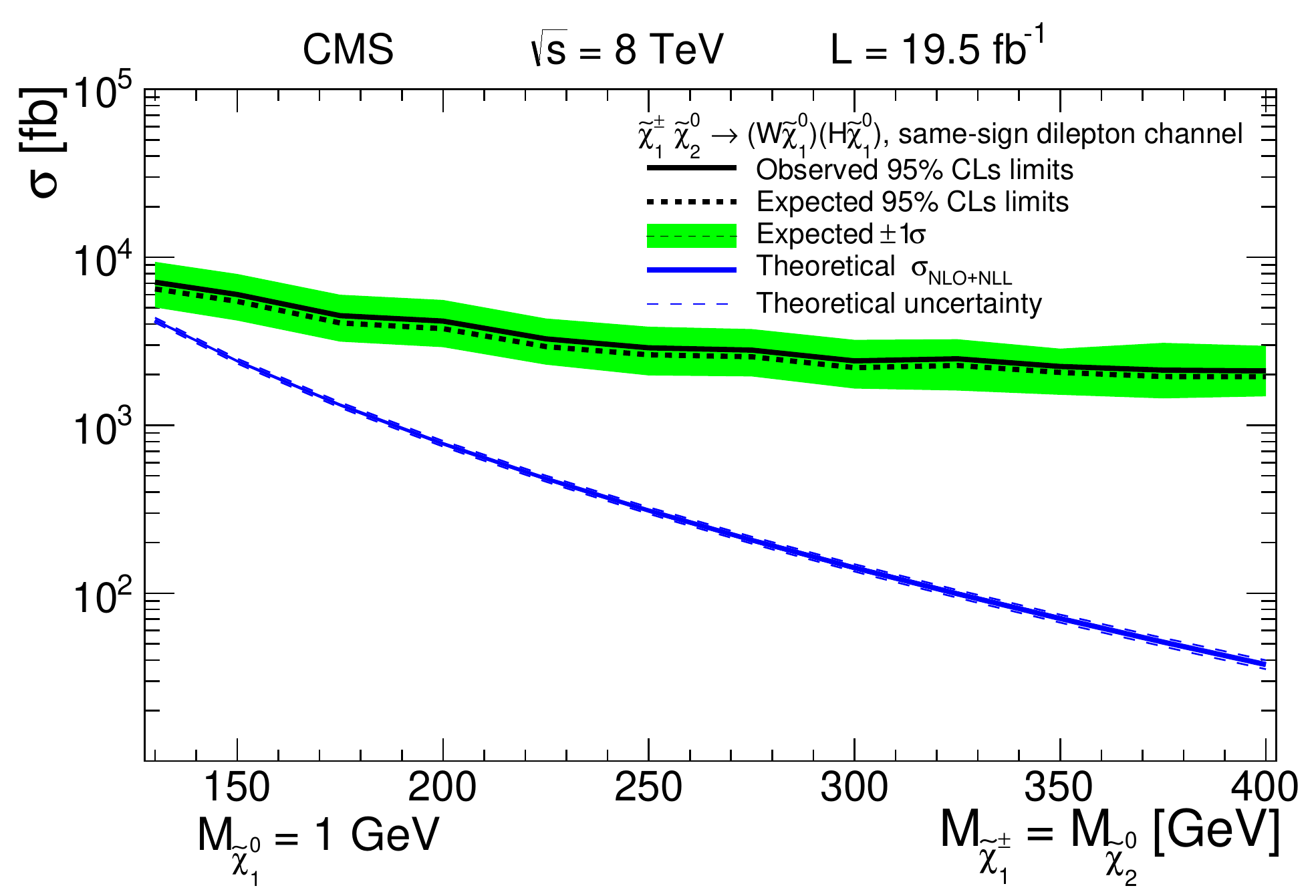}
\includegraphics[width=0.49\textwidth]{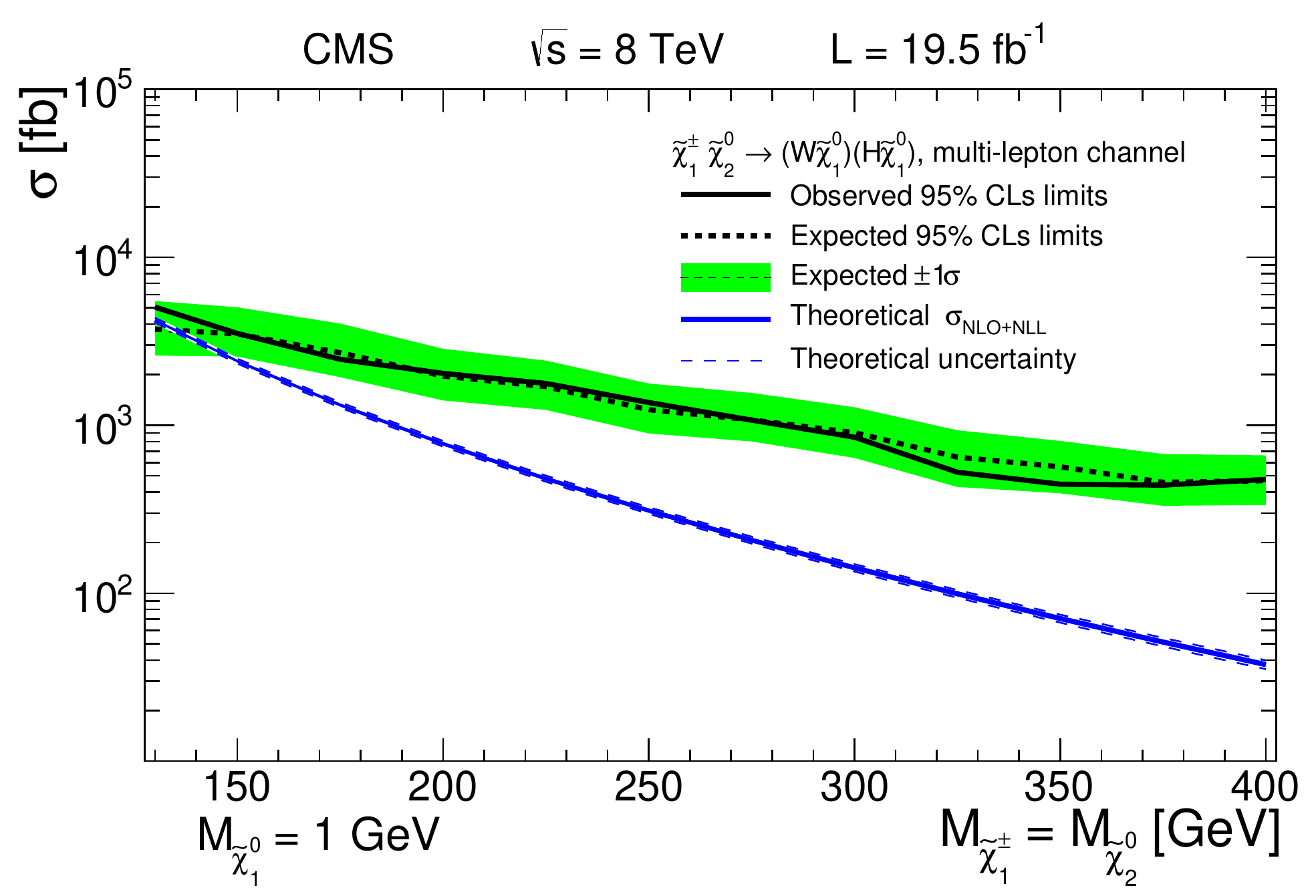}
\includegraphics[width=0.49\textwidth]{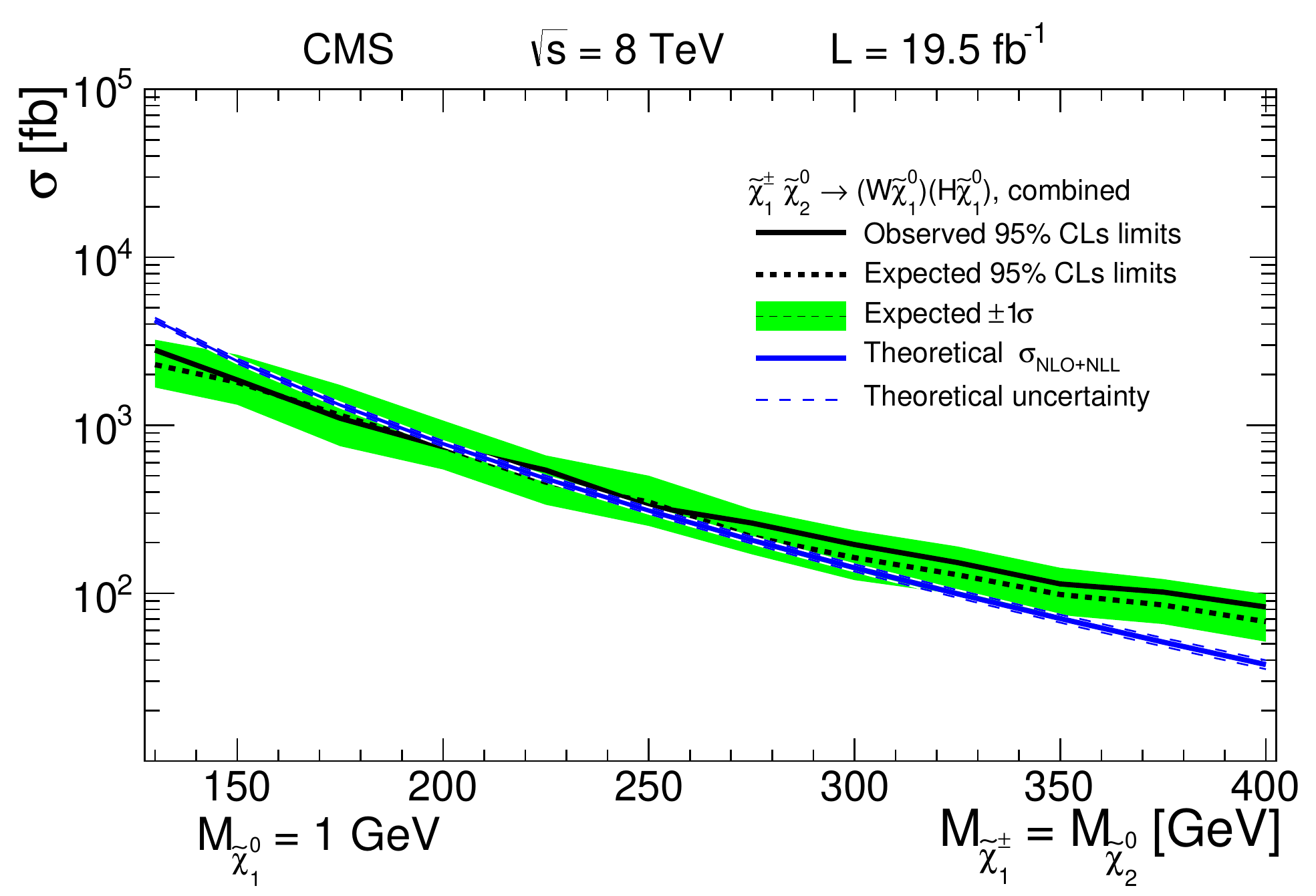} \caption{ The interpretations of the results from (upper left) the single-lepton search,
(upper right) the same-sign dilepton search, (lower left) the multilepton search, and (lower right) the
combination of the three searches. The black curves show the expected (dashed) and observed (solid)
limits on the \chipmo\chitn\ cross section times $\mathcal{B}(\chipmo\chitn\to\PW\PH+\MET)$.  The green band
shows the one-standard-deviation variation of the expected limit due to experimental uncertainties.
The solid blue curve shows the theoretical prediction for the cross section, with the dashed blue bands
indicating the uncertainty of the cross section calculation.
\label{fig:interpretations1d}
}
\end{figure*}

In Fig.~\ref{fig:interpretations1d}, the cross section upper limits for the \whmet\ signal model are presented as a function of \mchi,
for a fixed mass $\mlsp=1$\GeV, both individually from the three search regions and their combination.

\cleardoublepage \section{The CMS Collaboration \label{app:collab}}\begin{sloppypar}\hyphenpenalty=5000\widowpenalty=500\clubpenalty=5000\input{SUS-13-006-authorlist.tex}\end{sloppypar}
\end{document}

%% file: SUS-13-006-authorlist.tex
\textbf{Yerevan Physics Institute,  Yerevan,  Armenia}\\*[0pt]
V.~Khachatryan, A.M.~Sirunyan, A.~Tumasyan
\vskip\cmsinstskip
\textbf{Institut f\"{u}r Hochenergiephysik der OeAW,  Wien,  Austria}\\*[0pt]
W.~Adam, T.~Bergauer, M.~Dragicevic, J.~Er\"{o}, C.~Fabjan\cmsAuthorMark{1}, M.~Friedl, R.~Fr\"{u}hwirth\cmsAuthorMark{1}, V.M.~Ghete, C.~Hartl, N.~H\"{o}rmann, J.~Hrubec, M.~Jeitler\cmsAuthorMark{1}, W.~Kiesenhofer, V.~Kn\"{u}nz, M.~Krammer\cmsAuthorMark{1}, I.~Kr\"{a}tschmer, D.~Liko, I.~Mikulec, D.~Rabady\cmsAuthorMark{2}, B.~Rahbaran, H.~Rohringer, R.~Sch\"{o}fbeck, J.~Strauss, A.~Taurok, W.~Treberer-Treberspurg, W.~Waltenberger, C.-E.~Wulz\cmsAuthorMark{1}
% ...
\vskip\cmsinstskip
% ...
1:~~Also at Vienna University of Technology, Vienna, Austria\\
2:~~Also at CERN, European Organization for Nuclear Research, Geneva, Switzerland\\
% ...
